\documentclass[%
reprint,
bibnotes,
amsmath,amssymb,
aps,superscriptaddress
]{revtex4-2}
\usepackage{booktabs}
\usepackage{tabularx}
\usepackage{graphicx}
\usepackage{dcolumn}
\usepackage{array,tabularx}
\usepackage[table]{xcolor} 
\newcolumntype{C}{>{\centering\arraybackslash}X}
\newcolumntype{P}{>{\centering\arraybackslash}p{1.2cm}}
\usepackage{bm}
\usepackage{graphicx,bbm,amsthm}
\usepackage{tabularx,booktabs}
\usepackage[pdfpagemode=UseNone,pdfstartview=FitH,colorlinks=true,linkcolor=blue,urlcolor=blue,anchorcolor=blue,citecolor=blue]{hyperref}
\renewcommand{\theequation}{\arabic{equation}}

\usepackage{multirow}
\usepackage{braket}
\usepackage{array}
\usepackage{amsmath}
\usepackage{tabularx,float}
\begin{document}
	
	\title{Theory and Experiment of Chirality-induced Magnetic Nonreciprocity Manifested by Coupling Phase}
	
	\author{Jiguang~Yao}
	\affiliation{Department of Physics and Astronomy, University of Manitoba, Winnipeg, Canada R3T 2N2}
	
	\author{Ying~Yang}
	\affiliation{School of Applied and Engineering Physics, Cornell University, Ithaca, New York 14853, USA }
	
	\author{Chenyang~Lu}
	\affiliation{Department of Physics and Astronomy, University of Manitoba, Winnipeg, Canada R3T 2N2}
	
	\author{Lihua~Zhong}
	\affiliation{Department of Physics and Astronomy, University of Manitoba, Winnipeg, Canada R3T 2N2}
	\affiliation{State Key Laboratory of Infrared Physics, Shanghai Institute of Technical Physics, Chinese Academy of Sciences, Shanghai 200083, China }

	\author{Xiaolong~Fan}
    \affiliation{The Key Laboratory for Magnetism and Magnetic Materials of the Ministry of Education, Lanzhou University, Lanzhou 730000, China}
	
	\author{Desheng~Xue}
    \affiliation{The Key Laboratory for Magnetism and Magnetic Materials of the Ministry of Education, Lanzhou University, Lanzhou 730000, China}
	
	\author{C.-M.~Hu}
	\email{hu@physics.umanitoba.ca; \\URL: http://www.physics.umanitoba.ca/$\sim$hu}
	\affiliation{Department of Physics and Astronomy, University of Manitoba, Winnipeg, Canada R3T 2N2}
	
	\date{\today}
	
\begin{abstract}
	
Magnetic interactions have long served as the most robust and widely used approach for realizing nonreciprocity, with an externally applied magnetic field breaking time-reversal symmetry (TRS) and chiral photon–magnon interactions introducing spatial asymmetry. In this work, we investigate the chirality mechanisms essential for magnetic nonreciprocity from a unified experimental and theoretical perspective. We begin by examining conventional chiral interactions that generate chiral electromagnetic fields through specially designed structures, and then place particular emphasis on synthetic chirality enabled by nontrivial phase accumulation in traveling-wave-mediated coupling systems. We establish a microscopic theoretical framework that maps field polarization onto the phase of a complex coupling strength and validate it with systematic experiments, thereby providing a consistent formalism that describes both conventional and synthetic chirality. Notably, we highlight the symmetry properties and the unique features of synthetic chirality that distinguish it from conventional nonreciprocal mechanisms.

\end{abstract}

\maketitle

\section{\label{sec:level1}Introduction}

Nonreciprocity enables precise one-way control of waves and information, making it possible to protect, isolate, and route signals beyond the limits of reciprocal physics. It plays a central role in technologies ranging from optical communication \cite{jalas2013and}, microwave engineering \cite{kord2020microwave}, and emerging quantum information platforms \cite{barzanjeh2025nonreciprocity}. Although nonreciprocity can be realized based on magnetic-free mechanisms, such as in the cases of parametric time modulation \cite{fang2012realizing, estep2014magnetic, sounas2017non} and nonlinear systems \cite{fan2012all,sounas2018broadband,rosario2018nonreciprocity}, the most prevalent and robust approach across the microwave-to-optical spectrum \cite{lax1962microwave} still relies on harnessing magnetic interactions with an externally applied field breaking time-reversal symmetry (TRS) \cite{caloz2018electromagnetic}. 

In the magnetic nonreciprocity regime, TRS breaking is a necessary but insufficient condition; the realization of directional asymmetry additionally requires a spatial symmetry breaking \cite{nakamura2025nonreciprocal}. Spatial asymmetry is conventionally enabled by direction-dependent chiral interactions. In the context of magnetic materials being probed by electromagnetic waves, chirality means that the helicity of spin precession and the handedness of light cannot be superimposed on their mirror images, analogous to its original geometric definition. The chirality of spin precession is inherently determined by the applied field (right-handed when viewed along the magnetic field), while the chirality of light is locked to its propagation direction in specialized structures, known as spin-momentum locking \cite{bliokh2015quantum,van2016universal,yu2023chirality,suarez2025chiral}. Because chiral light interacts preferentially with spin, this direction-locked chirality leads to propagation-direction–dependent light–matter coupling and hence nonreciprocity under a fixed magnetic field. These direction-dependent chiral interactions under broken TRS constitute the basis of conventional magnetic nonreciprocity. Chiral light is conventionally generated using structures supporting chiral modes, mainly chiral waveguides and chiral cavities. Chiral waveguides range from the microstrip lines \cite{wen1969coplanar,hines1971reciprocal,elshafiey1996full,bayard2003electromagnetic,kuanr2009nonreciprocal}, photonic crystal waveguides   \cite{sollner2015deterministic,young2015polarization}, optical fibers \cite{petersen2014chiral}, surface acoustic waves \cite{xu2020nonreciprocal}, spoof surface plasmon polaritons \cite{wu2012novel,qian2025unidirectional} to polymer-based structures \cite{sun2024inverse}. Chiral cavities include metamaterial-based chiral cavities \cite{riso2023strong}, semiconductor microcavities \cite{suarez2023spin}, Fabry-Pérot cavities \cite{yuan2021stimulated,yang2023non}, and whispering gallery mode (WGM) resonators \cite{junge2013strong,tang2019chip,gu2022generation,yang2023non,lukin2023two,ardisson2025controllable}. When the handedness of chirality becomes direction-dependent, chiral waveguides and chiral cavities can support nonreciprocal phenomena such as unidirectional transmission \cite{wen1969coplanar,hines1971reciprocal,elshafiey1996full,bayard2003electromagnetic,kuanr2009nonreciprocal,wu2012novel,li2023unidirectional,sun2024inverse,qian2025unidirectional} and direction-selective strong coupling \cite{junge2013strong,tang2019chip,zhang2020broadband,zhong2022controlling,gu2022generation,yang2023non,bourhill2023generation,ardisson2025controllable}, respectively. The basic schematics and typical transmission spectra are summarized in Table~\ref{table 1}.   

\begin{table*}[t] 	\label{table 1}
	\centering
	\caption{\textbf{Summary of chirality-induced nonreciprocity in magnon-involved TRS-broken systems, including structural chirality and synthetic chirality.} The typical $|S_{21}|$ and $|S_{12}|$ spectra are calculated as a function of the frequency detuning ($\Delta_c=\omega-\omega_c$) at zero field detuning ($\Delta_m=\omega_m-\omega_c=0$), where $\omega_{c(m)}$ is the resonant frequency of the cavity (magnon) mode. In the chiral-waveguide system, a magnon mode (purple sphere) interacts with a waveguide (yellow strip) supporting propagation-direction-dependent chiral fields (red and blue circles). The corresponding spectrum is calculated using Eq.~\ref{eq5} with $\kappa_{mp}/2\pi=1$ MHz and $\kappa_{mq}/2\pi=0$ MHz, resulting in unidirectional transmission. In the chiral-cavity system, a magnon mode interacts with a chiral cavity (green sphere with internal helix representing handedness), the handedness of which is determined by the propagation direction of the traveling wave. The corresponding spectrum is calculated using Eq.~\ref{eq12} with $\kappa_{c}/2\pi=5$ MHz, $g_{21}/2\pi=30$ MHz and $g_{12}/2\pi=0$ MHz, showing direction-selective strong coupling. In the traveling-wave-mediated system, a synthetic chiral field (mixed-color circles) arises from the combined cavity field (green dashed arrow) and traveling-wave field (red and blue dashed arrows). The corresponding spectrum is calculated using Eqs.~\ref{eq18} and \ref{eq20} with $\kappa_{m}/2\pi=1$ MHz, $\kappa_{c}/2\pi=5$ MHz, $g/2\pi=30$ MHz, $\Psi_{21}=-\pi/2$, $\Psi_{12}=\pi/2$ and $\Phi_l=\pi/2$. For all three cases, the intrinsic damping rates of the magnon mode and the cavity mode are set to $\alpha_0/2\pi = 1$ MHz and $\beta_0/2\pi = 5$ MHz, respectively.} 
	
	\begin{tabular}{c c c c c}
		\hline\hline
		\noalign{\vspace{1mm}}
		Nonreciprocity system & Setup schematic & Typical spectrum & Key features & Reference \\ [1mm]
		\hline
		\noalign{\vspace{1mm}}
		\multicolumn{4}{l}{Nonreciprocity resulting from structural chirality} \\ [1mm]
		\hline
		\parbox[c]{3cm}{Chiral waveguide} &
		\parbox[c]{3.5cm}{
			\centering 
			\vspace{2mm}
			\includegraphics[width=3cm]{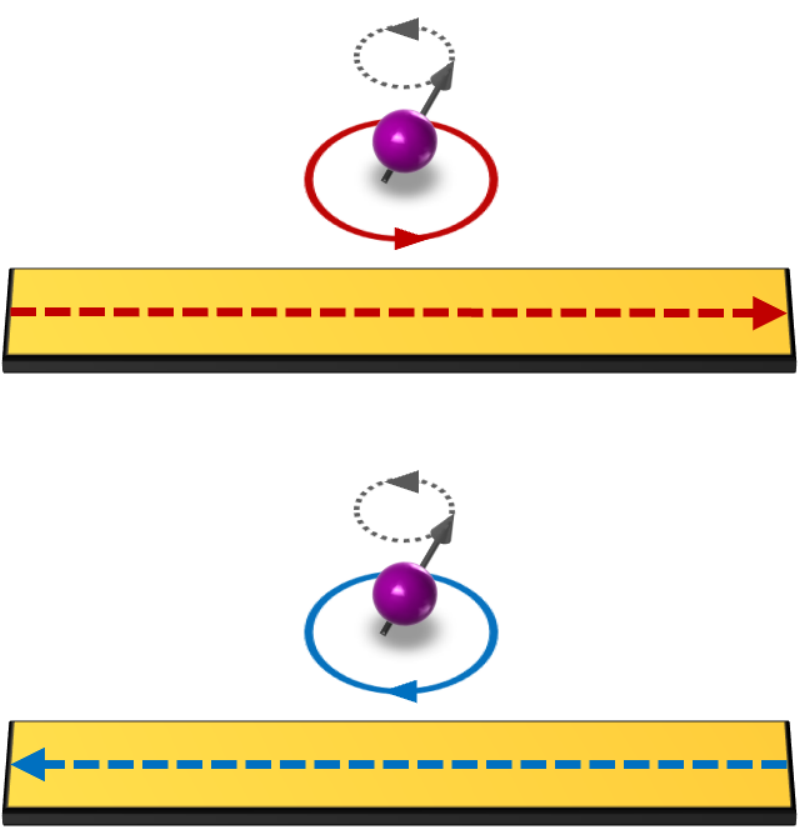}
		} \vspace{2mm} &
		\parbox[c]{3.8cm}{
			\centering
			\vspace{2mm}
			\includegraphics[width=3.5cm]{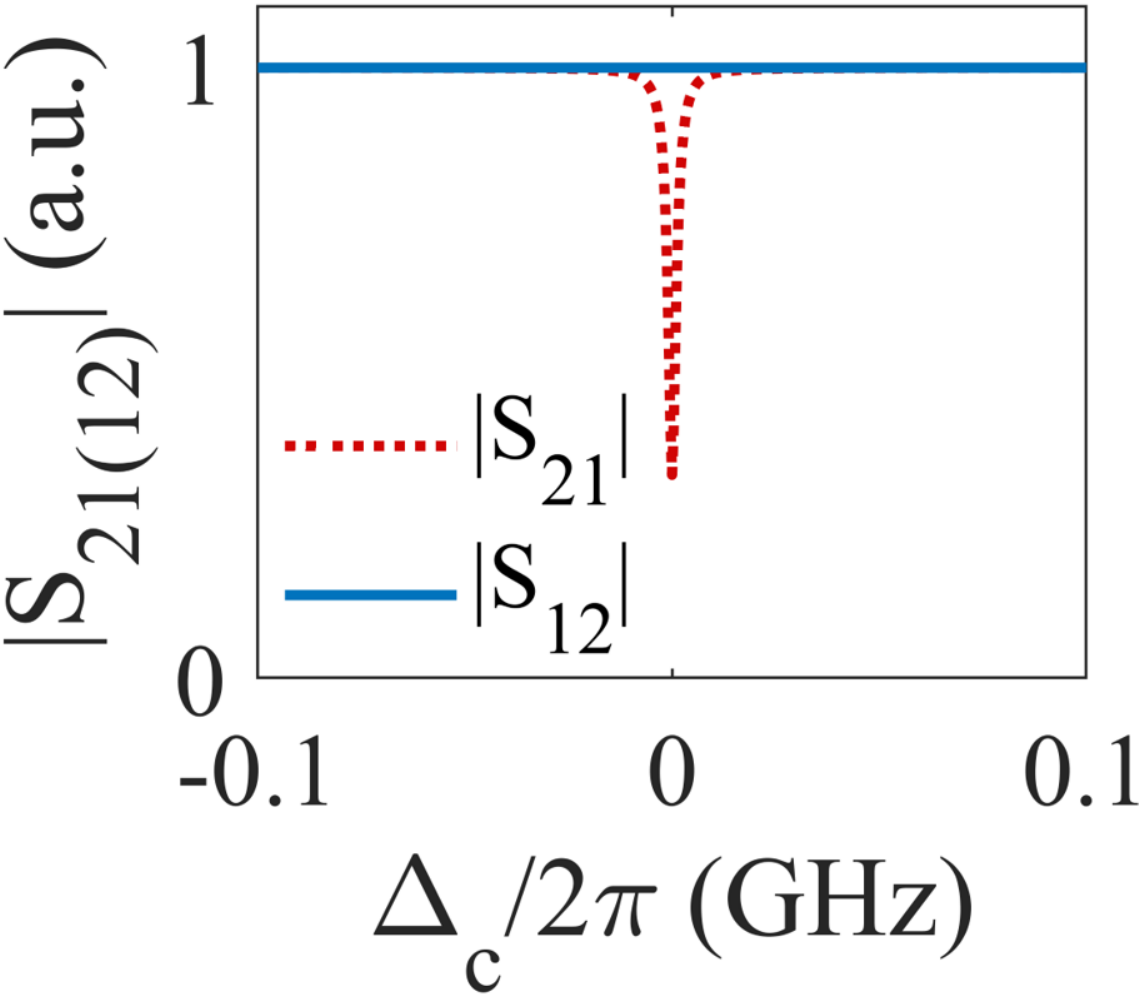}
		} \vspace{2mm} & \parbox[c]{3.5cm}{
			\begin{itemize}
				\item nonreciprocal coupling strength magnitude
				\item unidirectional transmission
			\end{itemize}
		}&\parbox[c]{3.2cm}{ \cite{wen1969coplanar,hines1971reciprocal,elshafiey1996full,bayard2003electromagnetic,kuanr2009nonreciprocal,wu2012novel,li2023unidirectional,sun2024inverse,qian2025unidirectional} } \\
		
		\parbox[c]{3cm}{Chiral cavity} &
		\parbox[c]{3.5cm}{
			\centering
			\vspace{2mm}
			\includegraphics[width=3cm]{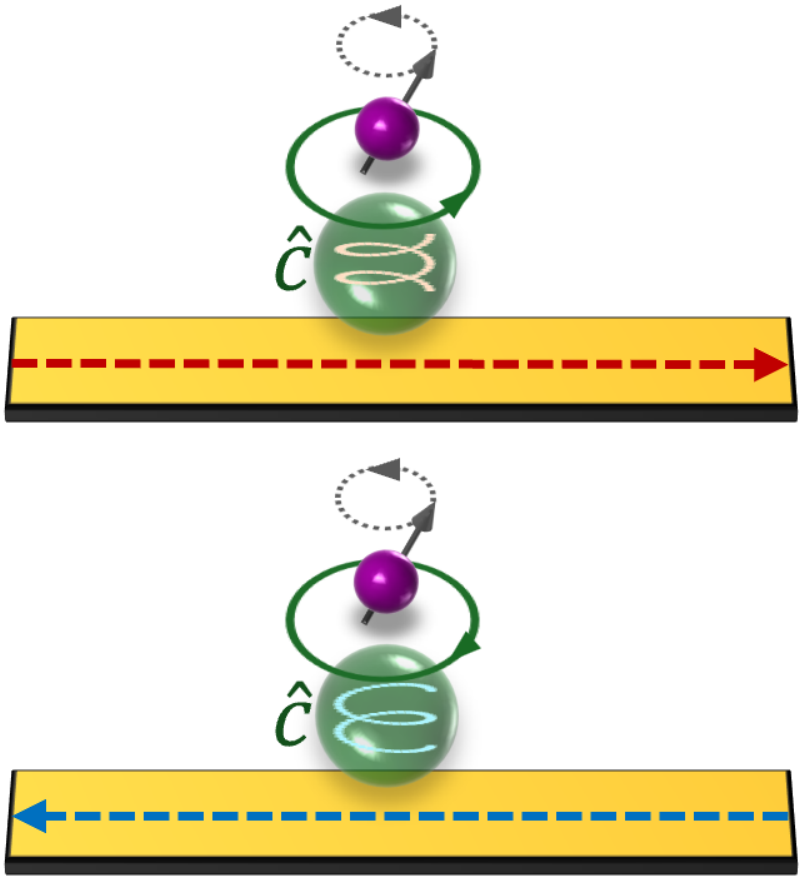}
		} \vspace{2mm}&
		\parbox[c]{3.8cm}{
			\centering
			\vspace{2mm}
			\includegraphics[width=3.5cm]{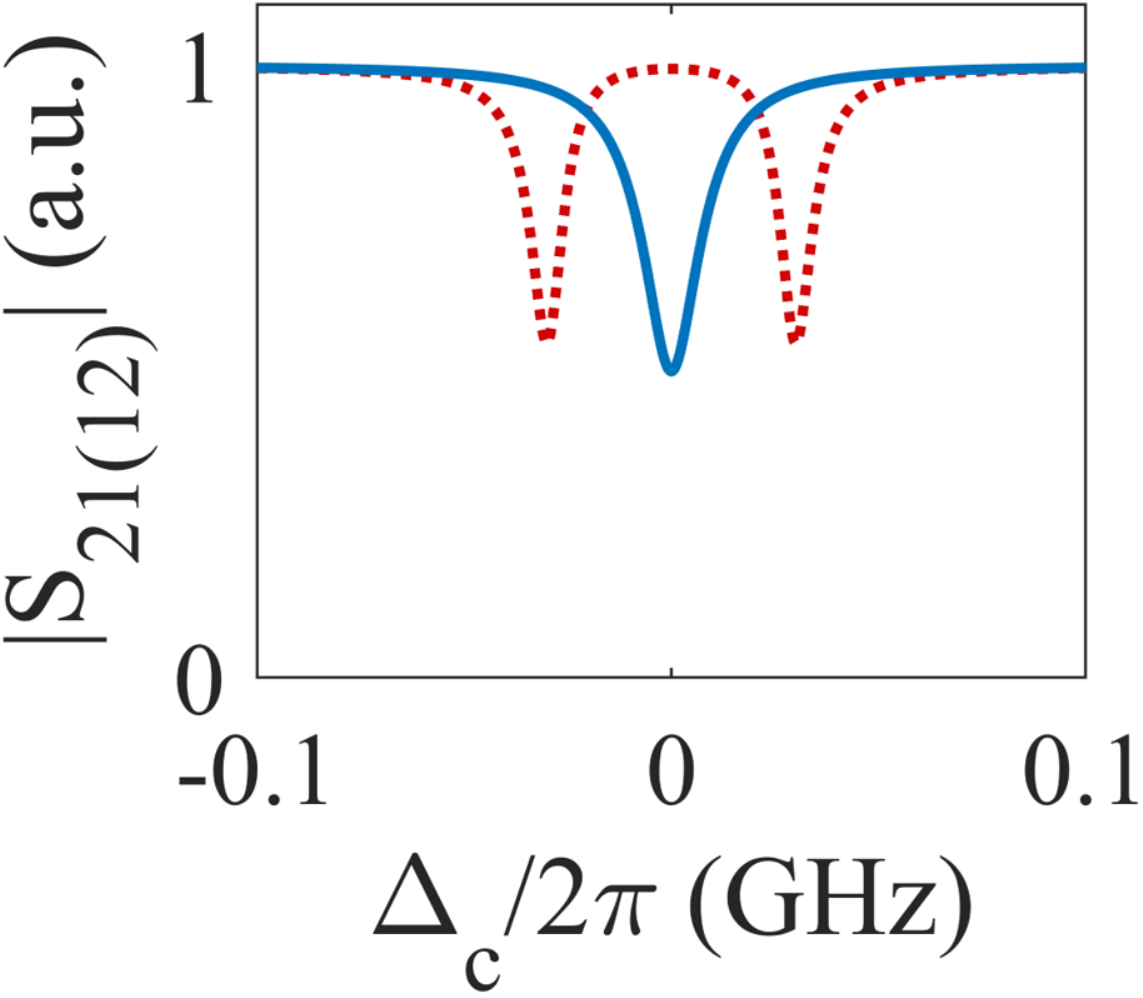}
		} \vspace{2mm} & \parbox[c]{3.5cm}{
			\begin{itemize}
				\item nonreciprocal coupling strength magnitude
				\item direction-selective strong coupling
			\end{itemize}
		}& \parbox[c]{3.2cm}{ \cite{junge2013strong,tang2019chip,gu2022generation,yang2023non,zhang2020broadband,zhong2022controlling,bourhill2023generation,ardisson2025controllable}\textsuperscript{a}} \\
		
		\hline
		\noalign{\vspace{1mm}}
		\multicolumn{4}{l}{Nonreciprocity resulting from synthetic chirality} \\ [1mm]
		\hline
		\parbox[c]{3cm}{Traveling-wave-mediated coupling} &
		\parbox[c]{3.5cm}{
			\centering
			\vspace{2mm}
			\includegraphics[width=3cm]{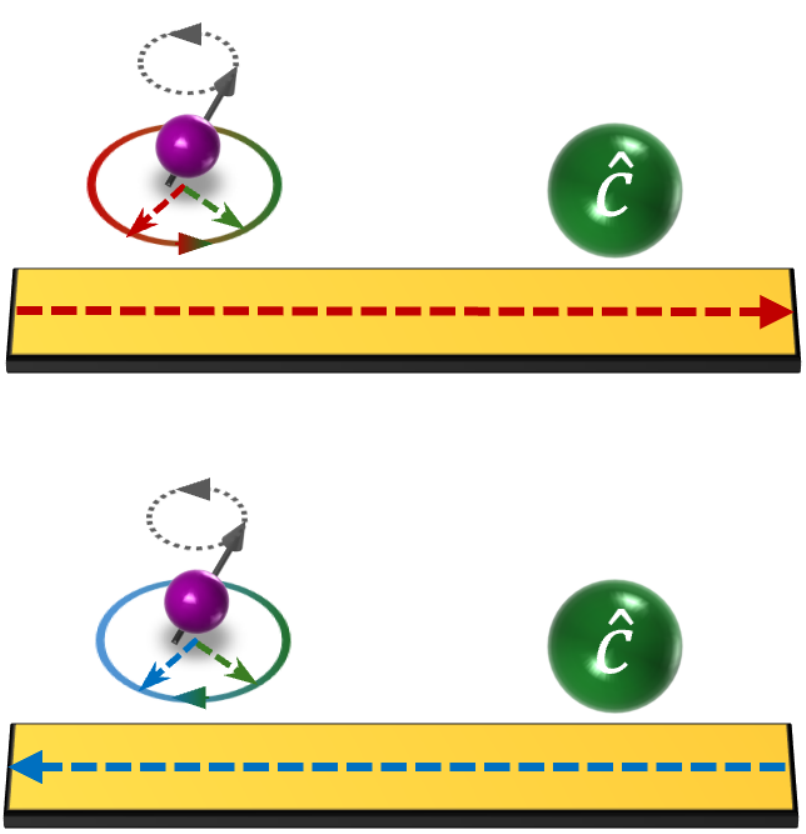}
		} \vspace{2mm} &	\parbox[c]{3.8cm}{ 
			\centering
			\vspace{2mm}
			\includegraphics[width=3.5cm]{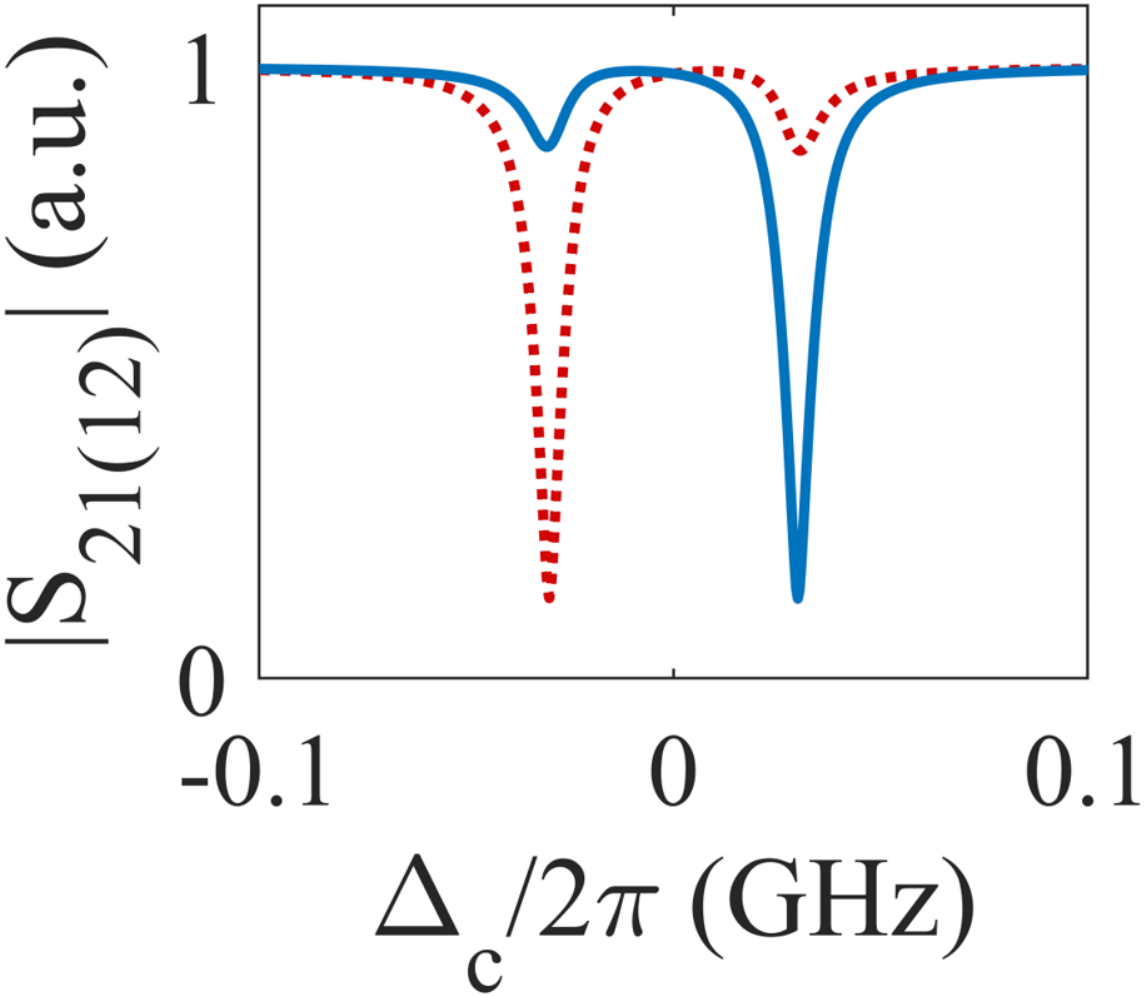}
		} \vspace{2mm} &  \parbox[c]{3.5cm}{
			\begin{itemize}
				\item directional synthetic phase
				\item unidirectional invisibility
				\item Mirror symmetry: $S_{21}(\Delta_m,\Delta_c)=S_{12}(-\Delta_m,-\Delta_c)$
			\end{itemize}
		}&\parbox[c]{3.2cm}{ \cite{wang2019nonreciprocity,qian2020manipulation,zhao2020broadband,shi2021mirror,wang2023realization,yao2025nonreciprocal} }  \\
		\hline\hline
	\end{tabular}
\vspace{2pt}
\begin{flushleft}
	\textsuperscript{a} Refs.~\cite{junge2013strong,tang2019chip,gu2022generation,yang2023non} correspond to atomic systems coupled to cavities with direction-dependent chiral electric fields.
\end{flushleft}

\end{table*}
On a different route, TRS or spatial symmetry can be broken by manipulating the coupling phase. Nontrivial phase accumulation is usually realized along a closed-loop system. A representative example is phase-modulation-induced synthetic gauge fields breaking TRS in magnetic-free nonreciprocity regimes in various platforms, ranging from photonic crystals \cite{fang2012realizing}, resonant circuits \cite{estep2014magnetic}, optomechanical systems \cite{fang2017generalized,  chen2021synthetic}, ultracold atoms \cite{lin2009synthetic,goldman2014light}, to superconducting qubits  \cite{roushan2017chiral}. Recently, such nontrivial phase manipulation has also been demonstrated in the context of magnon and cavity-photon interaction platforms known as cavity magnonics \cite{rameshti2022cavity}. In these demonstrations, phase accumulation becomes nontrivial only when magnon and cavity modes form a coupled loop system \cite{gardin2023manifestation, gardin2024engineering}, and thus has been neglected in simple coupled systems. In addition, ``common base'' can also be introduced into the coupled loop system to induce nontrivial phase, resembling reservoir engineering \cite{metelmann2015nonreciprocal,fang2017generalized} in magnetic-free regimes. It is realized based on traveling-wave-mediated coupling, which exhibits unparalleled nonreciprocal phenomena that have been attracting substantial attention, such as unidirectional invisibility  \cite{wang2019nonreciprocity,qian2020manipulation,zhao2020broadband,shi2021mirror,wang2023realization} and nonreciprocal slow-fast pulse control \cite{yao2025nonreciprocal}. However, the physical origin and role of the nontrivial phase differ fundamentally from those in magnetic-free systems. In this magnetic case, the phase accumulation stems inherently from the Zeeman coupling term between a spin and the microwave field via its polarization-dependent coupling (Eqs. \ref{eqB7} and \ref{eqB8}). Crucially, while TRS is broken directly by the applied field itself, the accumulated phase is primarily responsible for breaking spatial symmetry by forming a synthetic chirality from two linearly polarized fields associated with two coupled components (Table \ref{table 1}), thereby enabling the definitive directionality required for nonreciprocity.

The goal of this paper is to develop a unified picture of magnetic nonreciprocity, encompassing both conventional nonreciprocity arising from structural chirality and the emerging form enabled by nontrivial phase manipulation. We establish a microscopic theoretical framework that maps field polarization onto the phase of a complex coupling strength within a well-defined spin coordinate system. This allows us to describe nonreciprocity from the perspectives of both structural and synthetic chirality, and to clarify the unique features and new functionalities that the latter can offer. The overall structure of the paper is as follows. In Sec.~\ref{Nonreciprocity in single magnon mode}, we demonstrate the transmission of a magnon mode side-coupled to a microstrip. We present the standard procedure for incorporating microwave field polarization into the coupling strength to describe direction-dependent chiral interactions, which in turn give rise to nonreciprocity. In Sec.~\ref{Nonreciprocal coherent coupling}, we demonstrate that interaction with a chiral cavity whose handedness is determined by the propagation direction of traveling waves can lead to direction-selective strong coupling. In Sec.~\ref{Traveling wave-mediated nonreciprocal coupling}, we demonstrate a nonconventional chiral interaction enabled by nontrivial phase accumulation in a common-base-involved coupled loop system, which is based on the cooperative effects of direct coherent and indirect traveling-wave-mediated couplings. 
We pay particular attention to two cases distinguished by the cavity type, which couple to the traveling wave via either magnetic-dipole or electric-dipole interaction. Special attention is given to the unique features and symmetry properties of the system. In Sec.~\ref{conclusion}, conclusions and outlooks are given.  
 
\section{Nonreciprocity in single magnon mode} \label{Nonreciprocity in single magnon mode}

 As a starting point, we investigate the nonreciprocity induced by chiral interaction between the magnon mode and a chiral waveguide. This canonical configuration, widely used in magnetic isolators, can be traced back to seminal developments in the mid-20th century \cite{wen1969coplanar,hines1971reciprocal}.  Such chiral microwave fields can be naturally generated in microstrip and waveguide structures \cite{wen1969coplanar,hines1971reciprocal,elshafiey1996full,bayard2003electromagnetic,kuanr2009nonreciprocal,yu2020chiral} and further enhanced by engineered waveguide geometries \cite{wu2012novel,qian2025unidirectional}.
 \begin{figure}[bhtp]
 	\centering 
 	\includegraphics[width=8.8 cm]{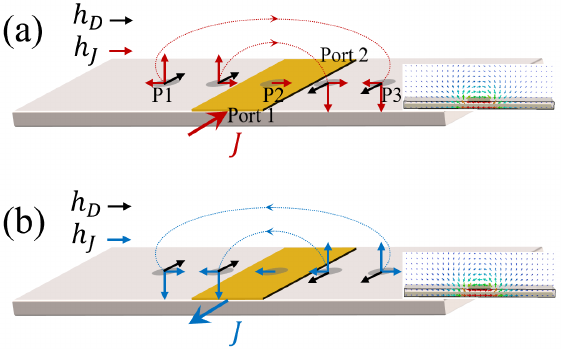}
 	\caption{Field polarization when the wave transfers (a) from port 1 to port 2 and (b) the opposite direction. The red and blue arrows represent the Oersted field $\mathbf{h}_J$ from the conduction current $J$, and the black arrows denote the microwave field $\mathbf{h}_D$ from the displacement current. Insets illustrate the simulation of $\mathbf{h}_J$.}
 	\label{fig1}
 \end{figure}
\begin{figure*}[htbp]
	\centering 
	\includegraphics[width=0.9\textwidth]{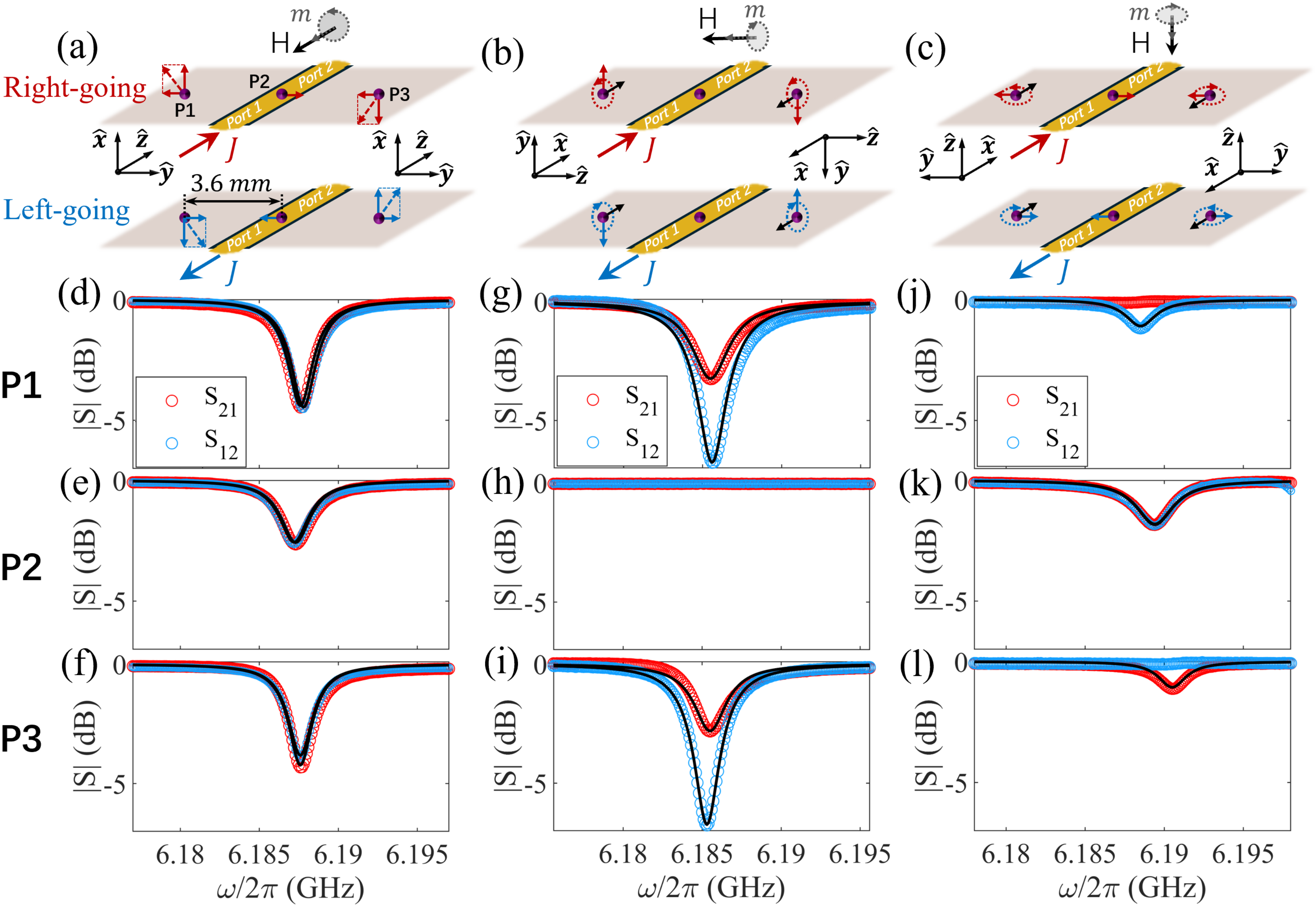}
	\caption{(a)-(c) Distributions of the effective field $\mathbf{h}_{\text{eff}}$ at $t=0$ for both propagation directions at three representative positions ($P1$, $P2$, $P3$) when the magnetic field ($\mathbf{H}$) is applied (a) along, (b) perpendicular across, and (c) normal to the microstrip. The arrows on the purple YIG sphere represent the effective microwave field, i.e., the projection of the microwave field in Fig. \ref{fig1} onto the plane perpendicular to the magnetization (chosen as $x-y$ plane), which further forms linearly (dashed arrows) or circularly/elliptically (dashed circles) polarized effective fields. The gray circles at the top represent the handedness of spin precession. (d)-(l) The measured transmission spectra $|S_{21}|$ (red circles) and $|S_{12}|$ (blue circles) at positions (d),(g),(j) $P1$, (e),(h),(k) $P2$, and (f),(i),(l) $P3$ under the three applied-field configurations. The black curves are calculated using Eq.~\ref{eq5} with the fitted parameters listed in Table~\ref{tableG1} in Appendix~\ref{Calibration of the parameters of YIG and DR}. }
	\label{fig2}
\end{figure*}

Here, we demonstrate nonreciprocity based on this typical chiral interaction in a side-coupled magnon-waveguide system by placing a YIG sphere beside a microstrip. The YIG sphere, with a diameter of 1 mm, is biased by an applied magnetic field $H$, which tunes the resonant frequency of the Kittel magnon mode as $\omega_m/2\pi=\gamma_e\mu_0(|H|+H_A)$. Here, $\gamma_e$ is the gyromagnetic ratio, $\mu_0$ is the vacuum permeability, and $H_A$ is the anisotropy field. The microstrip is 4.65 mm wide and 60 mm long, fabricated on a Rogers RT5880 substrate, and impedance-matched to 50 $\Omega$ at both ports. The chirality of the effective microwave field interacting with the YIG sphere is controlled by two factors: the position relative to the microstrip and the direction of the applied magnetic field.
On the one hand, the microwave field near the air-substrate interface is the superposition of the Oersted field $\mathbf{h}_J$, produced by the longitudinal conduction current, and $\mathbf{h}_D$, oriented along the microstrip and resulting from the displacement current \cite{bayard2003electromagnetic}. The microwave field of the traveling wave is given by $\mathbf{h}(t) = (\mathbf{h}_D+e^{i\pi/2}\mathbf{h}_J)e^{-i\omega t}$, where $\omega$ is the driving frequency and $\pi/2$ is the phase lag of $\mathbf{h}_J$ relative to $\mathbf{h}_D$ (see Appendix~\ref{Spin-momentum locking in microstrip} for details). As a result, the field becomes elliptically or circularly polarized at specific positions relative to the microstrip. The spatial distributions of the microwave magnetic field at $t=0$ for waves propagating from port 1 to port 2 and from port 2 to port 1 are shown in Figs.~\ref{fig1}(a) and (b), respectively, where the field polarization, and thus the resulting handedness, is locked to the propagation direction, evidencing spin–momentum locking.

On the other hand, the microwave field $\mathbf{h}(t)$ interacts with the magnon mode via its effective component $\mathbf{h}_{\text{eff}}(t)$ projected onto the plane perpendicular to the magnetization. To maintain a standard mathematical form for the magnon Hamiltonian and the Zeeman interaction (see Appendix~\ref{Interaction of magnon mode with traveling wave} for details), we define a spin coordinate system in which the $z$-axis is parallel to the spin angular momentum $\mathbf{S}_z$ [which is antiparallel to the magnetization and the applied field, as shown in Figs.~\ref{fig2}(a)-(c)]. The choice of the $x–y$ axes is arbitrary and is usually taken along a specific structural orientation or microwave field for simplicity. The effective field in this frame is
\begin{equation}\label{eq1}
	\mathbf{h}_{\text{eff}} (t) = (h_x\hat{x} + e^{i\varphi_r}h_y\hat{y})e^{-i\omega t},
\end{equation}
where $\varphi_r$ is the relative phase between the $x$- and $y$-components. Specifically, $\varphi_r=0$ corresponds to a linearly polarized field, while $\varphi_r=\frac{\pi}{2}$ and $\varphi_r=-\frac{\pi}{2}$ respectively correspond to left- and right-handed circularly polarized fields when viewed from the $+z$-axis toward the origin along the direction of the applied field. The chirality of $\mathbf{h}_{\text{eff}}(t)$, controlled by $\varphi_r$, varies with the orientation of the applied field, as shown in Figs.~\ref{fig2}(a)-(c).

 The interaction Hamiltonian between the microwave field and the magnon mode is written as
 \begin{equation}\label{eq2}
 		\hat{H}_{\text{int}} = \int dk \hbar\tilde{\lambda}_{mp} \hat{m}^\dagger\hat{p}_k 
 		+ \int dk \hbar\tilde{\lambda}_{mq} \hat{m}^\dagger\hat{q}_k + \text{h.c.}.
 \end{equation}
Here, $\hat{m}$ ($\hat{m}^{\dagger}$) is the annihilation (creation) operator of the magnon mode. $\hat{p}_k$ ($\hat{p}_k^{\dagger}$) and $\hat{q}_k$ ($\hat{q}_k^{\dagger}$) denote the annihilation (creation) operators for the right- and left-going traveling-wave modes, respectively. $\text{h.c.}$ denotes the Hermitian conjugate. The coupling strength $\tilde{\lambda}_{mp(q)}$ between the magnon mode and the right- (left-)going traveling wave is determined by the effective microwave field as (see Appendix~\ref{Interaction of magnon mode with traveling wave} for details)
\begin{equation}\label{eq3}
		\tilde{\lambda}_{m}=\lambda_m e^{i\Phi_m}\propto h_x+i e^{i\varphi_r} h_y,
\end{equation}
where $\lambda_m$ and $\Phi_m$ represent the magnitude and the phase of the coupling strength, respectively. For a circularly polarized field ($\varphi_r=\pm\frac{\pi}{2}$), the coupling phase is $\Phi_m = 0$; for a linearly polarized field ($\varphi_r=0$), $\Phi_m$ represents the angle between the $x$-axis and the microwave field and is inconsequential to the physical results (see Appendix~\ref{Interaction of magnon mode with traveling wave} for details).

The total Hamiltonian is then given by
\begin{equation}\label{eq4}
 \hat{H} = \hbar \tilde{\omega}_m \hat{m}^\dagger \hat{m} + \int dk \hbar \omega_k \hat{p}^\dagger_k \hat{p}_k + \int dk \hbar \omega_k \hat{q}^\dagger_k \hat{q}_k+ \hat{H}_{\text{int}}.
\end{equation} 
Here, the first term corresponds to the Kittel magnon mode with $\tilde{\omega}_m = \omega_m - i\alpha_0$, where $\omega_m$ and $\alpha_0$ represent the magnon resonant frequency and the intrinsic damping rate, respectively, while the following two terms correspond to the right- and left-going traveling-wave modes, respectively, with frequency $\omega_k$.  

Using input-output theory, the transmission coefficient is derived as (see Appendix~\ref{Transmission parameter of single magnon} for details)
\begin{equation}\label{eq5}
	\begin{split}
		S_{21(12)}(\omega)&=\frac{\omega-\omega_m+i\alpha_0+i\frac{\kappa_{mq(p)}-\kappa_{mp(q)}}{2} }
		{\omega-\omega_m+i\alpha_0+i\frac{\kappa_{mq}+\kappa_{mp}}{2}}, 
	\end{split}
\end{equation}
where
\begin{equation}\label{eq6} \kappa_{mp(q)}=2\pi \tilde{\lambda}_{mp(q)} \tilde{\lambda}_{mp(q)}^*
\end{equation}
is the extrinsic damping rate into the right- (left-)going traveling wave. The coupling phase $\Phi_{mp(q)}$ cancels out in the calculation of the extrinsic damping, meaning that only the magnitude of the coupling strength $\lambda_{mp(q)}$ matters. The counter-propagating waves generate different effective fields $\mathbf{h}_{\text{eff}}(t)$, resulting in different coupling-strength magnitudes $\lambda_{mp}\ne\lambda_{mq}$ according to Eq.~\ref{eq3}, and thus $\kappa_{mp}\ne\kappa_{mq}$ , which leads to nonreciprocity.

We perform transmission experiments by connecting the microstrip to a vector network analyzer (VNA) to investigate the position- and applied-field dependence of the nonreciprocity. The YIG sphere is sequentially positioned at three representative locations relative to the microstrip ($P1$, $P2$, and $P3$, as shown in Fig.~\ref{fig2}), where $P2$ is at the center of the microstrip, and $P1$ and $P3$ are symmetrically located on either side at the same distance $d \approx 3.6$ mm from the center. For each position, the applied field is set at three orientations: parallel to, perpendicular across, and normal to the microstrip. 

When $\mathbf{H}$ is parallel to the microstrip, the measured transmission spectra $|S_{21(12)}|$ (circles) at the three positions are shown in Figs.~\ref{fig2}(d)-(f) and fitted (black curves) based on Eq.~\ref{eq5} (see Appendix~\ref{Calibration of the parameters of YIG and DR} for the detailed fitting method and fitted parameters). In this case, the displacement current field $\mathbf{h}_D$ oscillates in the same direction as the applied field so it does not contribute to the spin precession. Therefore, the YIG sample only experiences the linearly polarized ($\varphi_r=0$) Oersted field $\mathbf{h}_J$ at all three positions, for both propagation directions. The coupling-strength magnitude is $\lambda_{mp}= \lambda_{mq} \propto h_J$, resulting in $\kappa_{mp}\simeq \kappa_{mq}$; the system is therefore reciprocal.  

For the other two applied-field settings, without loss of generality, we choose a spin coordinate system with $\mathbf{h}_D \parallel \hat{x}$ for simplicity. When $\mathbf{H}$ is perpendicular across the microstrip, the measured $|S_{21}|$ spectrum is shallower than $|S_{12}|$ at both $P1$ [Fig.~\ref{fig2}(g)] and $P3$ [Fig.~\ref{fig2}(i)], with fitted values of $\kappa_{mp}/2\pi=0.53$ MHz and $\kappa_{mq}/2\pi=0.93$ MHz, and $\kappa_{mp}/2\pi=0.4$ MHz and $\kappa_{mq}/2\pi=0.8$ MHz, respectively. The asymmetry in transmission spectra is caused by the effective field $\mathbf{h}_{\text{eff}}(t)$ being left- (right-) handed elliptically polarized for the right- (left-) going wave at both P1 and P3, corresponding to $\varphi_r=\pi/2$ ($-\pi/2$). The left-going wave then has a greater coupling of $\lambda_{mq} \propto |\mathbf{h}_D|+\mathbf{h}_J \cdot \hat{y}$ to the YIG than the right-going wave with  $\lambda_{mp} \propto |\mathbf{h}_D|-\mathbf{h}_J \cdot \hat{y}$, leading to more signal energy being transferred to the spin at resonance. 
At $P2$ [Figs.~\ref{fig2}(h)], $\mathbf{h}_D=0$ and $\mathbf{h}_J$ is aligned with the applied field. Consequently, the effective microwave field driving the YIG is zero, and no coupling is observed.

In contrast, when $\mathbf{H}$ is applied normal to the plane, the handedness of the effective field $\mathbf{h}_{\text{eff}}(t)$ at $P1$ is opposite to that at $P3$, as shown in Fig.~\ref{fig2}(c). Accordingly, Fig.~\ref{fig2}(j) shows that at $P1$, $|S_{21}|$ remains nearly flat around 0 dB, while $|S_{12}|$ shows a resonant dip, which is opposite to the behavior at $P3$, as shown in Fig.~\ref{fig2}(l). The fitted extrinsic damping rates are $\kappa_{mp}/2\pi=0$ MHz and $\kappa_{mq}/2\pi=0.15$ MHz for $P1$, and $\kappa_{mp}/2\pi=0.13$ MHz and $\kappa_{mq}/2\pi=0$ MHz for $P3$, indicating that the direction-dependent chiral fields are nearly purely circularly polarized, such that only the component with the same handedness as the YIG precession can interact with the magnon mode. At $P2$, where $h_D=0$, only a linearly polarized field interacts with the magnon mode, with fitted $\kappa_{mp}/2\pi=\kappa_{mq}/2\pi=0.3$ MHz, resulting in reciprocal transmission.  

\section{Nonreciprocal strong coupling} \label{Nonreciprocal coherent coupling}

Beyond a chiral waveguide, spins can interact with chiral cavities that support preferential handedness, enabling the realization of strong coupling \cite{artman1955measurement,huebl2013high,zhang2014strongly,goryachev2014high,tabuchi2014hybridizing,bai2015spin,li2019strong,hou2019strong}. The chirality of a cavity can also be locked to its propagation direction at special planes or lines within the cavity \cite{junge2013strong,yu2020circulating}, which can lead to a direction-selective strong coupling. This phenomenon has been widely demonstrated in the coupling between atoms and whispering gallery modes (WGMs) supporting direction-dependent chiral electric fields \cite{junge2013strong,tang2019chip,gu2022generation,yang2023non}. In addition, it has been demonstrated in magnetic systems through either specialized cavity design \cite{zhang2020broadband,bourhill2023generation,ardisson2025controllable} or excitation engineering \cite{zhong2022controlling}. Ideally, a specific handedness is uniquely associated with a given propagation direction, giving rise to a Rabi splitting in one direction, while only a single resonance appears in the opposite direction. In some cases, however, the cavity chirality is not perfectly bound to the propagation direction, and the resulting response is characterized by unequal coupling strengths for the two counter-propagating modes rather than fully directional coupling. 

\begin{figure}[htbp]
	\centering 
	\includegraphics[width=8.8 cm]{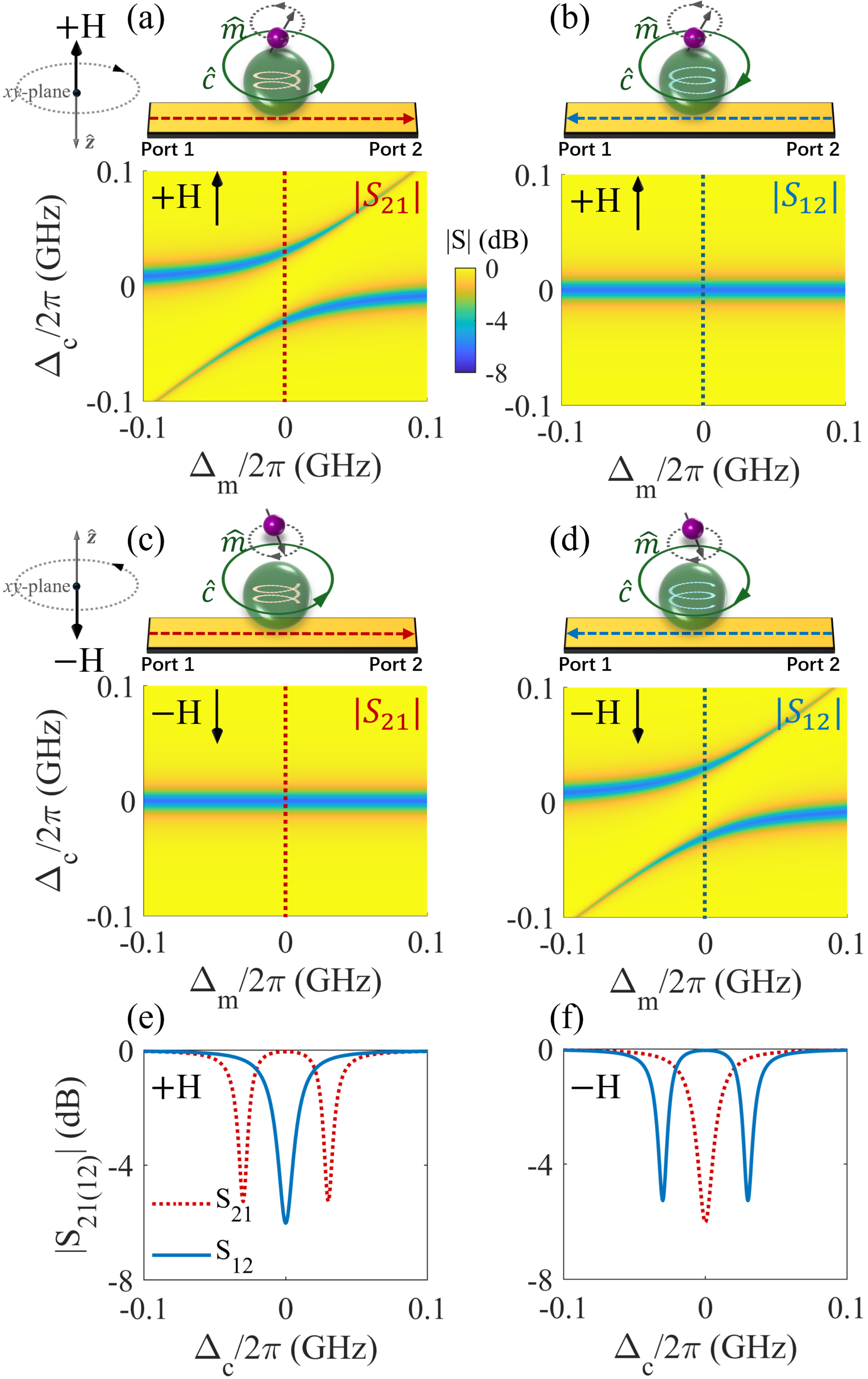}
	\caption{(a)-(d) The calculated (a),(c) $|S_{21}|$ and (b),(d) $|S_{12}|$ mappings as functions of $\Delta_m$ and $\Delta_c$ under (a),(b) positive and (c),(d) negative magnetic fields. In the relative schematics, the green and purple spheres represent the cavity mode $\hat{c}$ and the magnon mode $\hat{m}$, respectively. The green and dashed gray circles denote the handedness of the cavity and the spin precession, respectively. (e)(f) The calculated transmission spectra $|S_{21}|$ (red curves) and $|S_{12}|$ (blue curves) for (e) positive and (f) negative magnetic fields at $\Delta_m=0$, denoted as dashed lines in panels (a)-(d). All the results are calculated using Eq.~\ref{eq12} with $\alpha_0/2\pi=1$ MHz, $\beta_0/2\pi=5$ MHz, and $\kappa_{cp(q)}/2\pi=5$ MHz. In addition, we set $g_{21}/2\pi=30$ MHz, $g_{12}/2\pi=0$ MHz for the positive applied field, while $g_{21}/2\pi=0$ MHz, $g_{12}/2\pi=30$ MHz for the negative field.} 
	\label{fig3}
\end{figure}

To demonstrate the generality of our theory, we consider an ideal case of a chiral cavity that supports a direction-locked, purely chiral magnetic field when interacting with an open channel \cite{wang2025dynamic}. In our sense, a “chiral cavity” is not limited to a cavity supporting intrinsically chiral eigenmodes, but rather refers to a cavity system that supports chiral local fields through reasonable physical mechanisms, such as excitation engineering \cite{zhong2022controlling,joseph2024role}. The schematic is shown in Fig.~\ref{fig3}(a)-(d). The setup includes a cavity interacting with a magnon mode and traveling waves, and a magnon mode interacting only with the cavity. As a consequence of the direction-dependent cavity chirality, the cavity–magnon coupling becomes direction-selective. In this section, we establish a general theoretical model that captures this mechanism, focusing specifically on how the propagation-direction–dependent cavity chirality maps onto the cavity–magnon coupling strength, while omitting details of the cavity design.

Including all the components and interactions, the total Hamiltonian of the system reads
\begin{equation} \label{eq7}
		\begin{split}
	\hat{H}_{\text{tot}} =&\hbar \tilde{\omega}_c \hat{c}^\dagger\hat{c}+\hbar \tilde{\omega}_m \hat{m}^\dagger\hat{m}+\int dk \hbar \omega_k \hat{p}^\dagger_k \hat{p}_k + \int dk \hbar \omega_k \hat{q}^\dagger_k \hat{q}_k \\& +\hat{H}_{cp,q}+\hat{H}_{cm}.
	\end{split}
\end{equation}
Here, the first term corresponds to the cavity mode $\tilde{\omega}_c=\omega_c-i\beta_0$, where $\omega_c$ and $\beta_0$ represent the cavity resonant frequency and the intrinsic damping rate, respectively. The second term corresponds to the magnon mode. The third and fourth terms correspond to the right- and left-going waves, respectively. The last two terms represent the cavity-traveling-wave and the cavity-magnon interactions, respectively. 
The interaction Hamiltonian between the cavity mode and the traveling wave is written as
\begin{equation} 
	\hat{H}_{cp,q}=\int dk \hbar \tilde{\lambda}_{cp}\hat{c}^\dagger \hat{p}_k +\int dk \hbar \tilde{\lambda}_{cq}\hat{c}^\dagger \hat{q}_k +\text{h.c.},
\end{equation}
where $\tilde{\lambda}_{cp(q)}=\lambda_{cp(q)}e^{i\Phi_{cp(q)}}$ is the coupling strength between the cavity mode and the right- (left-)going traveling photon modes. Here, the cavity contains no TRS-breaking mechanism and therefore does not introduce any nonreciprocity. Furthermore, we assume an idealized interface where the coupling magnitudes to the two counter-propagating modes are identical for simplicity, which leads to $\lambda_{cp}=\lambda_{cq}=\lambda_{c}$. 
Therefore, we have identical extrinsic damping rates $\kappa_{cp}=\kappa_{cq}=\kappa_{c}$ for both traveling-wave propagation directions, where $\kappa_{c}=2\pi\tilde{\lambda}_{c}\tilde{\lambda}_{c}^*$.

The cavity mode also interacts with the magnon mode via the Hamiltonian
\begin{equation} 
	\hat{H}_{cm}=\hbar (\tilde{g}\hat{c} \hat{m}^\dagger+\tilde{g}^*\hat{c}^\dagger \hat{m}),
\end{equation} 
where the coupling strength is given by
\begin{equation} \label{eq10}
	\tilde{g}=ge^{i\Phi_g}\propto h_x+ie^{i\varphi_r}h_y, 
\end{equation}
which is determined by the cavity field
\begin{equation} \label{eq11}
 \mathbf{h}_c=(h_x \hat{x} + e^{i\varphi_r} h_y \hat{y})e^{-i\omega t},
 \end{equation}
defined in a spin coordinate system with $\mathbf{H} \parallel -\hat{\mathbf{z}}$. Specifically, when the cavity field is circularly polarized ($\varphi_r=\pm \pi/2$), $\Phi_g=0$. In addition, when $\mathbf{h}_c$ is propagation-direction dependent, $g$ becomes nonreciprocal, with $g_{21(12)}$ representing the coupling strength for waves propagating from port 1(2) to port 2(1). 

Using input-output theory, the transmission coefficients are derived as (see Appendix~\ref{Transmission parameter of nonreciprocal coherent coupling} for details):
 \begin{equation} \label{eq12}
 	S_{21(12)}(\omega)=\frac{(\omega-\tilde{\omega}_m)(\omega-\tilde{\omega}_c)-g_{21(12)}^2}
 	{(\omega-\tilde{\omega}_m)(\omega-\tilde{\omega}_c+i\kappa_{c})-g_{21(12)}^2},
 \end{equation} 
showing that nonreciprocal transmission results directly from the inequality $g_{21} \neq g_{12}$.

For simplicity, we demonstrate a purely circularly polarized case by setting $h_x=h_y=h$. Under a positive applied field $+\mathbf{H}$, $\mathbf{h}_c$ is assumed to be right-handed (viewed along the applied field) when the wave propagates from port 1 to port 2, while it becomes left-handed when the wave propagates from port 2 to port 1, as shown in the schematics in Figs.~\ref{fig3}(a) and (b). The former has the same chirality as the spin precession, corresponding to $\varphi_r=-\pi/2$, and gives $g \propto 2h$, resulting in strong coupling when $g>\alpha_0$ and $g>\beta_0$. In contrast, the latter has a chirality opposite to the spin precession, corresponding to $\varphi_r=\pi/2$ and $g = 0$, thus decoupling from the magnon mode. Accordingly, the calculated results using Eq.~\ref{eq12} in Figs.~\ref{fig3}(a) and (b) show a level repulsion in the $|S_{21}|$ mapping and a bare cavity mode in the $|S_{12}|$ mapping, as functions of frequency detuning ($\Delta_c=\omega-\omega_c$) and field detuning ($\Delta_m=\omega_m-\omega_c$). At $\Delta_m=0$ [Fig.~\ref{fig3}(e)], $|S_{21}|$ displays two split resonances with equal amplitude, compared to a bare cavity spectrum in $|S_{12}|$. When the magnetic field is inverted, the cavity field polarization is mapped onto the new spin-coordinate system with $\varphi_r=+\pi/2 \ (-\pi/2)$ for waves propagating from port 1 (2) to port 2 (1), as shown in Figs.~\ref{fig3}(c) and (d). This leads to an exchange of $g_{21}$ and $g_{12}$, resulting in the reversal of $|S_{21}|$ and $|S_{12}|$, as shown in Figs.~\ref{fig3}(c), (d), and (f).   

Recently, a representative cavity magnonics experiment demonstrated a tunable two-port cavity in which the cavity field polarization can be controlled via excitation vector fields. While the study did not address nonreciprocity or direction-selective coupling, it adopted a theoretical framework similar to ours by expressing the effect of cavity field handedness through the effective cavity–magnon coupling strength \cite{joseph2024role}.

\section{Traveling wave-mediated nonreciprocal coupling} \label{Traveling wave-mediated nonreciprocal coupling}
\begin{figure}[bhtp]
	\centering 
	\includegraphics[width=8.8 cm]{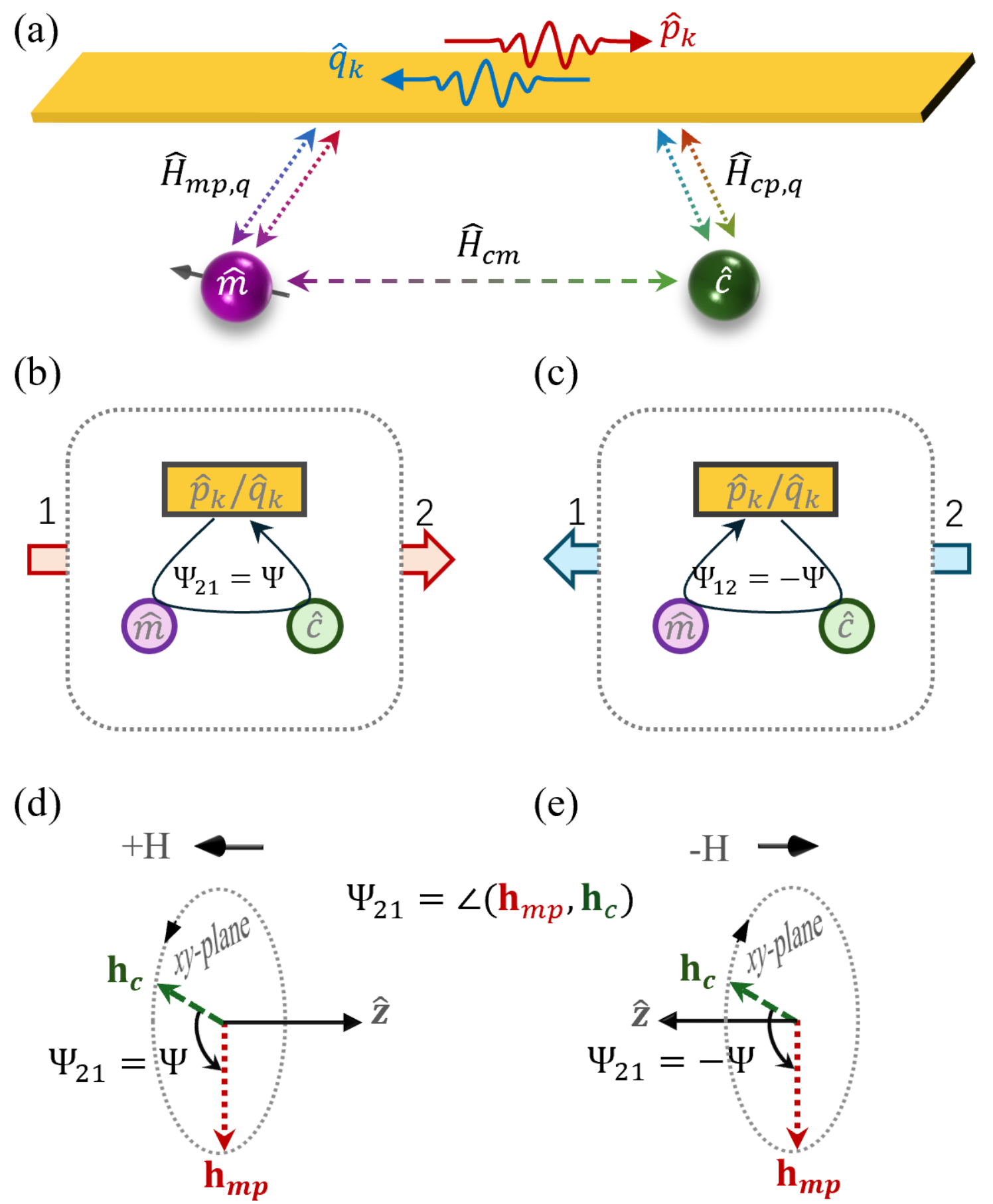}
	\caption{(a) Experimental setup: a magnon mode $\hat{m}$ and a cavity mode $\hat{c}$ are directly coupled via a Hamiltonian $\hat{H}_{cm}$. Both modes also couple to a common microstrip through the Hamiltonians $\hat{H}_{mp,q}$ and $\hat{H}_{cp,q}$, respectively. Consequently, $\hat{c}$ and $\hat{m}$ are indirectly coupled through the right- and left-going traveling-wave modes $\hat{p}_k$ and $\hat{q}_k$. The three interactions form a closed interaction loop. (b),(c) The synthetic phase $\Psi_{21(12)}$ accumulated along the closed interaction loop when the wave propagates from (b) port 1 to port 2, and from (c) port 2 to port 1. (d),(e) Synthetic phase $\Psi_{21}$ under (d) positive and (e) negative applied fields. Flipping the $z$-axis of the spin coordinate system reverses the sign of the synthetic phase. The red and green arrows represent the Oersted field $\mathbf{h}_{mp}$ generated by the right-going traveling wave and the cavity field $\mathbf{h}_c$, respectively.}
	\label{fig4}
\end{figure}

In this section, we demonstrate a form of nonreciprocity that results from a nonconventional chiral interaction based on closed-loop coupling systems instead of specialized architectures that support chiral light. Figure~\ref{fig4}(a) sketches the experimental configuration: a magnon mode and a cavity mode are coherently coupled through a direct field overlap, described by the Hamiltonian $\hat{H}_{cm}$. Simultaneously, both modes couple to a common microstrip that supports right- ($\hat{p}_k$) and left-going ($\hat{q}_k$) traveling waves, with the magnon-traveling-wave and cavity-traveling-wave interactions described by $\hat{H}_{mp,q}$ and $\hat{H}_{cp,q}$, respectively. These interactions give rise to a traveling-wave-mediated coupling \cite{lalumiere2013input,van2013photon} composed of coherent and dissipative components \cite{harder2018level,yang2019control,yu2019prediction,bhoi2019abnormal,wang2020dissipative,harder2021coherent,lu2025temporal}, both of which depend on the separation distance between the two modes.

The total Hamiltonian of the system is written as 
\begin{equation} 
	\hat{H}_{\text{tot}}=\hbar \tilde{\omega}_c \hat{c}^\dagger\hat{c}+\hbar \tilde{\omega}_m \hat{m}^\dagger\hat{m}+\int dk \hbar \omega_k (\hat{p}^\dagger_k \hat{p}_k+\hat{q}^\dagger_k \hat{q}_k)+\hat{H}_{\text{int}}.
\end{equation}
Here, the first two terms describe the cavity mode (magnetic component) and magnon mode. The third term represents the right- ($\hat{p}_k$) and left-going ($\hat{q}_k$) traveling-wave photon modes. The interaction Hamiltonian is decomposed as 
 \begin{equation} 
	\hat{H}_{\text{int}}=\hat{H}_{cm}+\hat{H}_{mp,q}+\hat{H}_{cp,q}.
\end{equation}
Here, the cavity-magnon interaction $\hat{H}_{cm}$ and the magnon-traveling-wave coupling $\hat{H}_{mp,q}$ originate from the interaction of the effective microwave magnetic fields with the spin, whereas the cavity-traveling-wave coupling $\hat{H}_{cp,q}$ arises from the electric- or magnetic-dipole interaction between the cavity mode and the propagating microwave fields.

All the microwave-field polarizations are defined in a spin coordinate system where the $z$-axis is aligned with the spin angular momentum $\mathbf{S}_z$ (opposite to the applied field), and the in-plane $x$-$y$ axes can be chosen arbitrarily. In this case, the applied field is set along the microstrip so that the YIG sphere only experiences a linearly polarized field. As discussed in Sec.~\ref{Nonreciprocity in single magnon mode}, for a linearly polarized field, the coupling phase of the magnon mode with the effective field is defined as the angle between the field direction and the
$x$-axis (see Appendix~\ref{Interaction of magnon mode with traveling wave} for details). Consequently, the coupling phase values depend on the chosen coordinate frame. Nevertheless, the transmission coefficients depend only on a specific combination of these phases, which remains invariant under coordinate rotations in the $xy$-plane (see Appendix \ref{Unitary transformation and gauge-invariant phase} for details). Thus, the predicted physical outcome is independent of the choice of $x$-$y$ axes. 

The interaction between the magnon mode and the cavity mode is written as
\begin{equation} 
	\hat{H}_{cm}=\hbar (\tilde{g}\hat{c} \hat{m}^\dagger+\tilde{g}^*\hat{c}^\dagger \hat{m}),
\end{equation}
where $\tilde{g}=ge^{i\Phi_g}$ is the direct coupling strength arising from the cavity–magnon field overlap. Depending on its interaction with the traveling wave, the cavity mode can be modeled as either a magnetic or an electric dipole, which ultimately provides the local cavity magnetic field $\mathbf{h}_c$ driving the magnons. The coupling phase $\Phi_g$ is determined by the angle between the cavity magnetic field $\mathbf{h}_c$ and the $x$-axis of the spin coordinate system. 

The magnon also interacts with the effective microwave magnetic fields from the right- and left-going traveling waves via \cite{lalumiere2013input,kannan2020generating,sinha2020non,kannan2020waveguide,wang2022giant,wang2025interpreting,wang2025dynamic}: 
\begin{equation}\label{eq16}
		\hat{H}_{mp,q} = \int dk \hbar e^{-i\Phi_l} \tilde{\lambda}_{mp} \hat{m}^\dagger\hat{p}_k 
		+\int dk \hbar e^{i\Phi_l} \tilde{\lambda}_{mq} \hat{m}^\dagger\hat{q}_k+ \text{h.c.}.\\
\end{equation}
Here, $\Phi_l=\frac{2\pi l}{\lambda}$ is the traveling phase, where $\lambda$ is the wavelength and $l$ is the separation distance between the YIG sphere and the cavity. The coupling strength is given by $\tilde{\lambda}_{mp(q)}=\lambda_{mp(q)}e^{i\Phi_{mp(q)}}$, where $\Phi_{mp(q)}$ is set by the angle between the Oersted field $\mathbf{h}_{mp(q)}$ generated by the right- (left-)going traveling wave at the YIG position and the $x$-axis of the spin coordinate system.  Moreover, $\Phi_{mp}-\Phi_{mq}=\pi \;(\mathrm{mod}\; 2\pi)$, because counter-propagating waves generate Oersted ﬁelds in opposite directions (see Appendix \ref{Transmission parameter of coupled system} for details).

The interaction between the cavity mode and the right- and left-going traveling waves reads
\begin{equation} \label{eq17}
	\hat{H}_{cp,q}=\int dk \hbar \tilde{\lambda}_{cp}\hat{c}^\dagger \hat{p}_k +\int dk \hbar \tilde{\lambda}_{cq}\hat{c}^\dagger \hat{q}_k+ \text{h.c.},
\end{equation}
where $\tilde{\lambda}_{cp(q)}=\lambda_{cp(q)}e^{i\Phi_{cp(q)}}$. Here, the coupling phases $\Phi_{cp}$ and $\Phi_{cq}$ are determined by the interaction type of the cavity with the traveling waves. For magnetic-dipole interactions, we set $\Phi_{cp}=0$ and $\Phi_{cq}=\pi$, accounting for the $\pi$-phase shift in the microwave field upon reversal of the propagation direction of the traveling wave. For electric-dipole interactions, we set $\Phi_{cp}=\Phi_{cq}=-\pi/2$, since the energy is first transferred to the electric component (which is intrinsically phase-shifted by $\pi/2$ relative to the magnetic component and the operator $\hat{c}$) of the cavity field, and the electric field of the traveling wave is independent of the propagation direction (see Appendix~\ref{Interaction of cavity mode with traveling wave} for details). 

Using input-output theory, the transmission coefficients of the system are derived as (see Appendix~\ref{Transmission parameter of coupled system} for details)
\begin{equation} \label{eq18}
	S_{21(12)}(\omega)=\frac{(\omega-\tilde{\omega}_m)(\omega-\tilde{\omega}_c)-\tilde{C}_{21(12)}}
	{(\omega-\tilde{\omega}_m+i\kappa_m)(\omega-\tilde{\omega}_c+i\kappa_c)-\tilde{D}},
\end{equation}
where $\tilde{D}=[\tilde{g}^*+\tilde{G}_le^{i(\Phi_{cp}-\Phi_{mp})}][\tilde{g}+\tilde{G}_le^{i(-\Phi_{cq}+\Phi_{mq})}]$. Here,  $\tilde{G}_l=-i\sqrt{\kappa_c\kappa_m}e^{i\Phi_l}$ describes the traveling-wave-mediated coupling strength, with its real and imaginary parts contributing to coherent and dissipative couplings, respectively.

The zeros of $S_{21(12)}$ are given by an effective two-mode Hamiltonian (see Appendix~\ref{Effective Hamiltonian} for details)
\begin{equation} \label{eq19}
	\frac{\hat{H}_{\mathrm{eff}}^{21(12)}}{\hbar}
	= \begin{pmatrix}
		\tilde{\omega}_c & -\sqrt{\tilde{C}_{21(12)}}\\[2pt]
		-\sqrt{\tilde{C}_{21(12)}} & \tilde{\omega}_m
	\end{pmatrix}.
\end{equation}
Here, the off-diagonal element $-\sqrt{\tilde{C}_{21(12)}}$ serves as an effective coupling strength. $\tilde{C}_{21(12)}$ is a function of all the coupling phases $\Phi_g$, $\Phi_{mp}$, $\Phi_{mq}$, $\Phi_{cp}$, and $\Phi_{cq}$, and the traveling phase $\Phi_l$, with the inherent relation  $\tilde{C}_{21}=\tilde{C}_{12}^*$. With physical constraints on $\Phi_{mp(q)}$ and $\Phi_{cp(q)}$, we can reduce $\tilde{C}_{21(12)}$ to a function of $\Phi_g$, $\Phi_{mp}$ and $\Phi_l$, written as (see Appendix \ref{Transmission parameter of coupled system} for details)
\begin{equation} \label{eq20}
	\tilde{C}_{21(12)}=g [g+2\sqrt{\kappa_c\kappa_m}\sin\Phi_l e^{i\Psi_{21(12)}}]
\end{equation}
for magnetic-dipole coupling, and 
\begin{equation} \label{eq21}
	\tilde{C}_{21(12)}=g [g-2\sqrt{\kappa_c\kappa_m}\cos\Phi_l e^{i\Psi_{21(12)}}]
\end{equation}
for electric-dipole interaction, where

\begin{subequations}\label{eq22}
	\begin{align}
	&\Psi_{21}=\Phi_{mp}-\Phi_g=\angle(\mathbf{h}_{mp},\mathbf{h}_c), 
		\\
	&\Psi_{12}=-\Psi_{21}=\Phi_{g}-\Phi_{mp}.
	\end{align}
\end{subequations}
Here, $\Psi_{21(12)}$ is equal to the phase difference between the Oersted field $\mathbf{h}_{mp}$ generated by the right-going traveling wave and the cavity magnetic field $\mathbf{h}_c$, and is defined as the synthetic phase. Eq. \ref{eq22}(b) is guaranteed by the relation $\tilde{C}_{21}=\tilde{C}_{12}^*$. It means that the phase accumulated along the closed interaction loop involving the cavity, magnon, and traveling-wave modes is direction-dependent, with its sign reversed upon propagation reversal, as shown in Figs.~\ref{fig4}(b) and (c). This is physically attributed to the specific order of interactions within the closed coupling loop.

 In addition, synthetic phase changes sign when the applied field is reversed, as illustrated in Figs.~\ref{fig4}(d) and (e). This is because reversing the magnetic field flips the spin-coordinate frame and thereby reverses the relative orientation between $\mathbf{h}_{mp}$ and $\mathbf{h}_c$. This is expressed as
 \begin{equation}\label{eq24}
 	\Psi_{21(12)}(\pm H)=-\Psi_{21(12)}(\mp H).
 \end{equation}
The combination of Eqs.~\ref{eq22}(b) and~\ref{eq24} gives $\Psi_{21}(\pm H)=\Psi_{12}(\mp H)$, which preserves the global time-reversal symmetry (TRS) in magnetic systems \cite{suarez2025chiral}:
 \begin{equation}\label{eq25} 
 	S_{21}(\omega,\pm H )=S_{12}(\omega,\mp H).
 \end{equation}

We emphasize that nonreciprocity in this coupling-loop configuration appears only when two conditions are met:  
(i) breaking global TRS by fixing the bias field, and  
(ii) a direction-dependent chiral interaction induced by a nontrivial synthetic phase. These two conditions together lead to nonreciprocity:
 \begin{equation}\label{eq26}
  			S_{21}(\omega,\pm H)\ne S_{12}(\omega,\pm H).
 \end{equation}
Here, condition (i) is inherently met by the fixed field $\pm H$. Condition (ii) requires that the synthetic phase be nontrivial: 
 \begin{equation}\label{eq27}
\Psi_{21(12)}  \ne n \pi (n \in \mathbb{Z}),
\end{equation}
according to Eqs.~\ref{eq18}, \ref{eq20} and~\ref{eq21}. This characteristic makes this system fundamentally different from the conventional structural chirality discussed in Secs.~\ref{Nonreciprocity in single magnon mode} and \ref{Nonreciprocal coherent coupling}, where the nonreciprocity results from the magnitude of the coupling strength rather than phase accumulation. Equation~\ref{eq27} guarantees that the traveling-wave Oersted fields and the cavity field are not collinear and thus can form a synthetic chiral field, giving rise to direction-dependent coupling and nonreciprocal transmission. Additionally, condition (ii) also requires $\sin \Phi_l \ne 0$ in Eq.~\ref{eq20} for magnetic-dipole interaction, and $\cos \Phi_l \ne 0$ in Eq.~\ref{eq21} for electric-dipole interaction. If these conditions do not hold, the dynamic phase (resembling $\varphi_r$ in Eq.~\ref{eq1}) of the Oersted field relative to the cavity field becomes zero, so synthetic chirality cannot be formed even if the two fields are not collinear. This predicts a distinct difference between magnetic- and electric-dipole interactions in terms of a distance dependence of nonreciprocity, as discussed in Appendix~\ref{Distance dependence of synthetic-chirality-induced nonreciprocity}. 
 
Last but not least, we point out that this system exhibits a unique mirror symmetry under the inversion of frequency detuning ($\Delta_c=\omega-\omega_c$) and field detuning ($\Delta_m=\omega_m-\omega_c$) (see Appendix~\ref{Mirror symmetry in synthetic-chirality-induced nonreciprocity} for details):
\begin{equation} \label{eq28}
	|S_{21}(\Delta_m,\Delta_c)|=|S_{12}(-\Delta_m,-\Delta_c)|.
\end{equation}
This symmetry emerges only when the traveling phase satisfies $\Phi_l=(n+\frac{1}{2})\pi~(n \in \mathbb{Z})$ for the magnetic-dipole cavity and $\Phi_l=n\pi~(n \in \mathbb{Z})$ for the electric-dipole cavity. 
It constitutes a distinctive fingerprint of the traveling-wave–mediated synthetic chirality and is absent in nonreciprocal responses based solely on structural chirality. Specifically, at $\Delta_m=0$, the mirror symmetry reduces to
\begin{equation} \label{eq29} |S_{21}(\Delta_c)|=|S_{12}(-\Delta_c)|.
	\end{equation} 
This provides a clear experimental hallmark of the synthetic-chiral coupling mechanism.

Below, we experimentally verify coupled systems in which cavities interact with traveling waves via magnetic- and electric-dipole interactions.

\subsection{Magnetic dipole interaction cavity}
\label{Traveling wave-mediated nonreciprocal coupling A}
\begin{figure}[htbp]
	\centering 
	\includegraphics[width=8.8 cm]{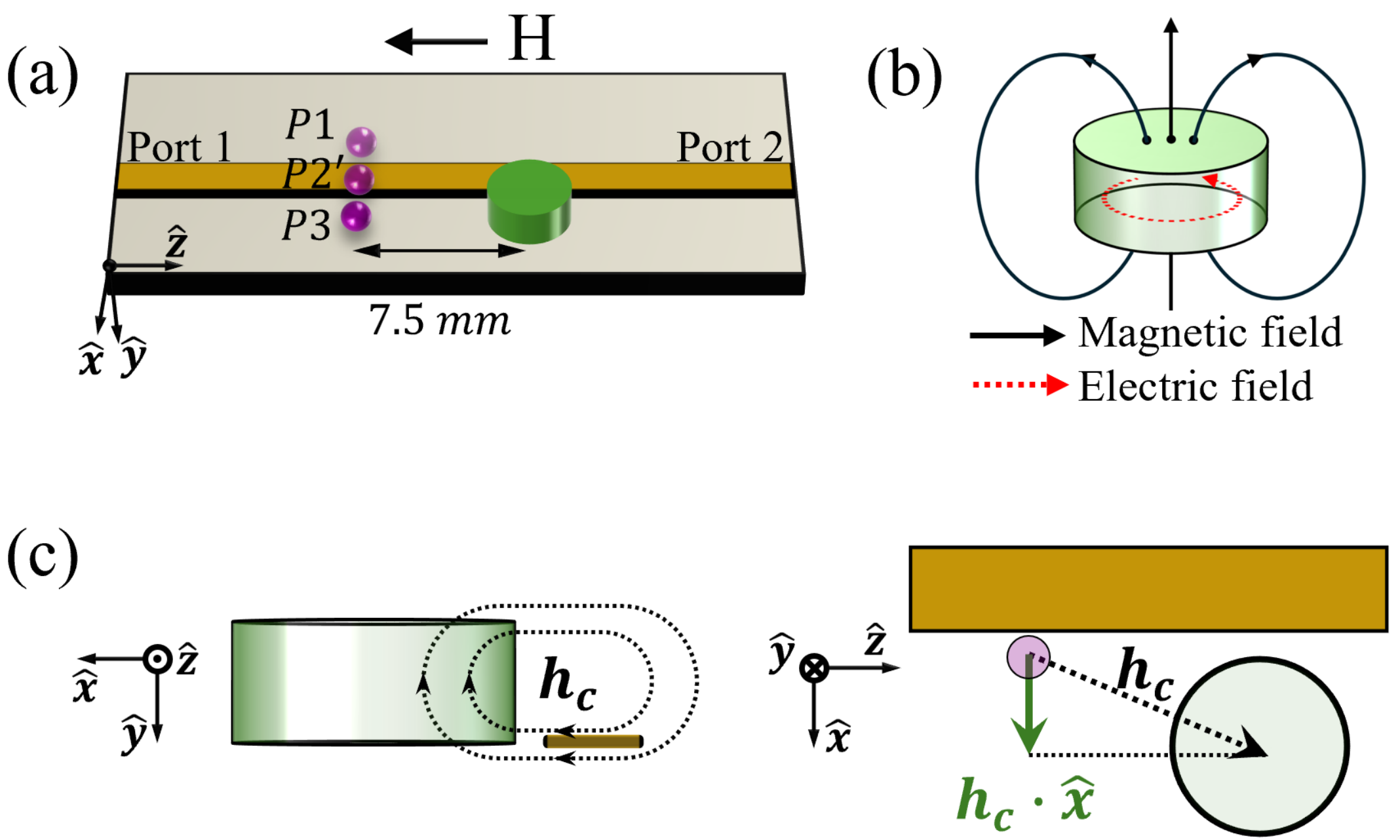}
	\caption{(a) Experimental setup: a YIG sphere, placed sequentially at $P1$, $P2'$, and $P3$, is coupled to a DR on a microstrip. The applied magnetic field is along the microstrip. (b) The TE$_{\delta01}$ mode (equivalent to a magnetic dipole) of DR.   The black curves and red dashed curves represent the magnetic and electric field, respectively. (c) The projections of the microwave field $\mathbf{h}_c$ from DR onto $xy$- and $xz$-planes, with its effective component (perpendicular to the magnetization) aligned with $x$-axis.}
	\label{fig5}
\end{figure}
In this section, we experimentally demonstrate a coupled system with a dielectric resonator (DR) as a representative magnetic-dipole-interaction resonator. The schematic in Fig.~\ref{fig5}(a) shows a YIG sphere coupled to a DR via a common microstrip line, with a horizontal center-to-center separation $l=7.5$ mm. The DR with a diameter of 9.1 mm, a height of 3.7 mm, and a relative permittivity of 34 is positioned on one side of the microstrip line. It supports a standing-wave resonance dominated by the TE$_{01\delta}$ mode \cite{yang2024anomalous,yao2025nonreciprocal}. This mode behaves effectively as a magnetic dipole \cite{guillon1985coupling,kajfez1986dielectric}, with energy oscillating between the magnetic and electric fields distributed inside the DR and in the near-field space around it. As shown in Fig.~\ref{fig5}(b), we orient the dipole upward, normal to the plane, without loss of generality. The corresponding cavity magnetic field $\mathbf{h}_c$ created by this dipole depends on the position of the YIG sphere. At the YIG positions $P1$, $P2'$ (slightly displaced by 0.7 mm from the center position of the microstrip toward the DR), and $P3$, the effective component of $\hat{\mathbf{h}}_c(\mathbf{r})$ that drives spin precession is linearly polarized and aligned along $\mathbf{\hat{x}}$, as depicted in Fig.~\ref{fig5}(c). Therefore, its direct interaction with the YIG sphere is non-chiral. In addition, to avoid the effect of conventional nonreciprocity caused by structural chirality, we set the applied magnetic field along the microstrip, a configuration in which the effective traveling-wave field that drives the magnon mode is also non-chiral ($\lambda_{mp}=\lambda_{mq}$), as demonstrated in Figs.~\ref{fig2}(d)-(f). 

\begin{figure*}[htbp]
	\centering 
	\includegraphics[width=0.9\textwidth]{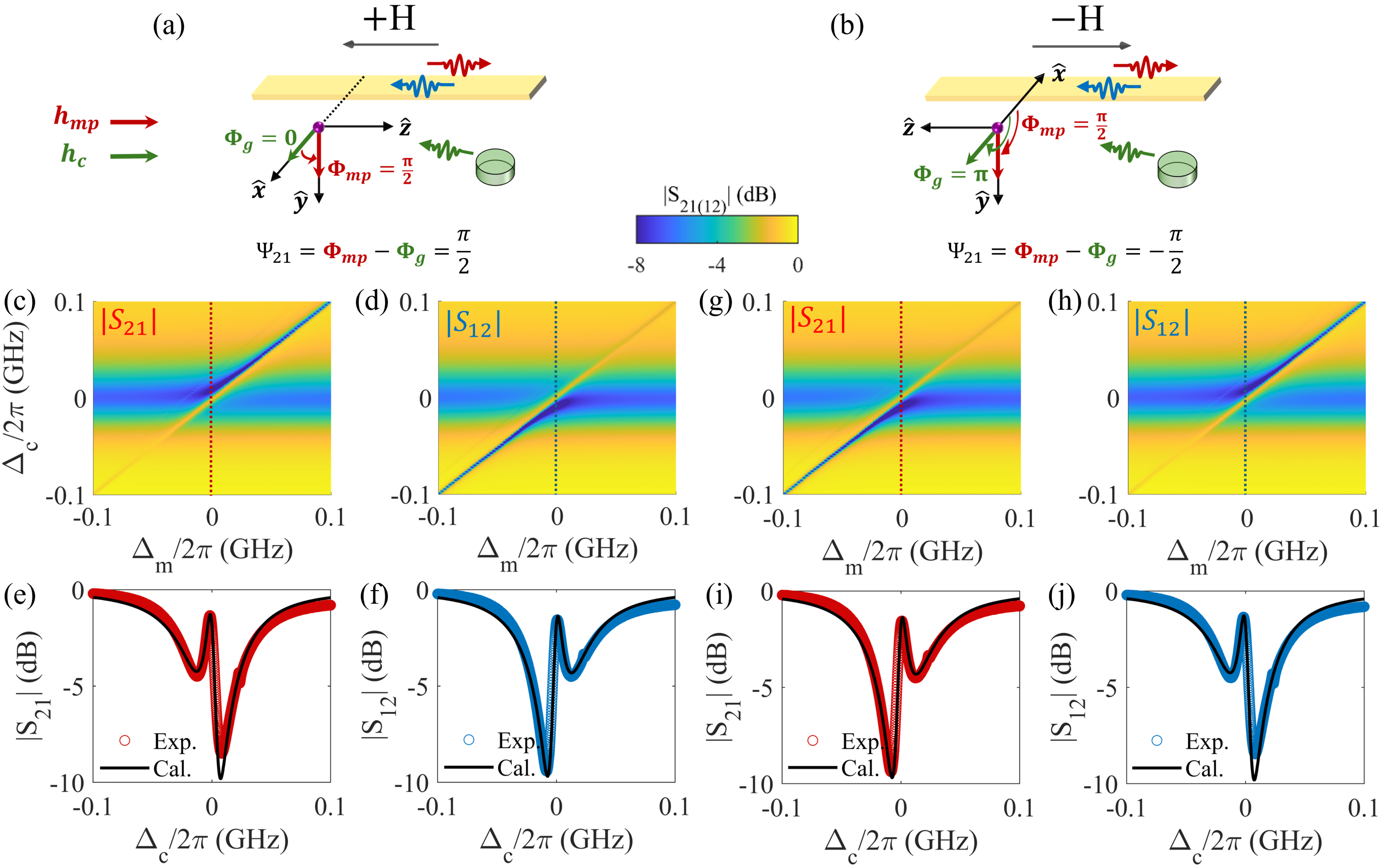}
	\caption{(a)(b) The coupling phases $\Phi_{mp}$, $\Phi_g$, and the resulting synthetic phase $\Psi_{21}$ in the given spin coordinate systems under (a) positive and (b) negative applied fields. The red and green arrows represent the Oersted field $\mathbf{h}_{mp}$ generated by the right-going traveling wave and the cavity field $\mathbf{h}_c$, respectively. At $P3$, the transmission mappings of (c),(g) $|S_{21}|$ and (d),(h) $|S_{12}|$ as functions of $\Delta_m$ and $\Delta_c$ under (c),(d) positive and (g),(h) negative magnetic fields. Transmission spectra $|S_{21}|$ (red circle) and $|S_{12}|$ (blue circle) for (e),(f) positive and (i),(j) negative magnetic fields at $\Delta_m=0$, denoted as dashed lines in panels (c),(d) and (g),(h). The black curves are calculated using Eqs.~\ref{eq18} and \ref{eq20} with fitted coupling strength $g/2\pi=9$ MHz. }
	\label{fig6}
\end{figure*}

\begin{figure*}[htbp]
	\centering 
	\includegraphics[width=0.95\textwidth]{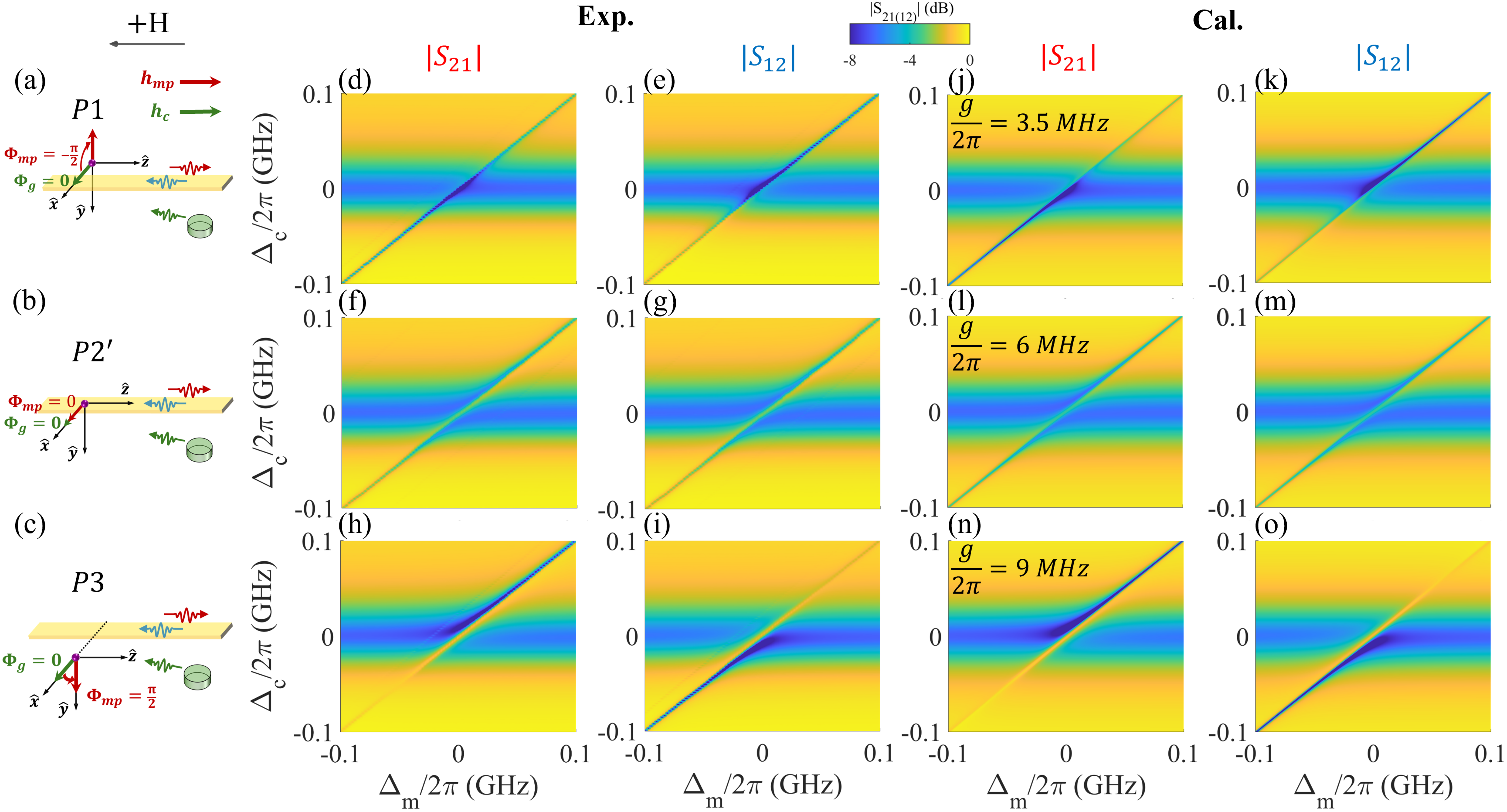}
	\caption{The coupling phases $\Phi_{mp}$ and $\Phi_g$ when YIG is set at (a) $P1$, (b) $P2'$ and (c) $P3$. The red and green arrows represent the Oersted field $\mathbf{h}_{mp}$ generated by the right-going traveling wave and the cavity field $\mathbf{h}_c$, respectively. Under the positive magnetic field, the measured (d),(f),(h) $|S_{21}|$ and (e),(g),(i) $|S_{12}|$ mappings at (d),(e) $P1$, (f),(g) $P2'$ and (h),(i) $P3$, respectively. (j)-(o) The corresponding calculated results using Eqs.~\ref{eq18} and \ref{eq20}. }
	\label{fig7}
\end{figure*}

The calibrations of the DR and the YIG sphere are implemented in independent side-coupled resonator-waveguide experiments. The DR mode resonates at a frequency of $\omega_c/2\pi=6.183$ GHz with an intrinsic damping rate $\beta_0/2\pi=18.74$ MHz, and is undercoupled to the traveling wave with an extrinsic damping rate $\kappa_c/2\pi=17.63$ MHz. All the calibrated damping rates of the YIG sphere are summarized in Fig.~\ref{fig18} in Appendix~\ref{Calibration of the parameters of YIG and DR}.
 
 As the first step, we investigate the applied-field dependence of the nonreciprocity at the YIG position P3, by setting $\Phi_l=\frac{2\pi l}{\lambda}\approx 0.46\pi$ with $\lambda=\frac{2\pi c}{\sqrt{\varepsilon_r}\omega_c}\approx32.7$ mm, where the relative permittivity is $\varepsilon_r=2.2$ for the RT5880 substrate and $c$ is the speed of light in vacuum. First, we deduce synthetic phase $\Psi_{21}$ by choosing the spin coordinate systems shown in Figs.~\ref{fig6}(a) and (b) for positive ($+\mathbf{H}$) and negative ($-\mathbf{H}$) applied fields, respectively, without loss of generality. We obtain $\Phi_{mp}=\pi/2$ for both field directions, and $\Phi_g=0$ and $\Phi_g=\pi$ for positive and negative fields, respectively. According to Eq.~\ref{eq22}, this leads to the synthetic phase $\Psi_{21}=\pi/2$ and $\Psi_{21}=-\pi/2$ for $+\mathbf{H}$ and $-\mathbf{H}$, respectively, satisfying the relation shown in Eq.~\ref{eq24}. Accordingly, $\Psi_{12}$ flips sign under both field directions, in accordance with the intrinsic relation prescribed by Eq.~\ref{eq22} (b). All the phases are summarized in Table~\ref{field dependent coupling phase}.  
  
 We note that the deduced synthetic phase $\Psi_{21(12)}$ satisfy the nonreciprocity condition in Eq.~\ref{eq27}. As a result, the measured $|S_{21}|$ [Fig.~\ref{fig6}(c)] and $|S_{12}|$ [Fig.~\ref{fig6}(d)] mappings as functions of $\Delta_m$ and $\Delta_c$ under a positive applied field show obvious nonreciprocity. In addition, considering $\Phi_l = 0.46\pi \approx \pi/2$ in our setup, the mirror symmetry described by Eq.~\ref{eq28} remains approximately valid and can be observed in Figs.~\ref{fig6}(c) and (d). Specifically, at $\Delta_m=0$, the measured (circles) $|S_{21}|$ spectrum in Fig.~\ref{fig6}(e) and the $|S_{12}|$ spectrum in Fig.~\ref{fig6}(f) are fitted (thin curves) using Eq.~\ref{eq18} and Eq.~\ref{eq20} with $g/2\pi=9$ MHz as the only fitting parameter. They also approximately satisfy the mirror symmetry described by Eq.~\ref{eq29}. When the field direction is reversed, the transmission nonreciprocity is also reversed, as displayed in Figs.~\ref{fig6}(g)-(j), reproducing the global TRS in Eq.~\ref{eq25}.   

\begin{table}[htbp]
	\centering
	\caption{$\Phi_{mp}$, $\Phi_g$ and $\Psi_{21(12)}$ at $+\mathbf{H}$ and $-\mathbf{H}$.}
	\begin{tabularx}{\columnwidth}{>{\centering\arraybackslash}X|>{\centering\arraybackslash}X>{\centering\arraybackslash}X>{\centering\arraybackslash}X>{\centering\arraybackslash}X}
		\hline\hline
		&$\Phi_{mp}$ & $\Phi_{g}$&$\Psi_{21}$ &$\Psi_{12}$ \\  
		\hline
		$+\mathbf{H}$ & $\pi/2$ & $0$& $\pi/2$ & $-\pi/2$ \\
		$-\mathbf{H}$ &  $\pi/2$ & $\pi$& $-\pi/2$ & $\pi/2$ \\
		\hline\hline
	\end{tabularx}
	\label{field dependent coupling phase}
\end{table} 

Fixing the positive applied field $+\mathbf{H}$, we further examine the spatial dependence of the nonreciprocity by changing the YIG position relative to the microstrip. This is because the synthetic phase $\Psi_{21(12)}$, which determines the nonreciprocity, is position-dependent. It mainly results from the position-dependent coupling phase $\Phi_{mp}$ of the YIG sphere with the traveling wave, which is determined by the polarization of the Oersted field $\mathbf{h}_{mp}$ with respect to the $x$-axis. As shown in Figs.~\ref{fig7}(a)–(c), three representative cases correspond to the YIG being placed at $P1$, $P2'$, and $P3$ of the microstrip. When the YIG moves from $P3$ to $P1$, $\Phi_{mp}$ changes from $\pi/2$ to $-\pi/2$, while the Oersted field near the center of the microstrip ($P2'$) is nearly aligned with the $x$-axis, corresponding to $\Phi_{mp}=0$. In contrast, the coupling-strength phase $\Phi_g$ remains zero at all positions. These effects further lead to a position-dependent synthetic phase $\Psi_{21(12)}$. All phase values are summarized in Table~\ref{coupling phase}.

\begin{table}[htbp]
	\centering
	\caption{$\Phi_{mp}$, $\Phi_g$ and $\Psi_{21(12)}$ at $P1$, $P2'$, and $P3$.}
	\begin{tabularx}{\columnwidth}{>{\centering\arraybackslash}X|>{\centering\arraybackslash}X>{\centering\arraybackslash}X>{\centering\arraybackslash}X>{\centering\arraybackslash}X}
		\hline\hline
		&$\Phi_{mp}$ & $\Phi_{g}$&$\Psi_{21}$ &$\Psi_{12}$ \\  
		\hline
		$P1$ & $-\pi/2$ & 0 & $-\pi/2$ & $\pi/2$\\
		$P2'$ &  $0$ & 0 &  $0$ &  $0$ \\  
		$P3$ &  $\pi/2$ & 0 &  $\pi/2$ &  $-\pi/2$\\  
		\hline\hline
	\end{tabularx}
	\label{coupling phase}
\end{table} 

Figures~\ref{fig7}(d)-(i) show the mappings of the measured $|S_{21}|$ and $|S_{12}|$ as functions of $\Delta_c$ and $\Delta_m$ when the YIG is placed at the three positions ($P1$, $P2'$, and $P3$). At $P2'$, $\Psi_{21}=\Psi_{12}=0$, leading to the disappearance of nonreciprocity [Figs.~\ref{fig7}(f) and (g)], as the condition required by Eq.~\ref{eq27} is not satisfied. The effective coupling strength $-\sqrt{\tilde{C}_{21(12)}}$ becomes a real number according to Eq.~\ref{eq20}, giving rise to a purely coherent coupling. Physically, this reflects the fact that the traveling-wave Oersted field is aligned with the cavity field and thus cannot be combined into a chiral field, leading to the absence of directional asymmetry. In contrast, nonreciprocal transmission is observed at $P1$ [Figs.~\ref{fig7}(d) and (e)] with $\Psi_{21}=-\pi/2$, and at $P3$ [Figs.~\ref{fig7}(h) and (i)] with $\Psi_{21}=\pi/2$. These experimental data are well reproduced by the calculated results using Eqs.~\ref{eq18} and~\ref{eq20} with $g$ as the only fitting parameter in Figs.~\ref{fig7}(j)-(o). The fitted coupling strength $g/2\pi$ decreases from 9 MHz to 6 MHz and then to 3.5 MHz as the YIG sphere moves from $P3$, $P2'$ to $P1$, consistent with the expectation that the direct coupling strength diminishes as the distance between the resonators increases. The reciprocal position $P2'$ slightly deviates from the center of the microstrip ($P2$), which may originate from a slight deviation of the DR field from the $x$-axis.

Finally, we note that a recently reported nonreciprocal slow-fast light phenomenon is also captured within the regime of our theory.
By setting $\Phi_l=\pi/2$ and $\Psi_{21}=\pi/2$, Eq.~\ref{eq18} reduces to Eq.~4 in Ref.~\cite{yao2025nonreciprocal}. Moreover, Eq.~5 in Ref.~\cite{yao2025nonreciprocal} represents the global TRS described by Eq.~\ref{eq25} and the nonreciprocity imposed by a fixed magnetic field described by Eq.~\ref{eq26}. In addition, the mirror symmetry described by Eq.~\ref{eq29} is broken in Fig.~3(a) of that paper, because the experimental $\Phi_l$ deviates the symmetry condition $\Phi_l=(n+\frac{1}{2})\pi~(n \in \mathbb{Z})$ (Eq.~\ref{eqG6} in Appendix~\ref{Mirror symmetry in synthetic-chirality-induced nonreciprocity}) for the magnetic-dipole cavity.
\subsection{Electric dipole interaction cavity}
\begin{figure}[htbp]
	\centering 
	\includegraphics[width=8.8 cm]{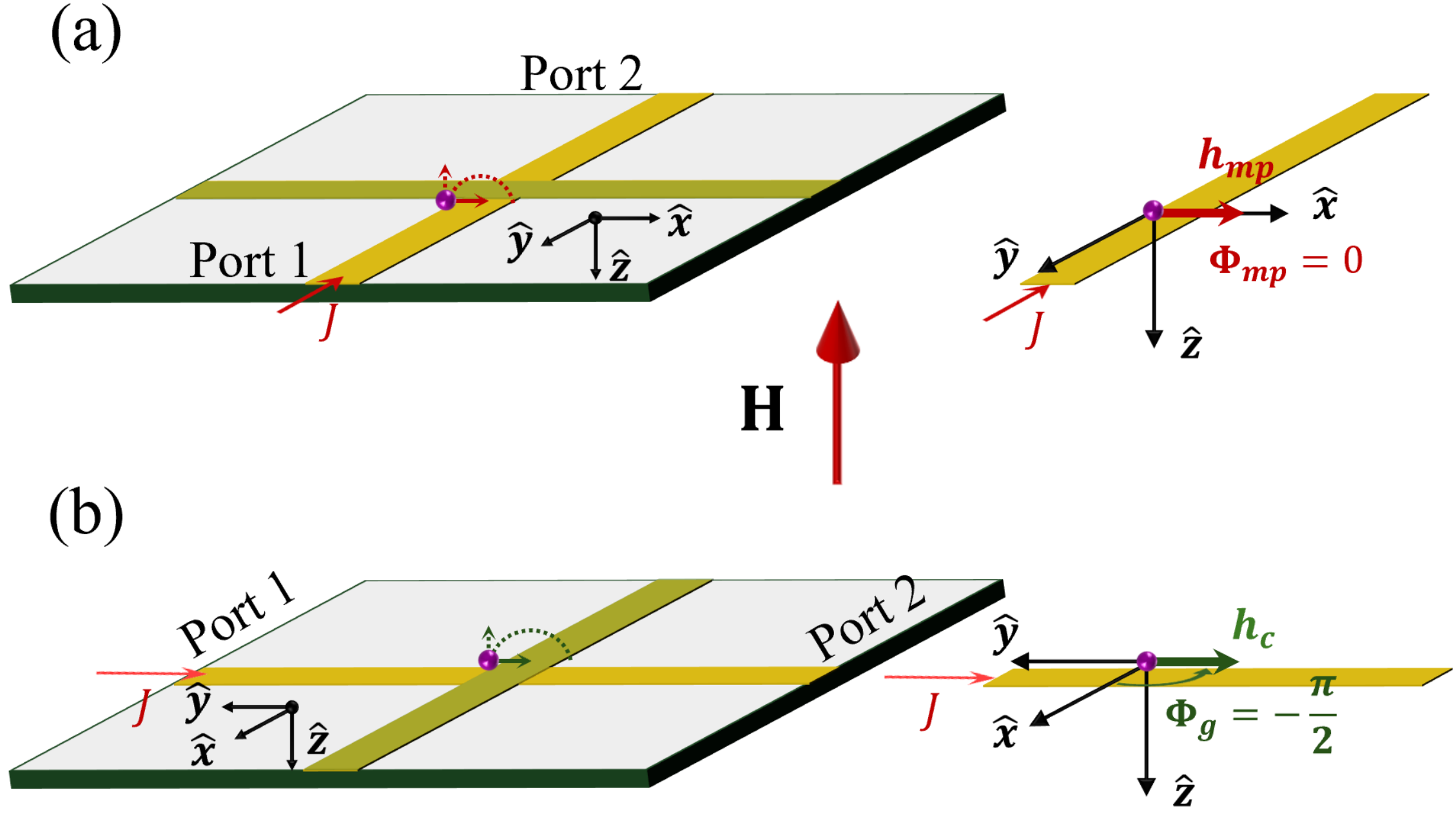}
	\caption{Schematic of the experimental setup in Ref.~\cite{wang2019nonreciprocity}. The microstrip (yellow strip) supports traveling-wave modes, and the cross-line cavity (green strip) supports a standing-wave cavity mode. A YIG sphere is set on the corner of the microstrip and the cavity. The applied field is normal to the substrate. The effective field polarization of (a) the Oersted field $\mathbf{h}_{mp}$ (red arrows) from the traveling wave and (b) the cavity field $\mathbf{h}_c$ (green arrows) when the wave is transferred from port 1 to port 2, corresponding to the coupling phases $\Phi_{mp}=0$ and $\Phi_g=-\frac{\pi}{2}$. }
	\label{fig8}
\end{figure}

We further demonstrate a case with a cross-line cavity as a representative electric-dipole resonator. Here, we take the experiment results in Ref.~\cite{wang2019nonreciprocity} as an example. The schematic of the setup is shown in Fig.~\ref{fig8}. The cross-line cavity supports both standing and traveling waves, and the YIG sphere is mounted slightly away from the cavity and off the microstrip center. The applied magnetic field is set normal to the plane, thus the effective microstrip fields lie in the substrate plane (the $xy$-plane in the given spin coordinate system shown in Fig.~\ref{fig8}). We note that the experimental results can be well reproduced by considering the field polarizations shown in Fig.~\ref{fig8}, where the effective Oersted field generated by the traveling wave is along the $\hat{x}$ axis [Fig.~\ref{fig8}(a)], and the effective cavity field is along the $-\hat{y}$ axis [Fig.~\ref{fig8}(b)] when the wave propagates from port 1 to port 2. These polarizations correspond to $\Phi_{mp}=0$ and $\Phi_g=-\pi/2$, thus $\Psi_{21}=\Phi_{mp}-\Phi_g=\pi/2$, so that $\Psi_{12}=-\pi/2$. 

\begin{figure}[htbp]
   	\centering 
   	\includegraphics[width=8.8 cm]{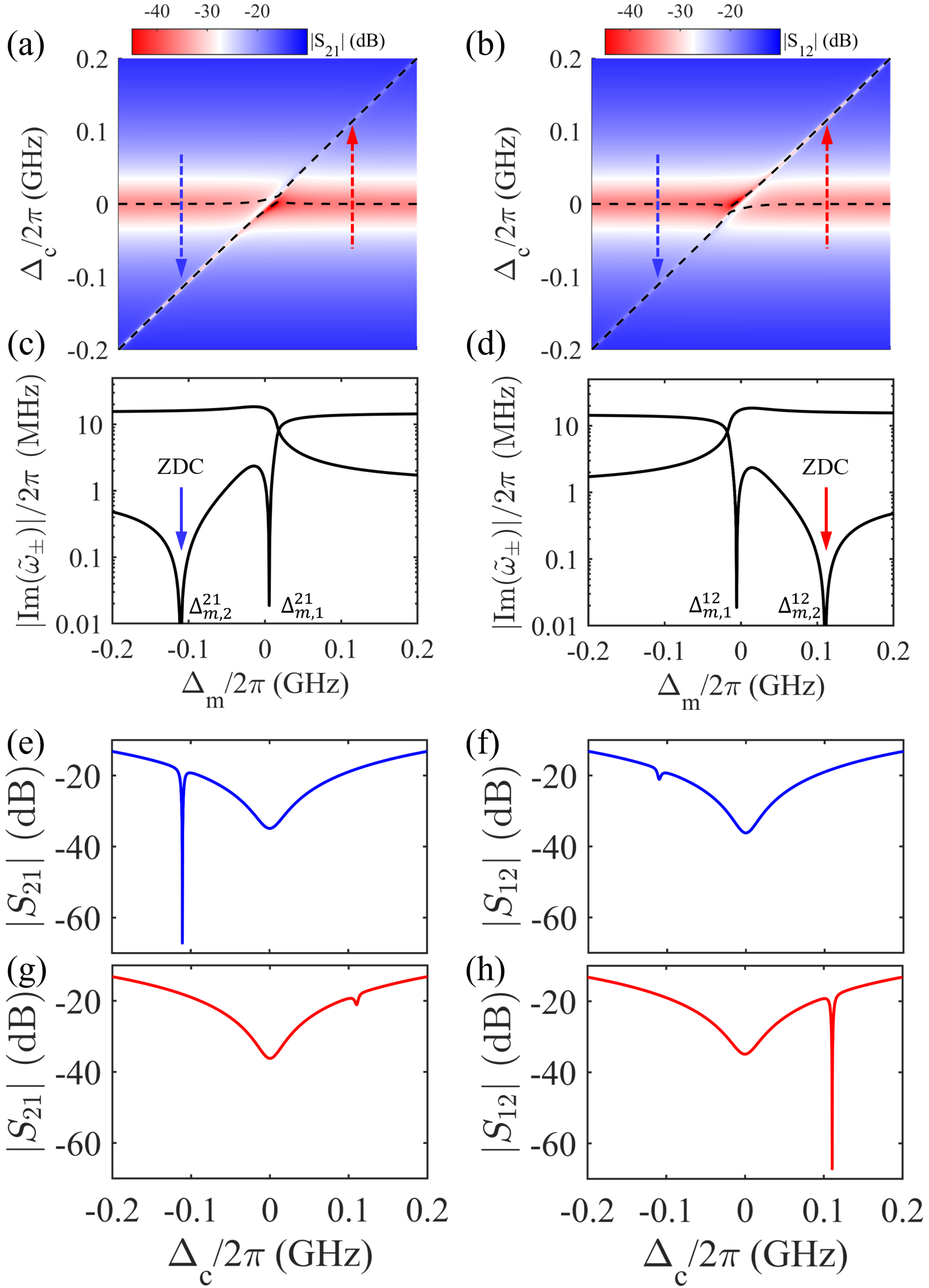}
   	\caption{The calculated (a) $|S_{21}|$ and (b) $|S_{12}|$ mappings using Eqs.~\ref{eq18} and \ref{eq21} as functions of $\Delta_m$ and $\Delta_c$. Blue and red arrows mark the field detuning $\Delta_m$ at which the spectra plotted in (e)–(h) are measured. The black dashed lines are the calculated real part of the  eigenvalues of Eq.~\ref{eq19}. (c)(d) The calculated imaginary part of the eigenvalues of Eq.~\ref{eq19}. The arrows mark the zero-damping conditions (ZDCs) at which the spectra plotted in (e)–(h) are calculated. The (e) $|S_{21}|$ and (f) $|S_{12}|$ spectra calculated at the $\Delta_m$ marked by the blue arrow. The (g) $|S_{21}|$ and (h) $|S_{12}|$ spectra calculated at the $\Delta_m$ marked by the red arrow. }
   	\label{fig9}
   \end{figure}

 Figures~\ref{fig9}(a) and (b) show the calculated $|S_{21}|$ and $|S_{12}|$ mappings as functions of $\Delta_c$ and $\Delta_m$, respectively, using Eqs.~\ref{eq18} and~\ref{eq21}, which present a clear nonreciprocity described by Eq.~\ref{eq26}. Here, the direct coupling strength $g/2\pi=2.1$ MHz is the only fitting parameter. Other parameters are mainly taken from Ref.~\cite{wang2019nonreciprocity} including the cavity mode frequency $\omega_c/2\pi=4.724$ GHz, the cavity intrinsic and extrinsic damping rates $\beta_0/2\pi=15$ MHz and $\kappa_c/2\pi=880$ MHz, respectively, and the magnon intrinsic damping rate $\alpha_0/2\pi=1.1$ MHz. Moreover, Ref.~\cite{wang2019nonreciprocity} adopts a magnon extrinsic damping rate of
 $\kappa_m/2\pi = 0.071~\mathrm{MHz}$, obtained from fitting a coupled system and the associated parameter relations, rather than from an independent magnon calibration, which could be slightly smaller than the experimental value (Table~\ref{tableG1} in Appendix~\ref{Calibration of the parameters of YIG and DR}). In addition, the transmission-coefficient equation for the coupled system is derived under the approximation of neglecting the coupling term between the magnon and the traveling wave (Eq.~S12 of Ref.~\cite{wang2019nonreciprocity}), which may not be preserved with a larger $\kappa_m$. In contrast, our new model in this paper requires no such approximation, and $\kappa_m$ is set to a larger value of $\kappa_m/2\pi = 1$ MHz. The traveling phase $\Phi_l$ is set to zero because the distance is negligible compared to the wavelength $\lambda= \frac{2\pi c}{\sqrt{\varepsilon_r}\omega_c} \approx 31.3$ mm, where the relative permittivity is $\varepsilon_r=3.55$ for the RO4003C substrate and $c$ is the speed of light in vacuum.   
 
 Figures~\ref{fig9}(c) and (d) display the imaginary parts of the eigenvalues of the effective Hamiltonian in Eq.~\ref{eq19} (see Appendix~\ref{Effective Hamiltonian}), which reproduce the two zero-damping conditions (ZDCs) at $\Delta_{m,1}^{21}=\omega_m-\omega_c\approx2g\sqrt{\kappa_c\kappa_m}/\beta_0$ 
 and $\Delta_{m,2}^{21}\approx -2g\sqrt{\kappa_c\kappa_m}/\alpha_0$ 
 for $S_{21}$, and $\Delta_{m,1}^{12}\approx -2g\sqrt{\kappa_c\kappa_m}/\beta_0$  
and $\Delta_{m,2}^{12}\approx2g\sqrt{\kappa_c\kappa_m}/\alpha_0$  
for $S_{12}$. These ZDCs are obtained from the zeros of the imaginary parts of the eigenvalues of Eq.~\ref{eq19} \cite{wang2019nonreciprocity}, under the approximation $g^2\kappa_c\kappa_m\gg \alpha_0\beta_0(\alpha_0\beta_0+g^2)$. The calculated transmission spectra $|S_{21}|$ and $|S_{12}|$ at the marked ZDC ($\Delta_{m,2}^{21}$) indicated by the blue arrow in Fig.~\ref{fig9}(c) are shown in Figs.~\ref{fig9}(e) and (f), reproducing unidirectional invisibility. Similarly, $|S_{21}|$ and $|S_{12}|$ at the marked ZDC ($\Delta_{m,2}^{12}$) with red arrow in Fig.~\ref{fig9}(d) are shown in Figs.~\ref{fig9}(g) and (h).   

Notably, with $\Phi_l=0$ (satisfying Eq.~\ref{eqG7} in Appendix~\ref{Mirror symmetry in synthetic-chirality-induced nonreciprocity}), a mirror symmetry described by Eq.~\ref{eq28} is observed in Figs.~\ref{fig9}(a) and (b). Specifically, Figs.~\ref{fig9}(e) and (h) show a clearer mirror-symmetry feature, given by $S_{21}(\Delta_c,\Delta_{m,2}^{21})=S_{12}(-\Delta_c,\Delta_{m,2}^{12})$, where $\Delta_{m,2}^{21}=-\Delta_{m,2}^{12}$. Similarly, Figs.~\ref{fig9}(f) and (g) show $S_{12}(\Delta_c,\Delta_{m,2}^{21})=S_{21}(-\Delta_c,\Delta_{m,2}^{12})$.
 
Although the calculated results replicate the measured data in Fig.~2 of Ref.~\cite{wang2019nonreciprocity}, the intrinsic structural chirality of the cross-line cavity could also contribute to the observed nonreciprocity when the magnetic field is applied normal to the microstrip plane, potentially affecting the accurate determination of the coupling strength. A more reliable verification of our theory therefore requires applying the field along the microstrip, where such structural effects are minimized. 

\section{Conclusion} \label{conclusion}
In summary, we establish a generalized theoretical framework to explain nonreciprocity resulting from chiral interactions in magnon-involved TRS-breaking systems. This is realized by mapping the polarization of microwave fields onto the phases of complex coupling strengths. We validate our theory in three representative platforms: a side-coupled magnon traveling-wave system, a chiral-cavity-involved coherent coupling system, and a traveling-wave-mediated coupling system. We show that the nonreciprocity in the first two systems originates from conventional structural chirality, whereas the last system exhibits a synthetic chirality enabled by a coupling-loop configuration, which is fundamentally distinct from conventional mechanisms.

Notably, compared with the conventional approach that relies on the design of chiral structures, the synthetic chirality results from the combined effect of two simple linearly polarized fields, coordinated through the interplay between direct field-overlap coupling and indirect base-mediated coupling. In conventional cases, nonreciprocity manifests as coupling strengths for counter-propagating waves having unequal magnitudes due to structure-induced chiral fields, whereas the novel synthetic chiral interaction case is realized through directional phase accumulation originating from the direction-dependent polarizations of microwave fields. This method offers a novel paradigm for use in realizing chiral interaction and nonreciprocity. As a result, the novel nonreciprocity exhibits characteristics absent in conventional mechanisms, such as mirror symmetry and zero-damping conditions, facilitating unparalleled applications like unidirectional invisibility and nonreciprocal slow-fast pulse propagation.    

From a broader perspective, the theoretical framework of the structural and synthetic chiral interactions demonstrated in this work is highly adaptable. Extending this mechanism to YIG thin films naturally bridges our findings with spintronics, which may add new degree of freedom for the remote control of spin currents \cite{bai2017cavity} by utilizing the resource of synthetic chirality. More fundamentally, the underlying physics is universal and can be applied to any platform involving engineered magnon interactions, such as spin-wave systems, magnon–phonon hybrid systems, cavity magnomechanical systems, and magnon–superconducting-circuit interfaces. The unconventional nonreciprocity method we demonstrated is unique, and has the potential to unlock unprecedented engineering of chiral interactions and nonreciprocity, benefiting broad areas such as nonreciprocal device design, chiral sensing, and nontrivial quantum network construction.  
\begin{acknowledgments}
This work has been funded by NSERC Discovery Grants and NSERC Discovery Accelerator Supplements (C.-M. H.). We thank Greg E. Bridges, Yi-Pu Wang and Jiongjie Wang for helpful discussions, and Bentley Turner for careful revisions and helpful comments.
\end{acknowledgments}

\section*{DATA AVAILABILITY}

The data that support the findings of this Letter are openly available (DOI to be generated once final article is accepted).

\appendix
\section{Spin-momentum locking in microstrip}\label{Spin-momentum locking in microstrip}

\renewcommand{\theequation}{\thesection\arabic{equation}}

\begin{figure}[htbp]
	\centering 
	\includegraphics[width=8.8 cm]{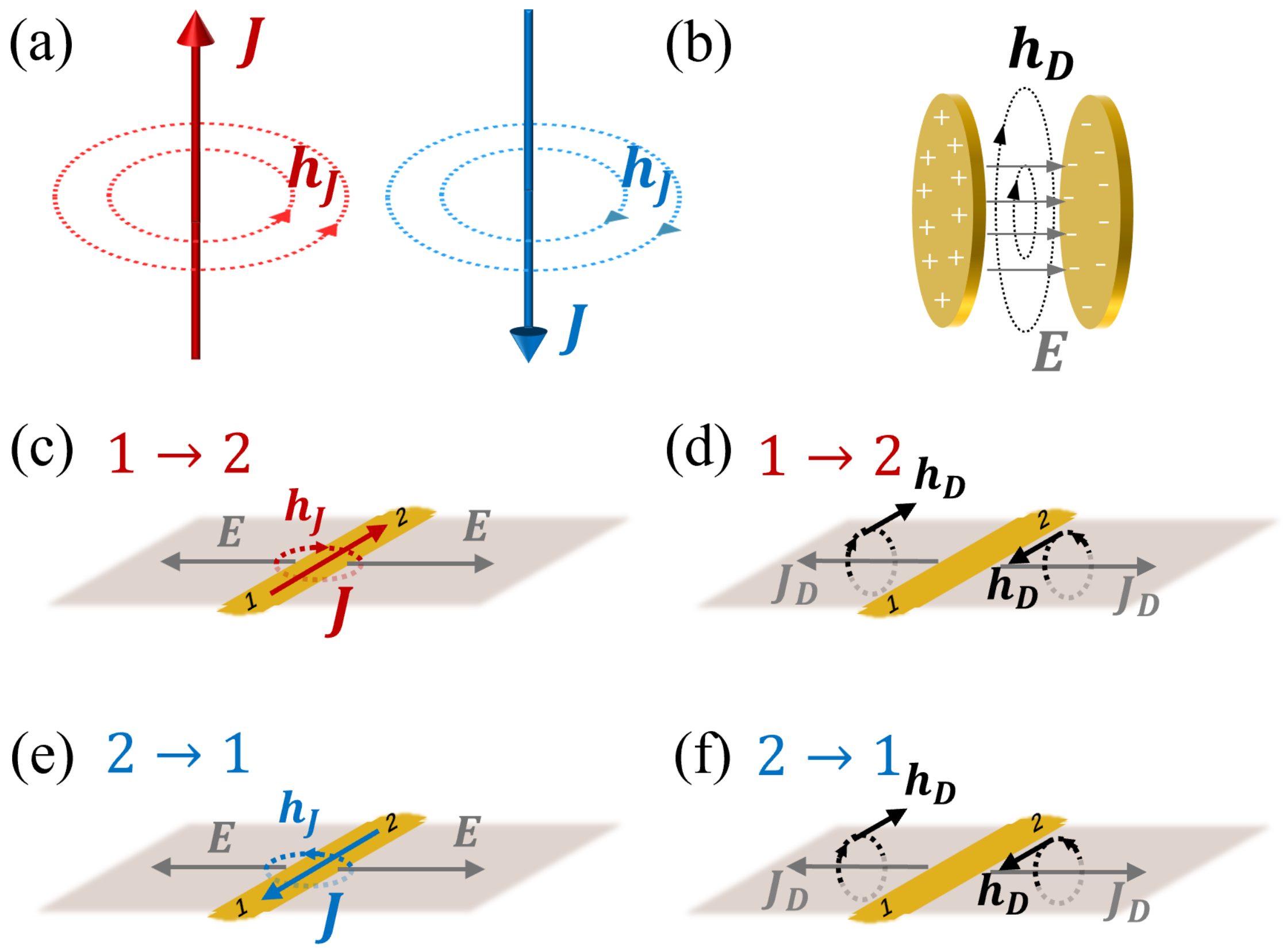}
	\caption{(a) Schematic of Oersted field $\mathbf{h_J}$ produced by conduction current $\mathbf{J}$ propagating in opposite directions. (b) Schematic of induced magnetic field $\mathbf{h_D}$ produced by the displacement current resulting from an oscillating electric field $\mathbf{E}$. The (c),(e) conduction currents and (d),(f) displacement currents in the microstrip when the traveling wave propagates from (c),(d) port 1 to port 2 and (e),(f) the opposite direction. }
	\label{fig10}
\end{figure} \label{Spin-momentum locking in microstrip}
In this section, we explain the spin-momentum locking effect in the microstrip. According to Maxwell's equation $\nabla \times \mathbf{B} =\mu_0 (\mathbf{J}+\mathbf{J_D})= \mu_0 \mathbf{J} + \mu_0 \varepsilon_0 \frac{\partial \mathbf{E}}{\partial t}$, the induced magnetic field arises from the conduction current along with the displacement current due to the oscillating electric field, as shown in Figs.~\ref{fig10}(a) and (b).

 These conduction and displacement currents can be excited simultaneously in one-dimensional microwave waveguide structures. For example, in a coplanar waveguide (CPW), the displacement current flows between the signal line and the ground planes, maintaining a $\pi$/2 phase shift relative to the conduction current \cite{bayard2003electromagnetic}. This phase quadrature serves as the physical foundation for the magnetic isolator \cite{wen1969coplanar,hines1971reciprocal,elshafiey1996full,bayard2003electromagnetic,kuanr2009nonreciprocal,yu2020chiral}. In a microstrip, although the displacement current is predominantly normal to the plane within the dielectric substrate, the fringing electric fields at the strip edges induce significant in-plane components. This introduces an oscillating electric field $\mathbf{E}\propto e^{-i\omega t}$ perpendicular to the transmission line [Figs.~\ref{fig10}(c) and (e)]. The oscillating electric field creates a displacement current $\mathbf{J_D}=\varepsilon_0 \frac{\partial \mathbf{E}}{\partial t}=-i\omega \varepsilon_0 \mathbf{E}$, which introduces a microwave field $\mathbf{h_D}$ that leads $\mathbf{h_J}$ by $\pi/2$ [Figs.~\ref{fig10}(d) and (f)]. As the traveling wave propagates from opposite directions, $\mathbf{h_J}$ reverses, while $\mathbf{h_D}$ remains unchanged because the oscillating electric field $\mathbf{E}$ is independent of the propagation direction. Hence, the combination of these two field components forms an elliptically or circularly polarized field whose handedness is determined by the traveling wave propagation direction, resulting in spin-momentum locking.  

\section{Interaction of magnon mode with traveling wave} \label{Interaction of magnon mode with traveling wave}
\begin{figure}[htbp]
	\centering 
	\includegraphics[width=8.8 cm]{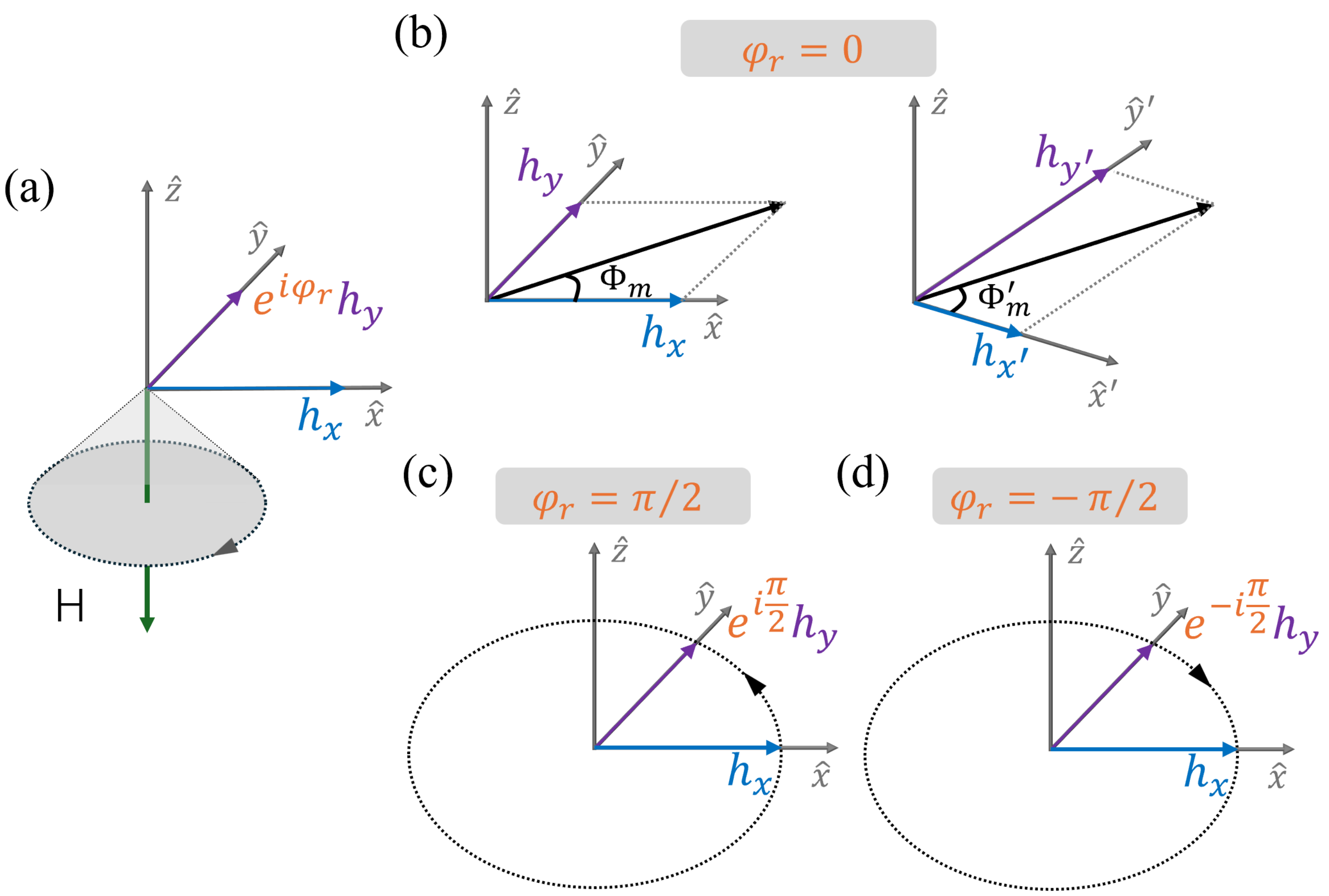}
	\caption{(a) The effective field $	\mathbf{h}_{\text{eff}} (t) = (h_x\hat{x} + e^{i\varphi_r}h_y\hat{y})e^{-i\omega t}$ at $t=0$ in spin coordinate system where $z$-axis antiparallel to the applied field, where $\varphi_r$ is the dynamic relative phase between $h_x$ and $h_y$. (b) When $\varphi_r=0$, the projections of the effective field in two coordinate systems $(x,y,z)$ and $(x',y',z)$. In this case, the coupling phase $\Phi_m(\Phi_{m'})$ represents the angle between $x (x')$-axis and the effective field, therefore is determined by the choice of the coordinate system. The effective field when (c) $\varphi_r=\pi/2$ (d) $\varphi_r=-\pi/2$, forming a (c) left-handed and (d) right-handed elliptically polarized (viewed from a point on the $+\hat{\mathbf{z}}$-axis looking toward the origin, along the direction of the magnetic field), respectively. } 
	\label{fig11}
\end{figure}

Consider an effective field $	\mathbf{h}_{\text{eff}} (t) = (h_x\hat{x} + e^{i\varphi_r}h_y\hat{y})e^{-i\omega t}$ defined in a spin coordinate system, where the $z$-axis is aligned with the spin angular momentum $\mathbf{S}_z$  
(opposite to the applied field), and the $x-y$ axes can be chosen arbitrarily. The effective field at $t=0$ is shown in Fig.~\ref{fig11}(a). 
The quantization of the field is \cite{gerry2023introductory}
\begin{equation}\label{eqB1} \hat{\mathbf{h}}(\mathbf{r})=\frac{1}{\mu_0 c}\sum_{k} (\eta_x\hat{x}+\eta_y\hat{y})(\frac{\hbar \omega_k}{2 \varepsilon_0V})^{1/2} (\hat{a}_{k}e^{i\mathbf{k}\cdot\mathbf{r}}+\hat{a}_{k}^\dagger e^{-i\mathbf{k}\cdot\mathbf{r}}),
\end{equation}
where $\eta_x=\frac{h_x}{\mathcal{N}}$, and $\eta_y=\frac{e^{i\varphi_r}h_y}{\mathcal{N}}$ with $\mathcal{N}=\frac{1}{\max{|\mathbf{h(\mathbf{r})}|}}$ representing the normalization constant. This microwave field interacts with the spins in the magnetic material by the Zeeman interaction:
\begin{equation}\label{eqB2}
	\hat{H_I}=-\mu_0\sum_j  \mathbf{m_j}\cdot\hat{\mathbf{h}}_j. 
\end{equation}
Here, $\mathbf{m}_j=-\gamma \hbar \mathbf{s}_j$ is the magnetic moment at the j-th spin site, where $\gamma$ is the gyromagnetic ratio and $\mathbf{s}_j$ is spin angular momentum at the j-th site spin. $\mathbf{h}_j$ represents the microwave field at the j-th site. Hence,
\begin{equation}\label{eqB3}
	\hat{H}_I=\gamma\hbar\mu_0\sum_j \hat{\mathbf{s}}_j\cdot\hat{\mathbf{h}}_j=\mu_0g\mu_B\sum_j \hat{\mathbf{s}}_j\cdot\hat{\mathbf{h}}_j,
\end{equation}
where $g$ is Land\'e $g$-factor and $\mu_B$ is Bohr magneton.

 Here, $\hat{\mathbf{s}}_j=\hat{s}_j^x \hat{x}+\hat{s}_j^y \hat{y}$, with 
 \begin{subequations}
	\begin{align}
		&\hat{s}_j^x=\frac{\hat{s}_j^++\hat{s}_j^-}{2}=\sqrt{2S}\,\frac{\hat{m}_j+\hat{m}_j^\dagger}{2},\\
		&\hat{s}_j^y=\frac{\hat{s}_j^+-\hat{s}_j^-}{2i}=\sqrt{2S}\,\frac{\hat{m}_j-\hat{m}_j^\dagger}{2i},
	\end{align}
\end{subequations} 
where the spin raising and lowering operators $\hat{s}_j^{\pm}\equiv \hat{s}_j^x\pm i\hat{s}_j^y$ are obtained using Holstein-Primakoff transformations \cite{holstein1940field}:
 \begin{equation}\label{eqB5}
	\begin{split}
		&\hat{s}_j^+=(\sqrt{2S-\hat{m}_j^\dagger \hat{m}_j})\hat{m}_j \approx \hat{m}_j\,\sqrt{2S},\\
		&\hat{s}_j^-=(\sqrt{2S-\hat{m}_j^\dagger \hat{m}_j})\hat{m}_j^\dagger \approx \hat{m}_j^\dagger\,\sqrt{2S}.
	\end{split}
\end{equation}
The bosonic operators defined on each site can be transformed into spin-wave operators via Fourier transformation:
 \begin{equation}\label{eqB6}
	\begin{split}
		&\hat{m}_j^\dagger=\frac{1}{\sqrt{N}}\sum_{\mathbf{q}}e^{-i\mathbf{q}\cdot\mathbf{r}_j}m_\mathbf{q}^\dagger, \\
		&\hat{m}_j=\frac{1}{\sqrt{N}}\sum_{\mathbf{q}}e^{i\mathbf{q}\cdot\mathbf{r}_j}m_\mathbf{q}.
	\end{split}
\end{equation}
Using the identity $\sum_j^N e^{i(\mathbf{k}-\mathbf{q})\cdot \mathbf{r}_j}=N\delta_{\mathbf{q},\mathbf{k}}$, we have
\begin{equation} \label{eqB7}
	\begin{split}
		\hat{H}_I=&\mu_0g\mu_B\sum_j \hat{\mathbf{s}}_j\cdot\hat{\mathbf{h}}_j\\
		=&\frac{g\mu_B}{2c}\sqrt{\frac{\hbar \omega_k}{ \varepsilon_0V}} \sqrt{NS} \sum_{\mathbf{k}}[(\eta_x+i\eta_y)\hat{a}_k\hat{m}^{\dagger}_k+(\eta_x-i\eta_y)\hat{a}^{\dagger}_k\hat{m}_k]\\
		=&\sum_{\mathbf{k}}\hbar(\tilde{\lambda}_m\hat{m}^\dagger\hat{a}_k+\tilde{\lambda}_m^*\hat{m}\hat{a}^\dagger_k),
	\end{split}
\end{equation}
where the coupling strength is
\begin{equation} \label{eqB8}
	\begin{split} \tilde{\lambda}_m&=\lambda_me^{i\Phi_m} \\
		&=\frac{g\mu_B}{2\hbar c}\sqrt{\frac{\hbar \omega_k}{ \varepsilon_0V}} \sqrt{NS} (\eta_x+i\eta_y) \\
		&\propto h_x+ie^{i\varphi_r}h_y.
	\end{split}
\end{equation}
 Hence, we have the coupling strength magnitude $\lambda_m\propto \sqrt{h_x^2+h_y^2-2h_xh_y\sin\varphi_r}$ and the coupling phase $\Phi_m\propto \arctan\frac{h_y\cos\varphi_r}{h_x-h_y\sin\varphi_r}$. 

Specifically, when $\varphi_r=0$, the microwave field is linearly polarized. In this case, $\lambda_m \propto |h|$, and $\Phi_{m}\propto \arctan\frac{h_y}{h_x}$ equals the angle between the microwave field and the $x$-axis of the spin coordinate system. It is worth noting that $\Phi_m$ depends on the choice of the spin coordinate system and is therefore not unique. In Fig.~\ref{fig11}(b), we illustrate the same microwave field projected onto two different coordinate systems: $(x,y,z)$ and $(x',y',z)$. The magnitude of the coupling strength, $\lambda_m \propto \sqrt{h_x^2+h_y^2}=\sqrt{h^2_{x'}+h^2_{y'}}$, remains identical for both systems. However, the coupling phase $\Phi_m$ is determined solely by the choice of $x$- and $y$-axes. Despite this frame-dependence in $\Phi_m$, the physical predictions of the theory remain independent of the choice of $x$- and $y$-axes in the simple case of microwave-spin interaction. This is because the measurable transmission depends only on the extrinsic damping rate $\kappa_{mp(q)} = 2\pi\tilde{\lambda}_{mp(q)} \tilde{\lambda}_{mp(q)}^*$, where the coupling phase $\Phi_m$ drops out due to the multiplication of conjugate complex coupling strengths.

When $\varphi_r=\frac{\pi}{2}$ and $-\frac{\pi}{2}$, the microwave field is a left-handed and right-handed elliptically polarized (viewed from a point on the $+\hat{\mathbf{z}}$-axis looking toward the origin, along the direction of the magnetic field), respectively. In this case, the coupling strength $\lambda_m$ is a real number with coupling phase $\Phi_m=0$. Specially, when $h_x=h_y=h$, $\varphi_r=\frac{\pi}{2}$ corresponds to a left-handed circular polarization with $\lambda=0$; while $\varphi_r=-\frac{\pi}{2}$ corresponds to a right-handed circular polarization with $\lambda=2h$. This indicates that only the microwave field with a rotation direction matching the spin precession can interact with the magnon mode.  

\section{Interaction of cavity mode with traveling wave} \label{Interaction of cavity mode with traveling wave}
In this section, we derive the interaction between the cavity mode and the traveling mode as described in Eq.~\ref{eq17} in the main text.
We first illustrate the derivation using a dielectric resonator (DR) operating as a magnetic-dipole cavity, and subsequently extend the result to the electric-dipole cavity by examining the intrinsic phase relation between its electric and magnetic fields.

\begin{figure}[htbp]
	\centering 
	\includegraphics[width=8.8 cm]{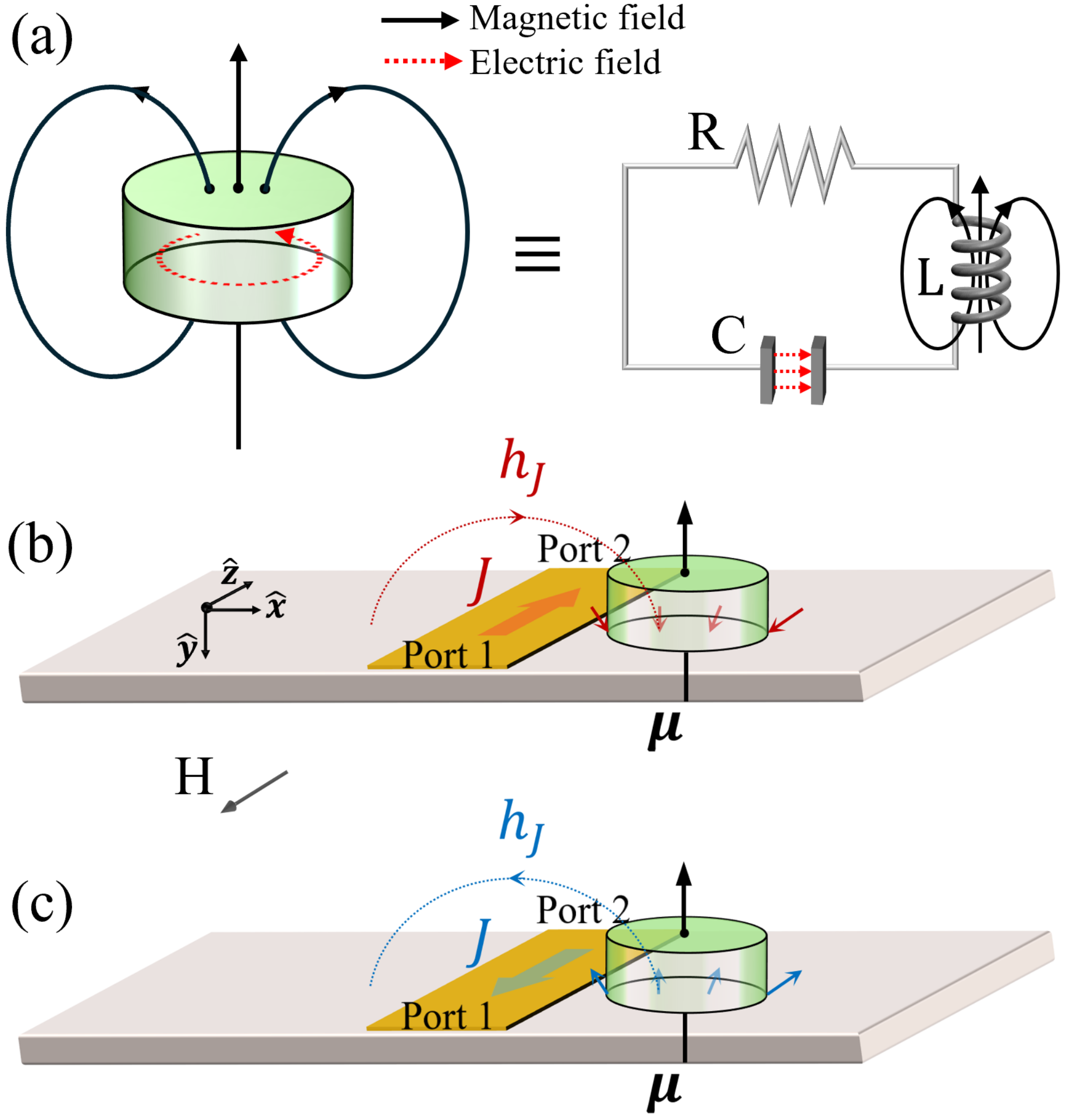}
	\caption{(a) The magnetic-dipole of DR modelled by an equivalent LCR circuit. The effective microwave field when the traveling-wave propagates from (b) port 1 to port 2 and (c) port 2 to port 1 interacts with the effective magnetic dipole $\mathbf{\mu}$ of the TE$_{\delta 01}$ mode of DR.}
	\label{fig12}
\end{figure}

The effective magnetic dipole of DR can be physically modelled by an equivalent LCR circuit, where the inductor creates an effective magnetic field and the capacitor creates an effective electric field, as depicted in Fig.~\ref{fig12}(a). The total energy stored in the inductance $L$ (carrying current $I$) and the capacitance $C$ (charged to voltage $V$) is given by $\hat{H}_c=\frac{1}{2}LI^2+\frac{1}{2}CV^2$.  
Quantizing the circuit by replacing $I$ and $V$ with operators yields \cite{hou2019strong,blais2021circuit}:
\begin{equation} \label{eqC1}
	\begin{split}
		\hat{I}&=\sqrt{\frac{\hbar \omega_c}{2L}}(\hat{c}^\dagger+\hat{c}),\\
		\hat{V}&=i\sqrt{\frac{\hbar \omega_c}{2C}}(\hat{c}^\dagger-\hat{c}),
	\end{split}
\end{equation}
where $\omega_c=\frac{1}{\sqrt{LC}}$ is the resonant frequency of the cavity, and $\hat{c}(\hat{c}^\dagger)$ is the annihilation (creation) operator of the cavity photon.

Without loss of generality, we assume the equivalent magnetic dipole is oriented upward and normal to the plane. The quantized magnetic dipole is thus given by:
\begin{equation} \label{eqC2}
\hat{\mathbf{{\mu}}} = N \hat{I} A(-\hat{\mathbf{y}})=NA\sqrt{\frac{\hbar \omega_c}{2L}}(\hat{c}^\dagger+\hat{c})(-\hat{\mathbf{y}}),
\end{equation}
Here, $N$ is the number of turns in the coil, and $A$ is the geometric area enclosed by a single turn of the coil. 
Alternatively, the dipole could be oriented downward. While this would generate an opposite cavity magnetic field $\mathbf{h}_c$ and alter the individual phases $\Phi_g$ and $\Phi_{cp(q)}$, it leaves the synthetic phase and the underlying physics unchanged.

The traveling wave’s microwave magnetic field $\hat{\mathbf{h}}$
couples to the cavity’s magnetic dipole through interaction Hamiltonian under the rotating-wave approximation:
\begin{equation} \label{eqC3}
		\begin{split}
	\hat{H}_I&=-\hat{\mathbf{\mu}} \cdot \mu_0\hat{\mathbf{h}}\\ 
	&=-\frac{NA}{c}\sqrt{\frac{\hbar \omega_c}{2L}} \sqrt{\frac{\hbar \omega_k}{2\epsilon_0V}}\sum_\mathbf{k}\eta_y (\hat{a}_k \hat{c}^\dagger+\hat{a}_k^\dagger \hat{c})\\
	&=\sum_k \hbar (\tilde{\lambda}_{c}\hat{c}^\dagger \hat{a}_k+\tilde{\lambda}_{c}^*\hat{c} \hat{a}_k^\dagger),
		\end{split}
\end{equation}
where $\tilde{\lambda}_{c}=\frac{\eta_y NA}{c}\sqrt{\frac{\hbar \omega_c}{2L}}\sqrt{\frac{\hbar\omega_k}{2\varepsilon_0 V}}=\lambda_{c}e^{i\Phi_{c}}$. Here, $\eta_y$ is the normalized $y$-component of the traveling-wave magnetic field. For traveling wave propagating from port 1 to port 2 [Fig.~\ref{fig12}(b)], $\eta_y=\frac{h_y}{\mathcal{N}}$, the effective microwave field is parallel to the magnetic dipole, leading to the coupling phase $\Phi_{cp}=0$; while for traveling wave propagating from port 2 to port 1 [Fig.~\ref{fig12}(c)], $\eta_y=-\frac{h_y}{\mathcal{N}}$, the effective field is anti-parallel, resulting in a coupling phase $\Phi_{cq}=\pi$. Physically, $\Phi_{cp(q)}$ tells us whether the magnetic dipole is parallel or antiparallel to the effective microwave field, and is therefore independent of the choice of spin coordinate system. Hence, $\Phi_{cp(q)}$ is considered as a trivial
phase that does not contribute to the synthetic chiral field acting on the YIG sphere or the synthetic phase $\Psi_{21(12)}$.

The above derivation applies when the DR acts as a magnetic-dipole cavity. When the traveling wave interacts with the cavity through an electric-dipole interaction, the interaction can be written as $\hat{\mathbf{d}}\cdot\hat{\mathbf{e}}$, where $\hat{\mathbf{d}}$ is the effective electric dipole and $\hat{\mathbf{e}}$ is the electric component of the traveling wave. For a traveling wave, the magnetic field $\hat{\mathbf{h}}$ is in phase with $\hat{\mathbf{e}}$. In contrast, for the cavity mode, the electric field $\mathbf{\hat{e}_c}$ (and hence the electric dipole $\hat{\mathbf{d}}$) and magnetic field $\mathbf{\hat{h}_c}$ (and hence the magnetic dipole $\hat{\mathbf{\mu}}$) are intrinsically phase-shifted by $\pi/2$, which can be equivalently regarded as the voltage component in Eq. \ref{eqC1}, with the relation $\hat{\mathbf{d}}\propto i (\hat{c}^\dagger-\hat{c})$, compared to $\hat{\mathbf{\mu}}\propto (\hat{c}^\dagger+\hat{c})$.  In addition, unlike the magnetic component $\hat{\mathbf{h}}$, which shifts by $\pi$ when the propagation direction is reversed in the magnetic-dipole cavity, the direction of $\hat{\mathbf{e}}$ is independent of the propagation direction, which leads to $\Phi_{cp}=\Phi_{cq}$. Therefore, the electric-dipole interaction $\hat{\mathbf{d}}\cdot \hat{\mathbf{e}}$ can be incorporated into Eq.~\ref{eqC3} as
\begin{equation} \label{eqC4}
	\begin{split}
		\hat{H}_I&=-\hat{\mathbf{d}} \cdot \hat{\mathbf{e}} \propto - i(\hat{c}^\dagger-\hat{c})(\hat{a}_k^\dagger+\hat{a}_k)\\
		&\propto (-i)\hat{c}^\dagger \hat{a}_k+(-i)^*\hat{c} \hat{a}_k^\dagger,
	\end{split}
\end{equation}
where the traveling-wave electric-field operator $\hat{\mathbf e}$ can be expanded as
$\hat{\mathbf e} \propto \sum_k (\hat a_k + \hat a_k^\dagger)$,
and is in phase with its magnetic component $\hat{\mathbf h}$. This yields $\tilde{\lambda}_c \propto -i = e^{-i\pi/2}$, 
and hence $\Phi_{cp}=\Phi_{cq}=-\pi/2$ for electric dipole cavity.

 \section{Transmission parameter of single magnon}\label{Transmission parameter of single magnon}
 In this section, we derive the transmission parameter of a single magnon mode side-coupled to a microstrip following similar procedure in Ref.~\cite{yang2020unconventional}. The total Hamiltonian is written as:
 \begin{equation}\label{eqD1}
 	\begin{split}
 		\hat{H}&=\hbar \tilde{\omega}_m \hat{m}^\dagger\hat{m}+\int dk \hbar \omega_k \hat{p}_k \hat{p}^\dagger_k + \int dk \hbar \omega_k \hat{q}_k \hat{q}^\dagger_k \\&+ \int dk \hbar(\tilde{\lambda}_{mp} \hat{m}^\dagger\hat{p}_k + \tilde{\lambda}_{mp}^* \hat{m} \hat{p}^\dagger_k)\\&
 		 + \int dk \hbar(\tilde{\lambda}_{mq} \hat{m}^\dagger\hat{q}_k+\tilde{\lambda}_{mq}^* \hat{m} \hat{q}^\dagger_k).
 	\end{split}
 \end{equation}
Here,  $\hat{m}$ ($\hat{m}^{\dagger}$) is the annihilation (creation) operator of the magnon mode.
$\tilde{\omega}_m=\omega_m-i\alpha_0$ is the Kittel magnon mode, with $\omega_m$ and $\alpha_0$ representing the magnon frequency and the intrinsic damping, respectively. The second and third terms correspond to the right- and left-going traveling wave modes, respectively, with frequency $\omega_k$ and wave vector $k$. $\hat{p}$ ($\hat{p}^{\dagger}$) and $\hat{q}$ ($\hat{q}^{\dagger}$) denote the relative annihilation (creation) operators. The last two terms are the interaction terms between the magnon and traveling waves with rotation wave approximation \cite{zhang2014strongly}, where $\tilde{\lambda}_{mp(q)}=\lambda_{mp(q)}e^{i\Phi_{mp(q)}}$ denote the complex coupling strength with right- (left-)going wave. 

\begin{figure}[b]
	\centering 
	\includegraphics[width=8.8 cm]{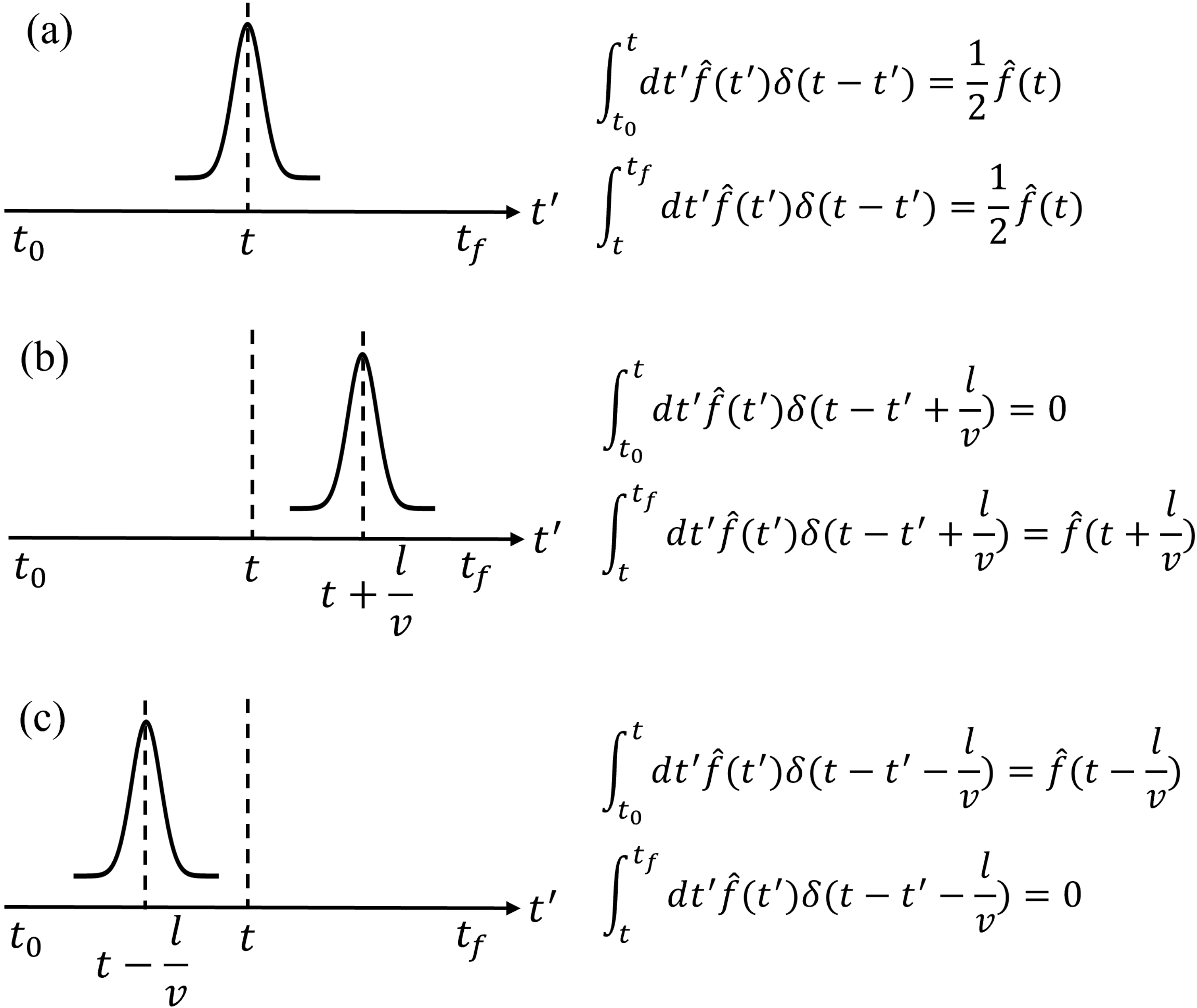}
	\caption{The integral of the Delta function centered at (a) $t'=t$, (b) $t'=t+\frac{l}{v}$, and (c) $t'=t-\frac{l}{v}$ over the intervals $[t_0,t]$ and $[t,t_f]$.}
	\label{fig13}
\end{figure}

Using Eq.~\ref{eqD1}, we get the Heisenberg equation of motion of the magnon mode $\hat{m} (t)$:
\begin{equation}\label{eqD2}
	\begin{split}
		\dot{\hat{m}}=-\frac{i}{\hbar}[\hat{m},\hat{H}]=-i\tilde{\omega}_m \hat{m}-i\int dk \tilde{\lambda}_{mp} \hat{p}_k -i\int dk \tilde{\lambda}_{mq} \hat{q}_k, 
	\end{split}
\end{equation}
and the Heisenberg equations of motion of the traveling waves $\hat{p}_k$ and $\hat{q}_k$:
\begin{equation}\label{eqD3}
	\begin{split}
		\dot{\hat{p}}_k&=-\frac{i}{\hbar}[\hat{p}_k,\hat{H}]=-i \omega_k \hat{p}_k -i \tilde{\lambda}_{mp}^* \hat{m}, \\
		\dot{\hat{q}}_k&=-\frac{i}{\hbar}[\hat{q}_k,\hat{H}]=-i \omega_k \hat{q}_k -i \tilde{\lambda}_{mq}^* \hat{m}.
	\end{split}
\end{equation}
This equation can be solved by considering the initial and final conditions of the right-going wave. The wave enters the cavity at an initial time $t_0 < t$ (before the interaction) and leaves the YIG at a final time $t_f > t$ (after the interaction), leading to:
\begin{equation}\label{eqD4}
	\begin{split}
		\hat{p}_k(t)&=e^{-i\omega_k(t-t_0)}\hat{p}_k(t_0)-i\int^t_{t_0} dt^{'}e^{-i\omega_k(t-t^{'})}\tilde{\lambda}_{mp}^* \hat{m}(t^{'}) , \\
		&=e^{-i\omega_k(t-t_f)}\hat{p}_k(t_f)-i\int^t_{t_f} dt^{'}e^{-i\omega_k(t-t^{'})}\tilde{\lambda}_{mp}^*\hat{m}(t^{'}). 
	\end{split}
\end{equation}
Integrating $\hat{p}_k$, we have:
\begin{subequations}\label{eqD5}
	\begin{align}
	\begin{split}
		\tilde{\lambda}_{mp}\int dk  \hat{p}_k(t)&=\tilde{\lambda}_{mp} \int dk e^{-i\omega_k(t-t_0)}\hat{p}_k(t_0)\\&-i \int^t_{t_0} dt^{'} \tilde{\lambda}_{mp}\tilde{\lambda}_{mp}^*\hat{m}(t^{'})\int dk e^{-i\omega_k(t-t^{'})}, 
	\end{split}
	\\
    \begin{split}
		\tilde{\lambda}_{mp}\int dk  \hat{p}_k(t)&=\tilde{\lambda}_{mp} \int dk e^{-i\omega_k(t-t_f)}\hat{p}_k(t_f)\\&
		-i \int^t_{t_f} dt^{'} \tilde{\lambda}_{mp}\tilde{\lambda}_{mp}^*\hat{m}(t^{'})\int dk e^{-i\omega_k(t-t^{'})}. 
	\end{split}
	\end{align}
\end{subequations}
We further define the input field $\hat{s}_{+1}$ and the output field $\hat{s}_{-2}$ as:
\begin{equation}\label{eqD6}
	\begin{split}
		\hat{s}_{+1}&=\frac{1}{\sqrt{2\pi}}\int dk e^{-i\omega_k(t-t_0)}  \hat{p}_{k}(t_0), \\
		\hat{s}_{-2}&=\frac{1}{\sqrt{2\pi}}\int dk e^{-i\omega_k(t-t_f)}  \hat{p}_{k}(t_f).
	\end{split}
\end{equation}
By employing the identity 
\begin{equation}\label{eqD7}
	\begin{split}
		\int^\infty_{-\infty} dk e^{-i\omega_k(t-t^{'})}=\frac{2\pi}{v} \delta(t-t'),
	\end{split}
\end{equation}
where $v$ is the wave speed, and considering the integrating range shown in Fig.~\ref{fig13}(a), Eq.~\ref{eqD5} can be written as
\begin{subequations}\label{eqD8}
	\begin{align}
	\begin{split}
		\tilde{\lambda}_{mp}\int dk  \hat{p}_k(t)
		&=e^{i\Phi_{mp}}\sqrt{\kappa_{mp}} \hat{s}_{+1} -i\frac{\kappa_{mp}}{2}\hat{m}(t),
	\end{split}
	\\
   \begin{split}
		\tilde{\lambda}_{mp}\int dk  \hat{p}_k(t)
		&=e^{i\Phi_{mp}}\sqrt{\kappa_{mp}}\hat{s}_{-2} +i\frac{\kappa_{mp}}{2}\hat{m}(t),
	\end{split}
	\end{align}
\end{subequations}
where $\kappa_{mp}=2\pi\tilde{\lambda}_{mp}\tilde{\lambda}_{mp}^*$ is the extrinsic damping rate of the magnon mode into the right-going wave.
This formula leads to the input-output relation for the right-going traveling wave:
\begin{equation}\label{eqD9}
	\begin{split}
		\hat{s}_{-2}&=\hat{s}_{+1}-ie^{-i\Phi_{mp}}\sqrt{\kappa_{mp}}\hat{m}(t). 
	\end{split}
\end{equation} 
Similarly, we repeat the steps of Eqs.~\ref{eqD4} to \ref{eqD8} to obtain the relations for the left-going wave $\hat{q}_k(t)$:

\begin{subequations}\label{eqD10}
	\begin{align}
	\begin{split}
		\tilde{\lambda}_{mq}\int dk  \hat{q}_k(t)
		&=e^{i\Phi_{mq}}\sqrt{\kappa_{mq}} \hat{s}_{+2} -i\frac{\kappa_{mq}}{2}\hat{m}(t),
    \end{split}
	   \\
    \begin{split}
		\tilde{\lambda}_{mq}\int dk  \hat{q}_k(t)
		&=e^{i\Phi_{mq}}\sqrt{\kappa_{mq}} \hat{s}_{-1} +i\frac{\kappa_{mq}}{2}\hat{m}(t),
	\end{split}
	\end{align}
\end{subequations}
where
\begin{equation}\label{eqD11} \kappa_{mq}=2\pi\tilde{\lambda}_{mq}\tilde{\lambda}_{mq}^*
\end{equation}
is the extrinsic damping rate of the magnon mode into the left-going wave. And $\hat{s}_{+2}$ and $\hat{s}_{-1}$ are defined as
\begin{equation}\label{eqD12}
	\begin{split}
		\hat{s}_{+2}&=\frac{1}{\sqrt{2\pi}}\int dk e^{-i\omega_k(t-t_0)}  \hat{q}_{k}(t_0), \\
		\hat{s}_{-1}&=\frac{1}{\sqrt{2\pi}}\int dk e^{-i\omega_k(t-t_f)}  \hat{q}_{k}(t_f).
	\end{split}
\end{equation}
From Eq.~\ref{eqD10}, we get the input-output relation for the left-going traveling wave: 
\begin{equation}\label{eqD13}
	\begin{split}
		\hat{s}_{-1}&=\hat{s}_{+2}-ie^{-i\Phi_{mq}}\sqrt{\kappa_{mq}}\hat{m}(t). 
	\end{split}
\end{equation}
Combine Eqs.~\ref{eqD2}, \ref{eqD8}(a) and \ref{eqD10}(a), the equation of motion of the magnon mode is
\begin{equation}\label{eqD14}
	\begin{split}
		\dot{\hat{m}}&=-i(\omega_m-i\beta_0-i\frac{\kappa_{mp}+\kappa_{mq}}{2}) \hat{m}\\&
		- \begin{pmatrix} e^{i\Phi_{mp}}\sqrt{\kappa_{mp}} & e^{i\Phi_{mq}}\sqrt{\kappa_{mq}} \end{pmatrix} \begin{pmatrix} \hat{s}_{+1} \\ \hat{s}_{+2} \end{pmatrix}. \\
	\end{split}
\end{equation}
Setting $\hat{m}(t)=\hat{m}e^{-i\omega t}$, when $\hat{s}_{+2}=0$, we obtain
\begin{equation} \label{eqD15}
		\hat{m}=\frac{e^{i\Phi_{mp}}\sqrt{\kappa_{mp}}\hat{s}_{+1}}
		{\omega-\omega_0+i\beta_0+i\frac{\kappa_{mp}+\kappa_{mq}}{2}}.
\end{equation}
Similarly, when $\hat{s}_{+1}=0$, we have
\begin{equation} \label{eqD16}
		\hat{m}=\frac{e^{i\Phi_{mq}}\sqrt{\kappa_{mq}}\hat{s}_{+2}}
		{\omega-\omega_0+i\beta_0+i\frac{\kappa_{mp}+\kappa_{mq}}{2}}.
\end{equation}
Substituting Eq.~\ref{eqD15} into Eq.~\ref{eqD9}, we derive the transmission parameter using $S_{21}=\frac{\hat{s}_{-2}}{\hat{s}_{+1}}$ by setting $\hat{s}_{+2}=0$:
 \begin{equation}\label{eqD17}
\begin{split}
		S_{21}(\omega)&=\frac{\omega-\omega_m+i\alpha_0+i\frac{\kappa_{mq}-\kappa_{mp}}{2} }
		{\omega-\omega_m+i\alpha_0+i\frac{\kappa_{mq}+\kappa_{mp}}{2}}.\\
			\end{split}
	\end{equation}
Similarly, $S_{12}=\frac{\hat{s}_{-1}}{\hat{s}_{+2}}$ is derived by substituting Eq.~\ref{eqD16} into Eq.~\ref{eqD13} with $\hat{s}_{+1}=0$:
 \begin{equation}\label{eqD18}
	\begin{split} 
		S_{12}(\omega)
		&=\frac{\omega-\omega_m+i\alpha_0+i\frac{\kappa_{mp}-\kappa_{mq}}{2} }
		{\omega-\omega_m+i\alpha_0+i\frac{\kappa_{mq}+\kappa_{mp}}{2}}.\\
	\end{split}
\end{equation}
 \section{Transmission parameter of nonreciprocal strong coupling}\label{Transmission parameter of nonreciprocal coherent coupling}
  In this section, we derive the transmission parameter of nonreciprocal strong coupling.  The system is composed of a magnon mode coherently coupled with a cavity mode that supports a chiral field. The cavity mode simultaneously interacts with a traveling wave, with its chirality determined by the wave’s propagation direction.
  The total Hamiltonian of the coupled system is
	\begin{equation}\label{eqIV1}
		\begin{split}
			\hat{H}=&\hbar \tilde{\omega}_c \hat{c}^\dagger\hat{c}+\hbar \tilde{\omega}_m \hat{m}^\dagger\hat{m}\\&+\int dk \hbar \omega_k \hat{p}_k \hat{p}^\dagger_k  + \int dk \hbar \omega_k \hat{q}_k \hat{q}^\dagger_k  +\hbar (\tilde{g}\hat{c}\hat{m}^\dagger+\tilde{g}^*\hat{c}^\dagger\hat{m}) \\&+ \int dk \hbar( \tilde{\lambda}_{cp} \hat{c}^\dagger\hat{p}_k+\tilde{\lambda}^*_{cp} \hat{c} \hat{p}^\dagger_k) + \int dk \hbar( \tilde{\lambda}_{cq} \hat{c}^\dagger\hat{q}_k+\tilde{\lambda}^*_{cq} \hat{c} \hat{q}^\dagger_k).
		\end{split}
	\end{equation}
Here, the first four terms represent the cavity,  the magnon, the right-going and left-going traveling waves, respectively. The fifth term describes the interaction between the cavity and the magnon mode, where $\tilde{g}=ge^{i\Phi_g}$ is the complex coupling strength. The last two terms are the interaction terms between the cavity mode and traveling waves using the rotation wave approximation \cite{zhang2014strongly}, where $\tilde{\lambda}_{cp(q)}=\lambda_{cp(q)}e^{i\Phi_{cp(q)}}$ denote the complex coupling strength with right- (left-)going wave. 
 The Heisenberg equations of motion of the cavity mode $\hat{c} (t)$ and magnon mode $\hat{m} (t)$ are written as
\begin{equation}\label{eqE2}
	\begin{split}
		\dot{\hat{c}}&=-\frac{i}{\hbar}[\hat{c},\hat{H}]=-i\tilde{\omega}_c \hat{c}-i\tilde{g}^*\hat{m}-i\int dk \tilde{\lambda}_{cp}  \hat{p}_k  -i\int dk \tilde{\lambda}_{cq}  \hat{q}_k , \\
		\dot{\hat{m}}&=-\frac{i}{\hbar}[\hat{m},\hat{H}]=-i\tilde{\omega}_m \hat{m}-i\tilde{g}\hat{c}.
	\end{split}
\end{equation}
And the Heisenberg equations of motion of the traveling waves $\hat{p}_k$ and $\hat{q}_k$ read
\begin{equation}\label{eqIV3}
	\begin{split}
		\dot{\hat{p}}_k&=-\frac{i}{\hbar}[\hat{p}_k,\hat{H}]=-i \omega_k \hat{p}_k -i \tilde{\lambda}^*_{cp} \hat{c}, \\
		\dot{\hat{q}}_k&=-\frac{i}{\hbar}[\hat{q}_k,\hat{H}]=-i \omega_k \hat{q}_k -i \tilde{\lambda}^*_{cq} \hat{c}.
	\end{split}
\end{equation}
Following the calculation process from Eq.~\ref{eqD4} to Eq.~\ref{eqD11} in Appendix \ref{Transmission parameter of single magnon}, we have
\begin{subequations}\label{eqE4}
	\begin{align}
		\begin{split}
			\tilde{\lambda}_{cp}\int dk  \hat{p}_k(t)&=e^{i\Phi_{cp}}\sqrt{\kappa_{c}} \hat{s}_{+1}-i\frac{1}{2}\kappa_{c}\hat{c}(t), 
		\end{split}
		\\
		\begin{split}
			\tilde{\lambda}_{cp}\int dk  \hat{p}_k(t)&=e^{i\Phi_{cp}}\sqrt{\kappa_{c}} \hat{s}_{-2}+i\frac{1}{2} \kappa_{c} \hat{c}(t), 
		\end{split}
	\end{align}
\end{subequations}
and
\begin{equation}\label{eqE5}
	\begin{split}
		\hat{s}_{-2}&=\hat{s}_{+1}-ie^{-i\Phi_{cp}}\sqrt{\kappa_{c}}\hat{c}(t), 
	\end{split}
\end{equation} 
as well as
\begin{subequations}\label{eqE6}
	\begin{align}
		\begin{split}
			\tilde{\lambda}_{cq}\int dk  \hat{q}_k(t)&=e^{i\Phi_{cq}}\sqrt{\kappa_{c}} \hat{s}_{+2}-i\frac{1}{2}\kappa_{c}\hat{c}(t),
		\end{split}
		\\
		\begin{split}
			\tilde{\lambda}_{cq}\int dk  \hat{q}_k(t)&=e^{i\Phi_{cq}}\sqrt{\kappa_{c}} \hat{s}_{-1}+i\frac{1}{2} \kappa_{c} \hat{c}(t), 
		\end{split}
	\end{align}
\end{subequations}
and
\begin{equation}\label{eqE7}
	\begin{split}
		\hat{s}_{-1}&=\hat{s}_{+2}-ie^{-i\Phi_{cq}}\sqrt{\kappa_{c}}\hat{c}(t).
	\end{split}
\end{equation}
Here, we define $\kappa_c=\kappa_{cp(q)}=2\pi\tilde{\lambda}_{cp(q)}\tilde{\lambda}_{cp(q)}^*$, assuming an idealized cavity–waveguide interface where the coupling magnitudes to the two counter-propagating modes are identical, with $\lambda_{cp}=\lambda_{cq}$.  

Plug Eqs.~\ref{eqE4}(a) and \ref{eqE6}(a) into Eq.~\ref{eqE2}, we get
	\begin{equation}\label{eqE8}
		\begin{split}
			\begin{pmatrix} \dot{\hat{c}} \\ \dot{\hat{m}} \end{pmatrix}=& -i \begin{pmatrix} \omega_c-i\beta_0-i\kappa_c & ge^{-i\Phi_g} \\ ge^{i\Phi_g} & \omega_m-i\alpha_0 \end{pmatrix} \begin{pmatrix} \hat{c} \\ \hat{m} \end{pmatrix} \\ & -i \begin{pmatrix} e^{i\Phi_{cp}}\sqrt{\kappa_{c}} & e^{i\Phi_{cq}}\sqrt{\kappa_{c}} \\ 0 & 0 \end{pmatrix} \begin{pmatrix} \hat{s}_{+1} \\ \hat{s}_{+2} \end{pmatrix}. 
		\end{split}
	\end{equation}
Combining Eqs.~\ref{eqE5}, \ref{eqE7} and \ref{eqE8}, and setting $\hat{m}(t)=\hat{m}e^{-i\omega t}$ and $\hat{c}(t)=\hat{c}e^{-i\omega t}$, we can derive the transmission parameter using $S_{21(12)}=\frac{\hat{s}_{-2(-1)}}{\hat{s}_{+1(+2)}}$ by setting $\hat{s}_{+2(+1)}=0$:
	\begin{equation}\label{eqIV12}
		\begin{split}
			S_{21(12)}(\omega)
			&=\frac{(\omega-\tilde{\omega}_m)(\omega-\tilde{\omega}_c)-g_{21(12)}^2}
			{(\omega-\tilde{\omega}_m)(\omega-\tilde{\omega}_c+i\kappa_{c})-g_{21(12)}^2},
		\end{split}
	\end{equation}
where $g_{21(12)}$ represents the magnitude of the coupling strength $\tilde{g}_{21(12)}=g_{21(12)}e^{i\Phi_{g_{21(12)}}}$ when the wave propagates from port 1(2) to port 2(1).

 \section{Transmission parameter of traveling wave-mediated nonreciprocal coupling system} \label{Transmission parameter of coupled system}
\begin{figure*}[htbp]
	\centering 
	\includegraphics[width=8.8 cm]{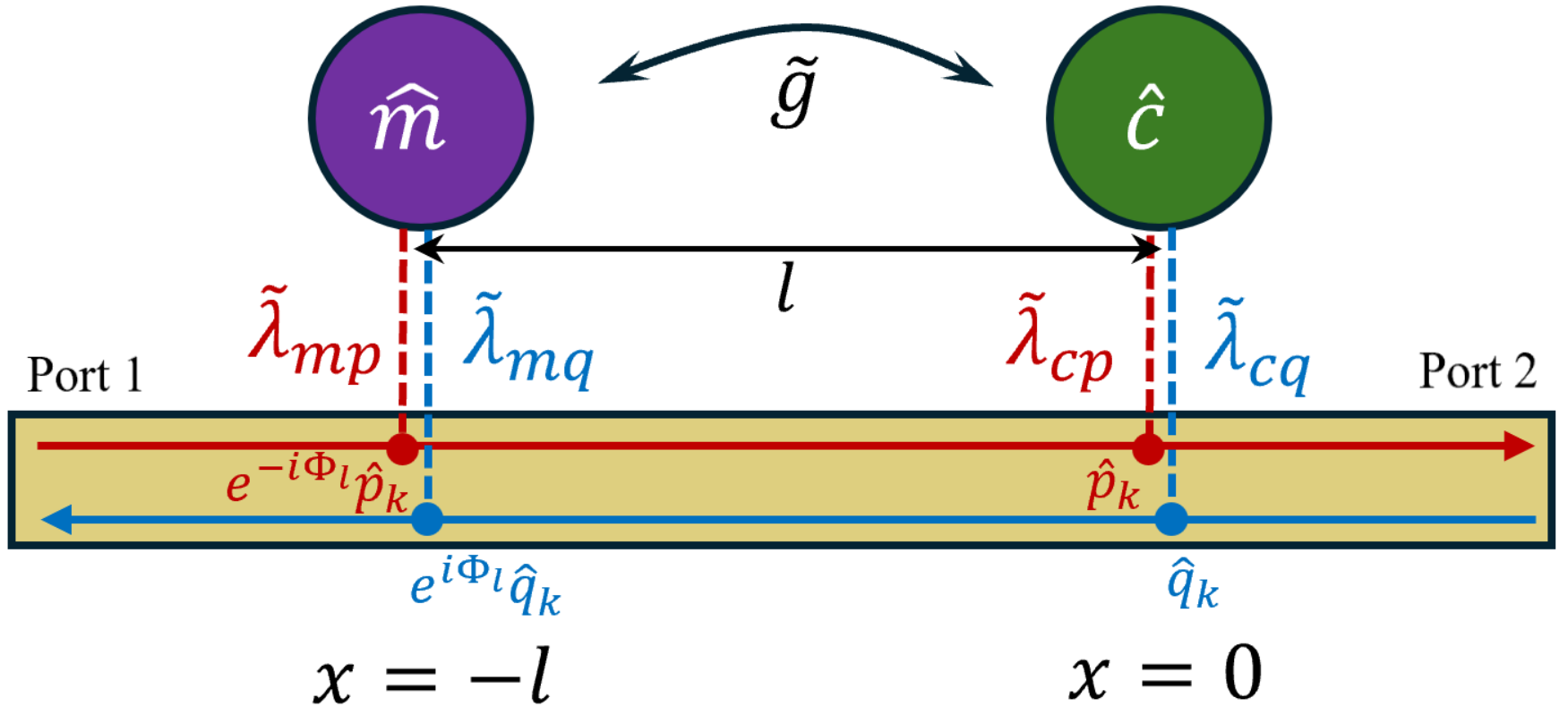}
	\caption{
		Schematic of a traveling-wave-mediated nonreciprocal coupling system.  The magnon mode $\hat{m}$ and cavity mode $\hat{c}$ are separated by a distance $l$ (located at $x=-l$ and $x=0$, respectively). With the cavity position as the reference, the right-going wave $\hat{p}_k$ acquires an advanced phase factor $e^{-i \Phi_l}$, while the left-going wave $\hat{q}_k$ incurs a delayed phase factor $e^{i \Phi_l}$ at the magnon position, where $\Phi_l=\frac{\omega_k l}{v}$ ($\omega_k$ is the frequency of the traveling-wave mode and $v$ is the wave speed). $\tilde{\lambda}_{mp(q)}$ and $\tilde{\lambda}_{cp(q)}$ are the coupling strengths of the magnon and cavity modes with the right- (left)-going wave, respectively, and $\tilde{g}$ is the direct coupling strength between $\hat{c}$ and $\hat{m}$. }
	\label{fig14}
\end{figure*}
In this section, we derive the transmission parameter for the traveling wave-mediated nonreciprocal coupling. The system consists of a YIG sphere and a cavity, separated by a distance 
$l$ and mounted beside a common microstrip. As illustrated in Fig. \ref{fig14}, the two modes are coupled both directly via field overlap and indirectly through traveling waves. The total Hamiltonian of the coupled system is \cite{lalumiere2013input,kannan2020generating,sinha2020non,kannan2020waveguide,wang2022giant,wang2025interpreting,wang2025dynamic}:
\begin{widetext}
	\begin{equation}\label{eqF1}
		\begin{split}
			\hat{H}=&\hbar \tilde{\omega}_c \hat{c}^\dagger\hat{c}+\hbar \tilde{\omega}_m \hat{m}^\dagger\hat{m}+\int dk \hbar \omega_k \hat{p}_k \hat{p}^\dagger_k  + \int dk \hbar \omega_k \hat{q}_k \hat{q}^\dagger_k +\hbar (\tilde{g}\hat{c}\hat{m}^\dagger+\tilde{g}^*\hat{c}^\dagger\hat{m}) \\&+ \int dk \hbar( \tilde{\lambda}_{cp} \hat{c}^\dagger\hat{p}_k+\tilde{\lambda}^*_{cp} \hat{c} \hat{p}^\dagger_k) + \int dk \hbar( \tilde{\lambda}_{cq} \hat{c}^\dagger\hat{q}_k+\tilde{\lambda}^*_{cq} \hat{c} \hat{q}^\dagger_k)\\&+ \int dk \hbar ( e^{-i\Phi_l}\tilde{\lambda}_{mp} \hat{m}^\dagger\hat{p}_k+e^{i\Phi_l}\tilde{\lambda}_{mp}^*\hat{m} \hat{p}^\dagger_k) + \int dk \hbar(e^{i\Phi_l} \tilde{\lambda}_{mq}  \hat{m}^\dagger\hat{q}_k+e^{-i\Phi_l} \tilde{\lambda}_{mq}^*  \hat{m} \hat{q}^\dagger_k).
		\end{split}
	\end{equation}
\end{widetext}
Here, the first four terms represent the cavity,  the magnon, the right-going and left-going traveling waves, respectively. The fifth term describes the interaction between the cavity and the magnon mode. The following two terms are the interaction terms between the cavity mode and traveling waves. And the last two terms are the interaction between the magnon mode and the traveling waves with the traveling phase $\Phi_l=\frac{2\pi l}{\lambda}$, where $\lambda$ is the wavelength of the traveling wave. 

Using the Hamiltonian, we obtain the Heisenberg equations of motion of the cavity mode $\hat{c} (t)$ and magnon mode $\hat{m} (t)$:
\begin{equation}\label{eqF2}
	\begin{split}
		\dot{\hat{c}}&=-\frac{i}{\hbar}[\hat{c},\hat{H}]\\&=-i(\omega_c-i\beta_0) \hat{c}-i\tilde{g}^*\hat{m}-i\int dk \tilde{\lambda}_{cp}  \hat{p}_k  -i\int dk \tilde{\lambda}_{cq}  \hat{q}_k , \\
		\dot{\hat{m}}&=-\frac{i}{\hbar}[\hat{m},\hat{H}]\\&=-i(\omega_m-i\alpha_0) \hat{m}-i\tilde{g}\hat{c}-i\int dk e^{-i\Phi_{l}} \tilde{\lambda}_{mp} \hat{p}_k \\&-i\int dk e^{i\Phi_{l}} \tilde{\lambda}_{mq} \hat{q}_k,
	\end{split}
\end{equation}
and the Heisenberg equations of motion of the traveling waves $\hat{p}_k$ and $\hat{q}_k$: 

\begin{equation}\label{eqF3}
	\begin{split}
		\dot{\hat{p}}_k&=-\frac{i}{\hbar}[\hat{p}_k,\hat{H}]=-i \omega_k \hat{p}_k -i \tilde{\lambda}^*_{cp} \hat{c}-i \tilde{\lambda}_{mp}^*  e^{i\Phi_{l}} \hat{m}, \\
		\dot{\hat{q}}_k&=-\frac{i}{\hbar}[\hat{q}_k,\hat{H}]=-i \omega_k \hat{q}_k -i \tilde{\lambda}^*_{cq} \hat{c}-i \tilde{\lambda}_{mq}^*  e^{-i\Phi_{l}} \hat{m}.
	\end{split}
\end{equation}
Considering the initial and final conditions at the initial time $t_0$ and the final time $t_f$, we have  
\begin{subequations}\label{eqF4}
	\begin{align}
		\begin{split}
		\hat{p}_k(t)&=e^{-i\omega_k(t-t_0)}\hat{p}_k(t_0)\\&-i\int^t_{t_0} dt^{'}e^{-i\omega_k(t-t^{'})}[\tilde{\lambda}^*_{cp}\hat{c}(t^{'})+\tilde{\lambda}_{mp}^*e^{i\Phi_{l}} \hat{m}(t^{'})] , 
		\end{split}
			\\
	\begin{split}
		\hat{p}_k(t)&=e^{-i\omega_k(t-t_f)}\hat{p}_k(t_f)\\&-i\int^t_{t_f} dt^{'}e^{-i\omega_k(t-t^{'})} [\tilde{\lambda}^*_{cp}\hat{c}(t^{'})+\tilde{\lambda}_{mp}^*e^{i\Phi_{l}} \hat{m}(t^{'})]. 
	\end{split}
	\end{align}
\end{subequations}

Here, $e^{i\Phi_l}=e^{i\omega_k \frac{l}{v}}$ is a function of traveling-wave frequency $\omega_k$, where $v$ is the wave speed. Using the identity in Eq.~\ref{eqD7} in Appendix \ref{Transmission parameter of single magnon}, we integrate $\hat{p}_k$ over all the $k$ modes:
\begin{subequations}\label{eqF5}
	\begin{align}
	\begin{split}
		\tilde{\lambda}_{cp}\int dk  \hat{p}_k(t)&=\sqrt{2\pi}\tilde{\lambda}_{cp} \hat{s}_{+1}\\&-i \int^t_{t_0} dt^{'} \tilde{\lambda}_{cp}\tilde{\lambda}^*_{cp}\hat{c}(t^{'})2\pi\delta(t-t^{'})\\&-i\int^t_{t_0} dt^{'}\tilde{\lambda}_{mp}^*\tilde{\lambda}_{cp} \hat{m}(t^{'}) 2\pi\delta(t-t^{'}-\frac{l}{c}) , 
		\end{split}
		\\
		\begin{split}
		\tilde{\lambda}_{cp}\int dk  \hat{p}_k(t)&=\sqrt{2\pi}\tilde{\lambda}_{cp} \hat{s}_{-2}\\&-i \int^t_{t_f} dt^{'} \tilde{\lambda}_{cp}\tilde{\lambda}^*_{cp}\hat{c}(t^{'})2\pi\delta(t-t^{'})\\&-i\int^t_{t_f} dt^{'}\tilde{\lambda}_{mp}^*\tilde{\lambda}_{cp} \hat{m}(t^{'}) 2\pi\delta(t-t^{'}-\frac{l}{c}).
	\end{split}
	\end{align}
\end{subequations}
Considering the integrating range shown in Figs.~\ref{fig13}(b) and (c), we have
\begin{subequations}\label{eqF6}
	\begin{align}
		\begin{split}
			\tilde{\lambda}_{cp}\int dk  \hat{p}_k(t)&=e^{i\Phi_{cp}}\sqrt{\kappa_{c}} \hat{s}_{+1}-i\frac{1}{2}\kappa_{c}\hat{c}(t)\\&-ie^{i(\Phi_{cp}-\Phi_{mp})}\sqrt{\kappa_{c}\kappa_{m}}e^{i\Phi_{l}}\hat{m}(t), 
		\end{split}
		\\
		\begin{split}
			\tilde{\lambda}_{cp}\int dk  \hat{p}_k(t)&=e^{i\Phi_{cp}}\sqrt{\kappa_{c}} \hat{s}_{-2}+i\frac{1}{2} \kappa_{c} \hat{c}(t). 
		\end{split}
	\end{align}
\end{subequations}

For linearly polarized traveling-wave fields, $\kappa_m=\kappa_{mp(q)}=2\pi\tilde{\lambda}_{mp(q)}\tilde{\lambda}_{mp(q)}^*$.
Combining Eq.~\ref{eqF6}(a) and (b), we get
\begin{equation}\label{eqF7}
	\begin{split}
		\hat{s}_{-2}&=\hat{s}_{+1}-i\begin{pmatrix}e^{-i\Phi_{cp}}\sqrt{\kappa_{c}} & e^{i(\Phi_{l}-\Phi_{mp})} \sqrt{\kappa_m}  \end{pmatrix} \begin{pmatrix} \hat{c} \\ \hat{m} \end{pmatrix}. \\
	\end{split}
\end{equation} 
Similarly, we have
\begin{subequations}\label{eqF8}
	\begin{align}
		\begin{split}
			\tilde{\lambda}_{cq}\int dk  \hat{q}_k(t)&=e^{i\Phi_{cq}}\sqrt{\kappa_{c}} \hat{s}_{+2}-i\frac{1}{2}\kappa_{c}\hat{c}(t),
		\end{split}
		\\
		\begin{split}
			\tilde{\lambda}_{cq}\int dk  \hat{q}_k(t)&=e^{i\Phi_{cq}}\sqrt{\kappa_{c}} \hat{s}_{-1}+i\frac{1}{2} \kappa_{c} \hat{c}(t)\\&   +ie^{i(\Phi_{cq}-\Phi_{mq})}\sqrt{\kappa_{c}\kappa_{m}}e^{-i\Phi_{l}}\hat{m}(t), 
		\end{split}
	\end{align}
\end{subequations}
which gives
\begin{equation}\label{eqF9}
	\begin{split}
		\hat{s}_{-1}&=\hat{s}_{+2}-i\begin{pmatrix} e^{-i\Phi_{cq}}\sqrt{\kappa_{c}} & e^{-i(\Phi_{l}+\Phi_{mq})} \sqrt{\kappa_m}  \end{pmatrix} \begin{pmatrix} \hat{c} \\ \hat{m} \end{pmatrix}.
	\end{split}
\end{equation}
We can also derive 
\begin{equation}\label{eqF10}
		\begin{split}
			\tilde{\lambda}_{mp}\int dk e^{-i\Phi_{l}} \hat{p}_k(t)
			&=e^{-i\Phi_{l}}e^{i\Phi_{mp}}\sqrt{\kappa_m}\hat{s}_{+1}-i\frac{1}{2}\kappa_{m}\hat{m}(t), 
		\end{split}
\end{equation}
and
\begin{equation}\label{eqF11}
		\begin{split}
			\tilde{\lambda}_{mq}\int dk e^{i\Phi_{l}} \hat{q}_k(t)
			&=e^{i\Phi_{l}}e^{i\Phi_{mq}}\sqrt{\kappa_{m}} \hat{s}_{+2}-i\frac{1}{2}\kappa_{m} \hat{m}(t) \\&-ie^{i(\Phi_{mq}-\Phi_{cq})}\sqrt{\kappa_{c}\kappa_{m}}e^{i\Phi_{l}} \hat{c}(t).
		\end{split}
\end{equation} 

Plug Eqs.~\ref{eqF6}(a), \ref{eqF8}(a), \ref{eqF10} and \ref{eqF11} into Eq.~\ref{eqF2}, we obtain the coupled equations:

\vspace{0.8em}
\begin{widetext}
\begin{equation}\label{eqF12}
	\begin{split}
		\begin{pmatrix} \dot{\hat{c}} \\ \dot{\hat{m}} \end{pmatrix}&=-i \begin{pmatrix} \omega_c-i\beta_0-i\kappa_c & ge^{-i\Phi_g}+\tilde{G}_le^{i(\Phi_{cp}-\Phi_{mp})} \\ ge^{i\Phi_g}+\tilde{G}_le^{i(-\Phi_{cq}+\Phi_{mq})} & \omega_m-i\alpha_0-i\kappa_m \end{pmatrix} \begin{pmatrix} \hat{c} \\ \hat{m} \end{pmatrix} -i \begin{pmatrix} e^{i\Phi_{cp}}\sqrt{\kappa_{c}} & e^{i\Phi_{cq}}\sqrt{\kappa_{c}} \\ e^{-i\Phi_{l}}e^{i\Phi_{mp}}\sqrt{\kappa_{m}} & e^{i\Phi_{l}}e^{i\Phi_{mq}}\sqrt{\kappa_{m}} \end{pmatrix} \begin{pmatrix} \hat{s}_{+1} \\ \hat{s}_{+2} \end{pmatrix},
	\end{split}
\end{equation}
\end{widetext}
where $\tilde{G}_l=-i\sqrt{\kappa_c\kappa_m}e^{i\Phi_l}$ is the traveling-wave mediated coupling strength. Combining Eqs.~\ref{eqF7}, \ref{eqF9} and \ref{eqF12}, we can derive the transmission parameter using $S_{21(12)}=\frac{\hat{s}_{-2(-1)}}{\hat{s}_{+1(+2)}}$ by setting $\hat{s}_{+2(+1)}=0$:

\begin{widetext}
\begin{equation}
	\begin{split}
		 S_{21}(\omega) &=\frac{(\omega-\tilde{\omega}_m)(\omega-\tilde{\omega}_c)-g [g+\tilde{G_l}e^{i(-\Phi_{cq}+\Phi_{mq}-\Phi_{g})}+\tilde{G_l}^*e^{i(-\Phi_{cp}+\Phi_{mp}-\Phi_{g})}] }
		{(\omega-\tilde{\omega}_m+i\kappa_m)(\omega-\tilde{\omega}_c+i\kappa_c)-[\tilde{g}^*+\tilde{G_l}e^{i(\Phi_{cp}-\Phi_{mp})}][\tilde{g}+\tilde{G_l}e^{i(-\Phi_{cq}+\Phi_{mq})}]},\\
	\end{split}
	\label{eqF13}
\end{equation}

	\begin{equation}
		\begin{split}
			S_{12}(\omega)
	        &=\frac{(\omega-\tilde{\omega}_m)(\omega-\tilde{\omega}_c)-g[g+\tilde{G_l}e^{i(\Phi_{cp}-\Phi_{mp}+\Phi_g)}+\tilde{G_l}^*e^{i(\Phi_{cq}-\Phi_{mq}+\Phi_g})] }
			{(\omega-\tilde{\omega}_m+i\kappa_m)(\omega-\tilde{\omega}_c+i\kappa_c)-[\tilde{g}^*+\tilde{G_l}e^{i(\Phi_{cp}-\Phi_{mp})}][\tilde{g}+\tilde{G_l}e^{i(-\Phi_{cq}+\Phi_{mq})}]}.
		\end{split}
		\label{eqF14}
	\end{equation}
\end{widetext}
Here, we set the terms in the numerators of $S_{21}$ and $S_{12}$ as $\tilde{C}_{21}=g [g+\tilde{G_l}e^{i(-\Phi_{cq}+\Phi_{mq}-\Phi_{g})}+\tilde{G_l}^*e^{i(-\Phi_{cp}+\Phi_{mp}-\Phi_{g})}]$ and $\tilde{C}_{12}=g [g+\tilde{G_l}e^{i(\Phi_{cp}-\Phi_{mp}+\Phi_{g})}+\tilde{G_l}^*e^{i(\Phi_{cq}-\Phi_{mq}+\Phi_{g})}]$. We notice that 
	\begin{equation}\label{eqF15}
		\begin{split}
			\tilde{C}_{21}=\tilde{C}_{12}^*&=g [g-2\sqrt{\kappa_c\kappa_m}\sin\frac{\Phi_1}{2}e^{i\frac{\Phi_2}{2}}]\\
		\end{split}
	\end{equation}
where $\Phi_1=\Phi_{cp}-\Phi_{cq}+\Phi_{mq}-\Phi_{mp}+2\Phi_l$ and $\Phi_2=-(\Phi_{cq}+\Phi_{cp})+\Phi_{mq}+\Phi_{mp}-2\Phi_g$.
\begin{figure}[!htbp]
	\centering 
	\includegraphics[width=8.8 cm]{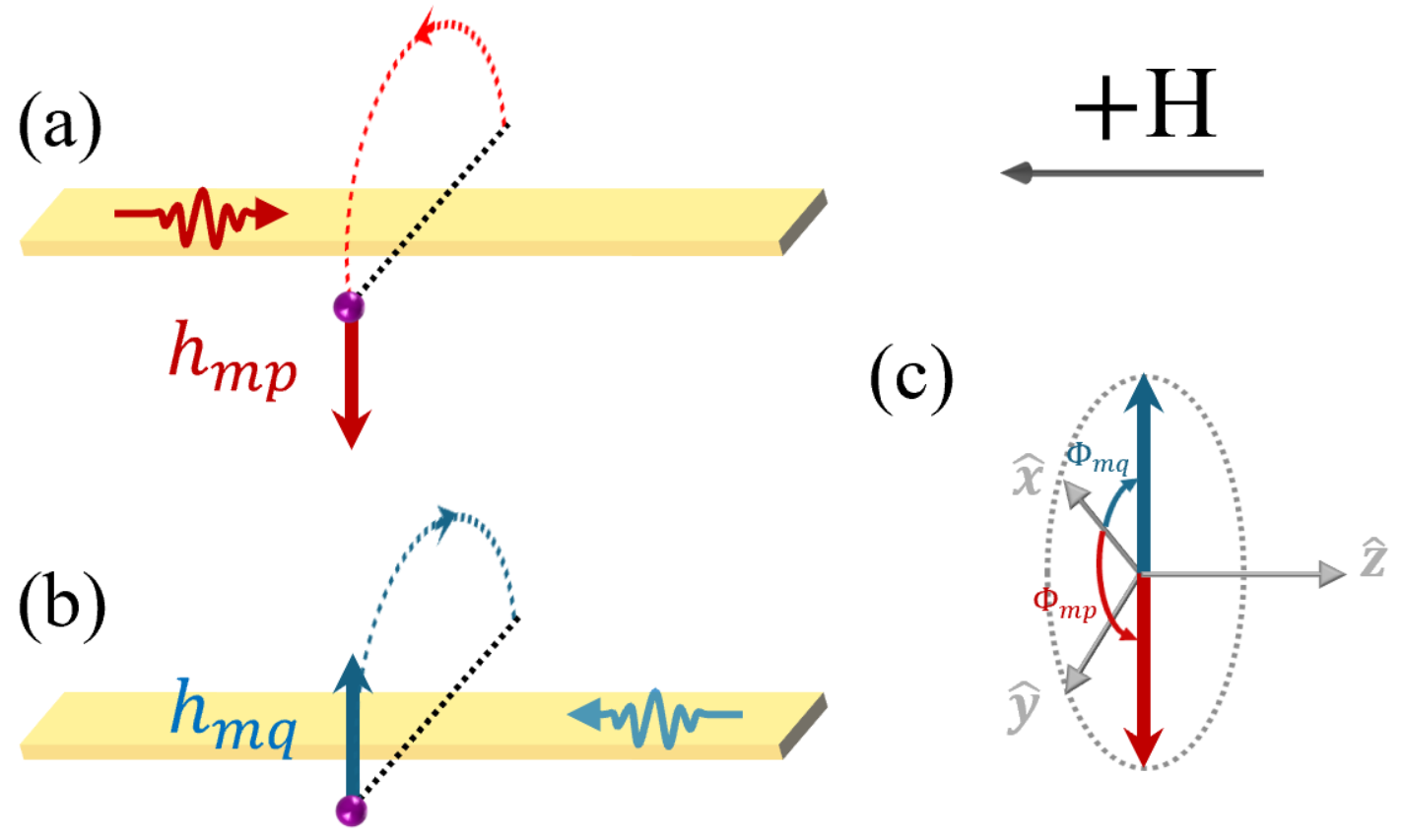}
	\caption{Schematic of the $\pi$ shift in traveling-wave-magnon coupling phases under propagation reversal. (a), (b) The Oersted field $\mathbf{h}_{mp(q)}$ [red (blue) arrow] generated by the right- (left)-going traveling wave. (c) The phase difference between $\mathbf{h}_{mp}$ and $\mathbf{h}_{mq}$ is $\Phi_{mp}-\Phi_{mq}=\pi\;(\mathrm{mod}\; 2\pi)$ accounting for the angular periodicity.}
	\label{fig15}
\end{figure}

This can be further simplified using physical constraint on the coupling phase relations. Firstly, counter-propagating waves generate Oersted fields in opposite directions, resulting in their coupling phases differing by $\pi$, leading to $\Phi_{mp}-\Phi_{mq}=\pi \;(\mathrm{mod}\; 2\pi)$ [Fig. \ref{fig15}]. Secondly, we have $\Phi_{cp}=0$ and $\Phi_{cq}=\pi$ for magnetic-dipole cavity and $\Phi_{cp}=-\pi/2$ and $\Phi_{cq}=-\pi/2$ for electric-dipole cavity.  Therefore, $\tilde{C}_{21}$ can be further reduced as

\begin{equation}\label{eqF16}
	\begin{split}
 \tilde{C}_{21}=\tilde{C}_{12}^*
  &=g [g+2\sqrt{\kappa_c\kappa_m}\sin\Phi_le^{i(\Phi_{mp}-\Phi_g)}]\\
 	\end{split}
\end{equation}
for magnetic-dipole cavity, and
\begin{equation}\label{eqF17}
	\begin{split}
		\tilde{C}_{21}=\tilde{C}_{12}^*
		&=g [g-2\sqrt{\kappa_c\kappa_m}\cos\Phi_le^{i(\Phi_{mp}-\Phi_g)}]\\
	\end{split}
\end{equation}
 for electric-dipole cavity.
 
 \section{Effective Hamiltonian} \label{Effective Hamiltonian}
 In our case, the coupling characteristics are dominated by the numerator of $S_{21(12)}$, where the spectra exhibit coupled dip resonances. The effective coupling structure can be analyzed by the zeros of $S_{21(12)}$, which requires
 \begin{equation} \label{eqH1}
 	(\omega-\tilde{\omega}_m)(\omega-\tilde{\omega}_c)-\tilde{C}_{21(12)}=0,
 \end{equation}
 according to Eq.~\ref{eq18}. This can be equivalently written as the characteristic equation of a matrix $\hat{H}_{eff}^{21(12)}$ as
 \begin{equation} \label{eqH2}
 	\text{det}(\omega I-\hat{H}_{eff}^{21(12)})=0,
 \end{equation}
 and the transmission zeros can be analyzed via the eigenvalue problem:
 \begin{equation} \label{eqH3}
 	(\omega I-\hat{H}_{eff}^{21(12)}) \left | \lambda_\pm  \right \rangle =0,
 \end{equation}
 where 
 \begin{equation} \label{eqH4}
 	\frac{\hat{H}_{\mathrm{eff}}^{21(12)}}{\hbar}
 	= \begin{pmatrix}
 		\tilde{\omega}_c & -\sqrt{\tilde{C}_{21(12)}}\\[2pt]
 		-\sqrt{\tilde{C}_{21(12)}} & \tilde{\omega}_m
 	\end{pmatrix}
 \end{equation}
 is defined as the effective Hamiltonian, and $\left | \lambda_\pm  \right \rangle$ are the eigenstates. Physically, it describes the coupling of two effective modes with complex frequencies $\tilde{\omega}_c=\omega_c-i\beta_0$ and $\tilde{\omega}_m=\omega_m-i\alpha_0$. Under this framework, the zeros of $S_{21}$ correspond to the eigenvalues of the effective Hamiltonian,  while the eigenstates characterize the mode configurations at these singular points.
 The eigenvalues are given by 
 \begin{equation} \label{H5}
 	\tilde{\omega}_{\pm}=\frac{(\tilde{\omega}_m+\tilde{\omega}_c)\pm \sqrt{(\tilde{\omega}_m-\tilde{\omega}_c)^2+4\tilde{C}_{21(12)}}}{2}.
 \end{equation}
 The experimentally relevant zeros located on the real-frequency axis are obtained when the imaginary part of the eigenvalues of the effective Hamiltonian becomes zero, at which the zero-damping conditions (ZDC) are derived.
 
 \section{Unitary transformation and gauge-invariant phase} \label{Unitary transformation and gauge-invariant phase}
 \begin{figure}[htbp]
 	\centering 
 	\includegraphics[width=8.8 cm]{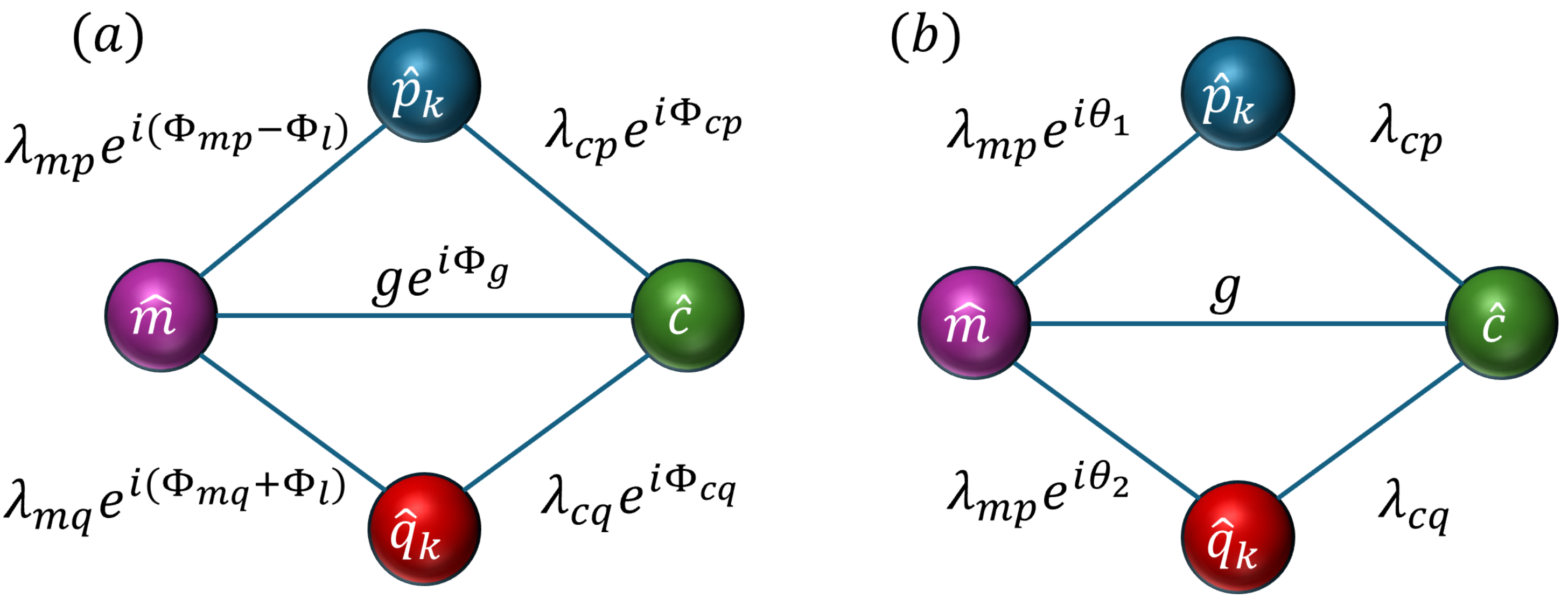}
 	\caption{ Schematic of the coupling phases of the traveling-wave-mediated coupling system. (a) The original system with all the coupling phases $\Phi_{mp,q}$, $\Phi_{cp,q}$ and $\Phi_g$, and the traveling phase $\Phi_l$. (b) The system after unitary transformation, with only two gauge-invariant phases $\theta_1$ and $\theta_2$ left.}
 	\label{fig16}
 \end{figure}
 In this section, we use unitary transformation to demonstrate the gauge-invariance of the synthetic phase in traveling-wave-mediated coupling system. Unitary transformation of the interaction Hamiltonian serves as powerful method to eliminate redundant phase to simplify coupling system \cite{gardin2023manifestation, gardin2024engineering}. Here, the interaction Hamiltonian from Eq. \ref{eqF1} is written as:
 \begin{equation}\label{eqH1}
 	\begin{split}
 		\hat{H}_{int}=&\hbar ge^{i\Phi_g}\hat{m}^\dagger\hat{c} + \int dk \hbar \lambda_{cp} e^{i\Phi_{cp}} \hat{c}^\dagger\hat{p}_k\\&  + \int dk \hbar \lambda_{cq} e^{i\Phi_{cq}} \hat{c}^\dagger\hat{q}_k + \int dk \hbar e^{i(\Phi_{mp}-\Phi_l)}\lambda_{mp} \hat{m}^\dagger\hat{p}_k\\& + \int dk \hbar e^{i(\Phi_{mq}+\Phi_l)} \lambda_{mq}  \hat{m}^\dagger\hat{q}_k+h.c..
 	\end{split}
 \end{equation}

 Utilizing the unitary transformations of $U_1=e^{-i\Phi_g\hat{m}^\dagger \hat{m}}$;  $U_2=e^{i\Phi_{cp}\int dk \hat{p}_k^\dagger \hat{p}_k}$; $U_3=e^{i\Phi_{cq} \int dk \hat{q}_k^\dagger \hat{q}_k}$ sequentially, the interaction Hamiltonian can be transformed as
  \begin{equation}\label{eqH2}
 	\begin{split}
 		\hat{H}_{int}=&\hbar g\hat{m}^\dagger \hat{c}+ \int dk \hbar \lambda_{cp} \hat{c}^\dagger\hat{p}_k + \int dk \hbar \lambda_{cq} \hat{c}^\dagger\hat{q}_k\\&+ \int dk \hbar e^{i\theta_1}\lambda_{mp} \hat{m}^\dagger\hat{p}_k + \int dk \hbar e^{i\theta_2} \lambda_{mq}  \hat{m}^\dagger\hat{q}_k\\&+h.c.,
 	\end{split}
 \end{equation}
 where $\theta_1=-\Phi_{cp}+\Phi_{mp}-\Phi_l-\Phi_g$ and $\theta_2=-\Phi_{cq}+\Phi_{mq}+\Phi_l-\Phi_g$ are the gauge-invariant phases.  
 
 The process can be illustrated with schematics in Fig. \ref{fig16}. In Fig. \ref{fig16}(a), the original interaction Hamiltonian has two coupling loops, and these constraints lead to two gauge-invariant phases $\theta_1$ and $\theta_2$ in Fig. \ref{fig16}(b) under unitary transformation \cite{gardin2023manifestation, gardin2024engineering}.
 
Furthermore, by recombining the gauge-invariant phases as: $\Phi_1=(\theta_2-\theta_1)/2=(\Phi_{cp}-\Phi_{cq}+\Phi_{mq}-\Phi_{mp}+2\Phi_l)/2$ and  $\Phi_2=(\theta_2+\theta_1)/2=[-(\Phi_{cq}+\Phi_{cp})+\Phi_{mq}+\Phi_{mp}-2\Phi_g]/2$, we obtain the $\Phi_{1,2}$ in Eq. \ref{eqF15}, which are also gauge-invariant. Hence, the effective coupling strength $\tilde{C}_{21(12)}$ and the transmission coefficient $S_{21(12)}$ depend on two gauge-invariant phases. With the physical constraints on $\Phi_{mp,q}$ and $\Phi_{cp,q}$ discussed in Appendix \ref{Transmission parameter of coupled system}, the two gauge-invariant phases $\Phi_{1,2}$ are further reduced to two phases with clear physical meanings: the traveling phase $\Phi_l$ (determined by the resonators distance) and synthetic phase $\Psi_{21}=\Phi_{mp}-\Phi_g$ (describe the synthetic chirality, as shown in Fig. \ref{fig4}(d) and (e) in the main text), as shown in Eqs. \ref{eqF16} and \ref{eqF17}.  

\section{Mirror symmetry in synthetic-chirality-induced nonreciprocity} \label{Mirror symmetry in synthetic-chirality-induced nonreciprocity}

In this section, we derive the unique mirror symmetry $|S_{21}(\Delta_m,\Delta_c)|=|S_{12}(-\Delta_m,-\Delta_c)|$ in synthetic-chirality-induced nonreciprocity. Using the definitions $\Delta_m=\omega_m-\omega_c$ and  $\Delta_c=\omega-\omega_c$, we rewrite Eq.~\ref{eq18} as functions of $\Delta_m$ and $\Delta_c$:
\begin{equation} \label{eqG1}
	\begin{split}
	&S_{21}(\Delta_m,\Delta_c)\\
	&=\frac{(\Delta_c-\Delta_m+i\alpha_0)(\Delta_c+i\beta_0)-\tilde{C}_{21}}{(\Delta_c-\Delta_m+i\alpha_0+i\kappa_m)(\Delta_c+i\beta_0+i\kappa_c)-\tilde{D}},
	\end{split}
\end{equation}
and
	\begin{equation} \label{eqG2}
		\begin{split}
			&S_{12}(-\Delta_m,-\Delta_c)\\
			&=\frac{(\Delta_c-\Delta_m-i\alpha_0)(\Delta_c-i\beta_0)-\tilde{C}_{12}}{(\Delta_c-\Delta_m-i\alpha_0-i\kappa_m)(\Delta_c-i\beta_0-i\kappa_c)-\tilde{D}},
		\end{split}
	\end{equation}
where $\tilde{D}=[\tilde{g}^*+\tilde{G_l}e^{i(\Phi_{cp}-\Phi_{mp})}][\tilde{g}+\tilde{G_l}e^{i(-\Phi_{cq}+\Phi_{mq})}]$.
Since $\tilde{C}_{21}=\tilde{C}^*_{12}$, we immediately notice that
\begin{equation} \label{eqG3}
 S_{21}(\Delta_m,\Delta_c)=S_{12}^*(-\Delta_m,-\Delta_c), 
\end{equation}
if and only if $\tilde{D} \in \mathbb{R}$.
Thus the necessary and sufficient condition for the mirror symmetry $|S_{21}(\Delta_m,\Delta_c)|=|S_{12}(-\Delta_m,-\Delta_c)|$ is that $\operatorname{Im}(\tilde{D})=0$, requiring:
\begin{widetext}
\begin{equation} \label{eqG4}
	\operatorname{Im}(\tilde{D})=2\sin(\frac{\psi_1+\psi_2}{2})[g\sqrt{\kappa_c\kappa_m}\cos(\frac{\psi_1-\psi_2}{2})+\kappa_c\kappa_m\cos(\frac{\psi_1+\psi_2}{2})]=0,
\end{equation}
\end{widetext}
where $\psi_1=\Phi_l+\Phi_{cp}-\Phi_{mp}+\Phi_g-\frac{\pi}{2}$ and $\psi_2=\Phi_l-\Phi_{cq}+\Phi_{mq}-\Phi_g-\frac{\pi}{2}$. A phase condition that guarantees Eq.~\ref{eqG4} for arbitrary parameter sets of $\kappa_c$, $\kappa_m$ and $g$ is $\frac{\psi_1+\psi_2}{2}=n\pi~(n \in \mathbb{Z})$, which leads to:
\begin{equation} \label{eqG5}
	\Phi_{cp}-\Phi_{cq}+\Phi_{mq}-\Phi_{mp}+2\Phi_l=(2n+1)\pi~(n \in \mathbb{Z}).
\end{equation}
For magnetic dipole interaction with $\Phi_{mq}-\Phi_{mp}=\pi\;(\mathrm{mod}\; 2\pi)$, $\Phi_{cp}=0$ and $\Phi_{cq}=\pi$, this reduces to 
\begin{equation} \label{eqG6}
	\Phi_l=(n+\frac{1}{2})\pi~(n \in \mathbb{Z}).
\end{equation}
For electric dipole interaction with $\Phi_{mq}-\Phi_{mp}=\pi\;(\mathrm{mod}\; 2\pi)$, $\Phi_{cp}=-\pi/2$ and $\Phi_{cq}=-\pi/2$, we instead require
\begin{equation} \label{eqG7}
	\Phi_l=n\pi~(n \in \mathbb{Z}).
\end{equation}

\section{Distance dependence of synthetic-chirality-induced nonreciprocity} \label{Distance dependence of synthetic-chirality-induced nonreciprocity}
\begin{figure}[h]
	\centering 
	\includegraphics[width=8.8 cm]{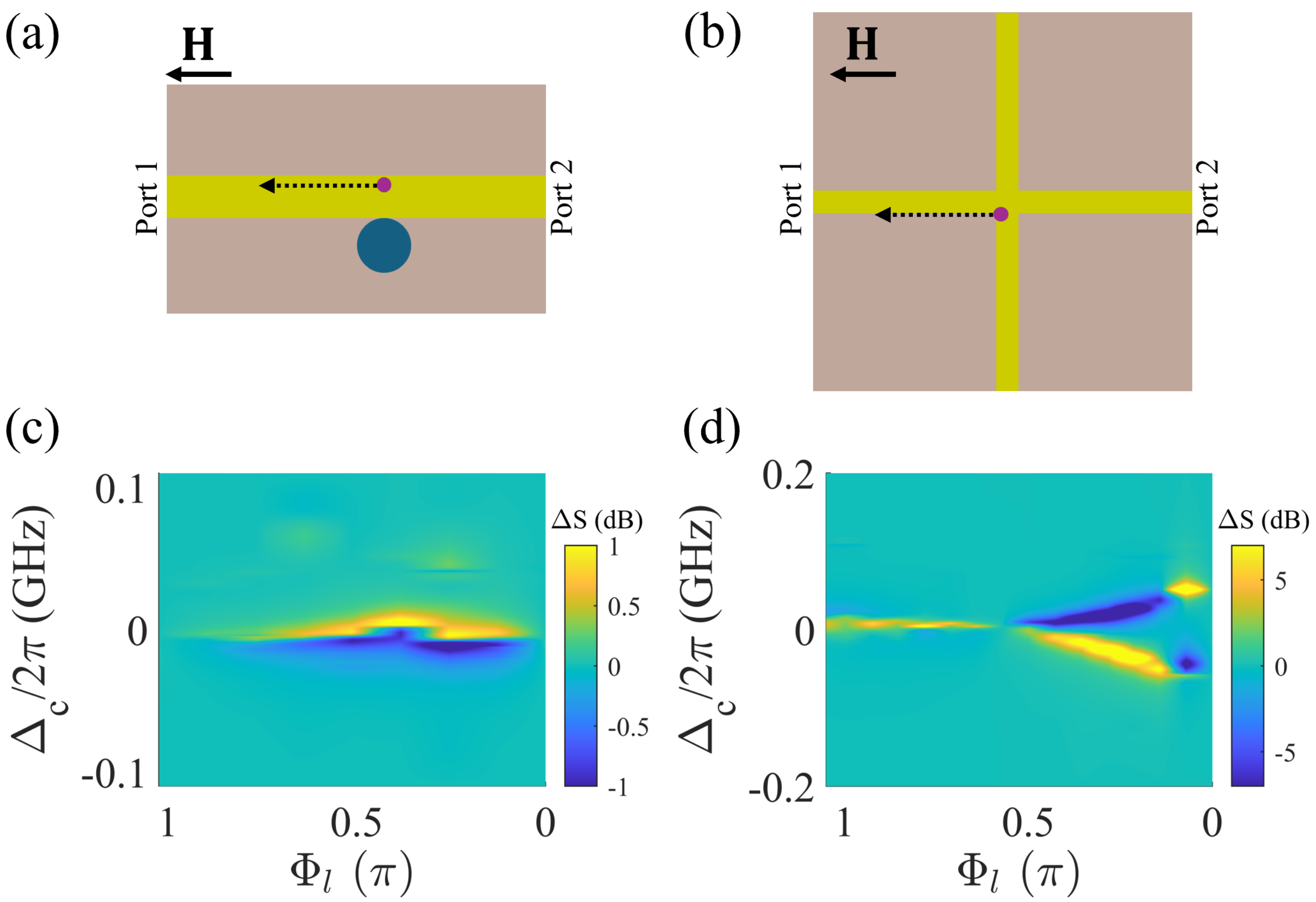}
	\caption{Simulation schematic of coupled (a) YIG-DR system and (b) YIG-cross-line system. The YIG is set at one side of microstrip. The DR cavity is set on a 1.575-mm-thick RT5880 microstrip substrate, whereas the cross-line cavity is implemented on a 0.813-mm-thick RO4003C substrate. (c)(d) At $\Delta_m=0$, the simulated $\Delta S=|S_{21}|-|S_{12}|$ mapping as functions of traveling phase $\Phi_l$ and $\Delta_c$ in (c) YIG-DR system and (d) YIG-cross-line system.}
	\label{fig17}
\end{figure}
In this section, we examine how the synthetic-chirality-induced nonreciprocity depends on the distance between the YIG sphere and the cavity. According to Eqs.~\ref{eq20} and \ref{eq21} in the main text, the nonreciprocity vanishes at $\Phi_l=n\pi$ for a magnetic-dipole cavity, and at $\Phi_l=(n+\tfrac{1}{2})\pi$ for an electric-dipole cavity.

To verify this prediction, we investigate coupled cavity–magnon systems including both the direct coupling and the traveling-wave-mediated interaction using CST MWS simulations, in a coupled YIG-DR (magnetic-dipole) system and a YIG-cross-line (electric-dipole) system. The DR cavity is set on a 1.575-mm-thick RT5880 microstrip substrate, whereas the cross-line cavity is implemented on a 0.813-mm-thick RO4003C substrate. The YIG sphere is positioned on the side of the microstrip to induce nonreciprocity while varying the YIG–cavity separation along the dashed arrows in Figs.~\ref{fig17}(a) and (b).

To better display the evolution, we define $\Delta S = |S_{21}| - |S_{12}|$ to express nonreciprocity at $\Delta_m=0$.
Figure~\ref{fig17}(c) shows the simulated $\Delta S$ mapping as functions of traveling phase $\Phi_l$ and $\Delta_c$. The DR–YIG system exhibits vanishing nonreciprocity at $\Phi_l=0$ and $\pi$, while the cross-line cavity shows suppression near $\Phi_l=\pi/2$, consistent with the expected magnetic--dipole cavity and electric-dipole cavity behaviors. The discontinuity near $\Phi_l = 0$ in the cross-line cavity may arise from the sharp variation of the cavity field.        

 \section{Calibration of the parameters of YIG and DR} 
 \label{Calibration of the parameters of YIG and DR}

\begin{figure*}[bht]
	\centering 
	\includegraphics[width=0.9\textwidth]{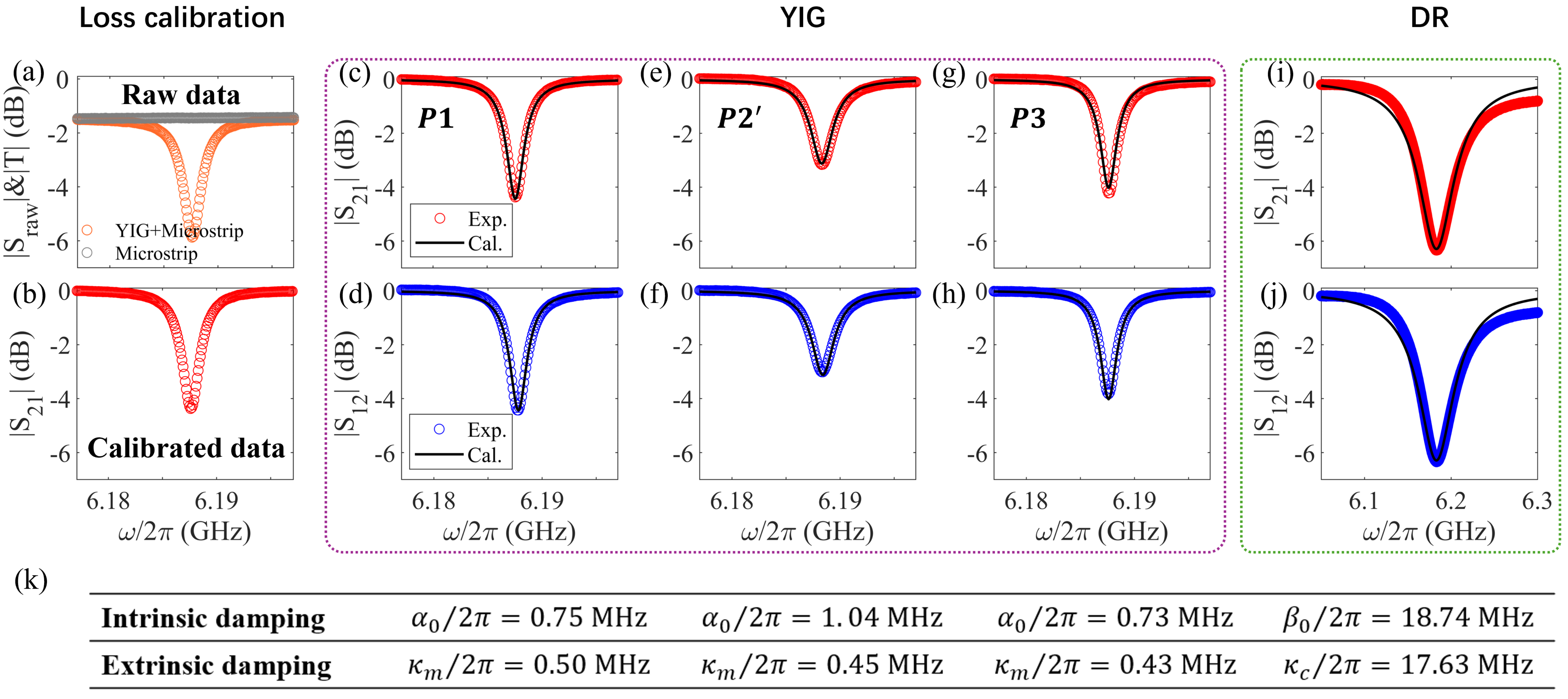}
	\caption{(a)(b)The microstrip loss calibration. (a) The orange circles represent the measured raw transmission amplitude $|S_{\text{raw}}|$ including the contribution from both the resonator system and the microstrip. The gray circles are the measured transmission amplitude $T$ of the microstrip with $|T|=0.84$ $(-1.5$ dB). (b) The calibrated transmission amplitude of the resonator system. (c)-(j) The measured (c), (e), (g), (i) $|S_{21}|$ and (d), (f), (h), (j) $|S_{12}|$ spectra for the side-coupled YIG sphere at (c), (d) $P1$, (e), (f) $P2'$, (g), (h) $P3$, and for (i), (j) the side-coupled DR. (k) The intrinsic and extrinsic damping rates of the YIG at $P1$, $P2'$, $P3$, and that of the DR, used to calculate the black curves in panels (c)-(j).}
	\label{fig18}
\end{figure*}

In this section, we present the calibration of the DR and YIG-sphere parameters, which are extracted from the DR-microstrip and YIG-microstrip side-coupled experiments, respectively. Our measured raw data include transmission coefficient $T$ of the microstrip, which is lossy with $|T|=0.84$ ($-1.5$ dB), shown by the gray circles in Fig.~\ref{fig18}(a), causing a reduced background in the transmission of the measured system. Therefore, we calibrate the measured resonator response by scaling the transmission coefficient amplitude of the microstrip using $|S|=|S_{\text{raw}}|/|T|$ throughout this paper, where $|S|$ corresponds to the transmission of the resonator alone, while $|S_{\text{raw}}|$ is the raw data including the microstrip and the resonator system. An example is shown in Figs.~\ref{fig18}(a) and (b).

The fitting formula is based on Eq.~\ref{eq5} in the main text. For convenience, it is reorganized as:
\begin{equation}\label{eqJ1}
 	\begin{split}
 		S_{21(12)}(\omega)&=\frac{\omega-\omega_m+i\alpha_{21(12)}}
 		{\omega-\omega_m+i(\alpha_{21(12)}+\kappa_{mp(q)})}, 
 	\end{split}
 \end{equation}
where $\alpha_{21(12)}=\alpha_0+\frac{\kappa_{mq(p)}-\kappa_{mp(q)}}{2}$. We obtain the fitted parameters $\alpha_{21(12)}$ and $\kappa_{mp(q)}$ by fitting $|S_{21(12)}|$ using Eq.~\ref{eqJ1}, and then calculate the intrinsic damping $\alpha_0$ as the average of $\alpha_0=\alpha_{21}-\frac{\kappa_{mq}-\kappa_{mp}}{2}$ and $\alpha_0=\alpha_{12}-\frac{\kappa_{mp}-\kappa_{mq}}{2}$. 

Following this fitting process, the fitted parameters in Fig.~\ref{fig3} in the main text are listed in Table~\ref{tableG1}. Here, the extrinsic damping rates depend on the local effective-microwave-field distribution, which varies with position. 

In addition, the circles in Figs.~\ref{fig18}(c)-(h) show the YIG spectra at $P1$, $P2'$, $P3$ in the coupling experiment in Sec.~\ref{Traveling wave-mediated nonreciprocal coupling A} of the main text. The black curves are calculated using Eq.~\ref{eq5} in the main text or Eq.~\ref{eqJ1} (with $\kappa_m=\kappa_{mq}=\kappa_{mp}$ due to linearly polarization) using the parameters listed in Fig.~\ref{fig18}(k). These parameters are also used to calculate the curves in Figs.~\ref{fig6}(e), (f) and (i), (j), and the transmission mappings in Fig.~\ref{fig7} in the main text. DR is set at a position supporting a standing wave and fitted using Eq.~\ref{eqJ1} by substituting $\alpha_0$ and $\kappa_m$ with $\beta_0$ and $\kappa_c$.
Here, we note that the intrinsic damping of the YIG sphere varies slightly with its position and the direction of the applied field. The position-related discrepancy may result from the traveling-wave modes of the YIG sphere that are not well captured by our present model. We point out that this traveling-wave mode may cause the discrepancy of the extrinsic damping rates into the right- and left-going traveling waves, thus the intrinsic damping deviates, but will not cause nonreciprocity \cite{yao2025nonreciprocal,yang2024anomalous}. An extreme case is the typical traveling-wave modes like a whispering-galley-mode, where the mode only dissipates into one direction, while remaining reciprocal at the same time. This is because the extrinsic damping rates into the two directions switch in the traveling-wave case while do not in the chiral interaction. In summary, both the chiral interaction and the traveling-wave resonator are characterized by unequal extrinsic damping, causing a difficulty in accurately describing the system with both effects involved, and potentially leading to inaccuracy in the fitting of damping rates. This would impact the fitted
coupling strength as \onecolumngrid
\vspace{10pt}
\begin{minipage}{\textwidth}
	\centering
	\makeatletter
	\def\@captype{table}
	\makeatother
	\caption{Fitting parameters of the magnon mode in Fig.~\ref{fig2} under different applied field directions at positions $P1$, $P2$, $P3$.} 
	\label{tableG1} 
	\vspace{5pt}
	\begin{tabularx}{\textwidth}{c*{9}{>{\centering\arraybackslash}X}}
		\hline\hline 
		& \multicolumn{3}{c}{H along microstrip} & \multicolumn{3}{c}{H perpendicular across microstrip} & \multicolumn{3}{c}{H normal to plane} \\
		\cline{2-10}
		& $\alpha_0/2\pi$ (MHz) & $\kappa_{mp}/2\pi$ (MHz) & $\kappa_{mq}/2\pi$ (MHz) & $\alpha_0/2\pi$ (MHz) & $\kappa_{mp}/2\pi$ (MHz)& $\kappa_{mq}/2\pi$ (MHz)& $\alpha_0/2\pi$ (MHz)& $\kappa_{mp}/2\pi$ (MHz)& $\kappa_{mq}/2\pi$ (MHz)\\
		\hline 
		$P1$ & 0.75 & 0.50 & 0.50 & 0.99 & 0.53 & 0.93 & 1.16 & 0 & 0.15 \\ 
		$P2$ & 0.97 & 0.33 & 0.33  & / & / & / & 1.28 & 0.30 & 0.30 \\ 
		$P3$ & 0.73 & 0.44 & 0.41 & 0.87 & 0.40 & 0.80 & 1.07 & 0.13 & 0 \\ 
		\hline\hline 
	\end{tabularx}
\end{minipage}
\vspace{10pt}
\twocolumngrid 
\noindent  well, but would not impact the nonreciprocity resulting from the chiral interaction. How to precisely disentangle the contributions of the two remains a challenge worthy of further investigation. Moreover, the applied-field-direction-related discrepancy may result from magnetocrystalline anisotropy \cite{zeng2020intrinsic}, which needs to be carefully verified in more accurate experiments.

\end{document}